\documentclass[useAMS,usenatbib]{mn2e}

\usepackage{graphicx}

\usepackage{txfonts}
\usepackage{natbib}
\usepackage{longtable,lscape}
\usepackage{multirow}
\usepackage{subfig}
\usepackage{rotating}

\newsavebox{\bigleftbox}
\newcommand\aj{AJ}
\newcommand\apj{ApJ}

\newcommand\aap{A\&A}
\newcommand\mnras{MNRAS}
\newcommand\apjl{ApJ}
\newcommand\pasp{PASP}
\newcommand\pasa{PASA}

\newcommand\araa{ARA\&A}

\def\x{\mbox{$\Delta_{\tiny{\mathrm{F275W,F814W}}}$}}
\def\y{\mbox{$\Delta_{\tiny{C~\mathrm{ F275W,F336W,F438W}}}$}}
\def\col{\mbox{${\mathrm{(F275W-F814W)}}$}}
\def\dx{\mbox{$\delta_{\tiny{\mathrm{F275W,F814W}}}$}}
\def\dy{\mbox{$\delta_{\tiny{C~\mathrm{ F275W,F336W,F438W}}}$}}

\title[]{The {\it Hubble Space Telescope} UV Legacy Survey of Galactic
  Globular Clusters. XIX. A Chemical Tagging of the Multiple Stellar
  Populations Over the Chromosome Maps}
\author[Marino et al.]
       {A.\,F.\,Marino$^{1,2}$,
         A.\,P.\,Milone$^{1}$,
         A.\,Renzini$^{3}$,
         F.\,D'Antona$^{4}$,
         J.\,Anderson$^{5}$,
         L.\,R.\,Bedin$^{3}$,\newauthor 
         A.\,Bellini$^{5}$,
         G.\,Cordoni$^{1}$,
         E.P.\,Lagioia$^{1}$,
         G.\,Piotto$^{1}$,
         M.\,Tailo$^{1}$
\\
$^{1}$Dipartimento di Fisica e Astronomia ``Galileo Galilei'', Univ. di Padova, Vicolo dell'Osservatorio 3, Padova, IT-35122\\
$^{2}$Centro di Ateneo di Studi e Attivita' Spaziali ``Giuseppe Colombo'' - CISAS, Via Venezia 15, Padova, IT-35131\\
$^{3}$Istituto Nazionale di Astrofisica - Osservatorio Astronomico di Padova, Vicolo dell'Osservatorio 5, Padova, IT-35122\\
$^{4}$Istituto Nazionale di Astrofisica - Osservatorio Astronomico di Roma, Via Frascati 33, I-00040 Monteporzio Catone, Roma, Italy\\
$^{5}$Space Telescope Science Institute, 3800 San Martin Drive, Baltimore,  MD 21218, USA\\
       }

\begin{document}
%\date{Accepted xxx December 15. Received xxx December 14; in original form xx October 11}
\date{Draft Version April, 9, 2019}

\pagerange{\pageref{firstpage}--\pageref{lastpage}} \pubyear{2019}

\maketitle
\label{firstpage}
 
\begin{abstract}
The {\it Hubble~Space~Telescope} UV Legacy Survey of Galactic Globular
Clusters (GCs) has investigated GCs and their stellar populations. In
previous papers of this series we have introduced a pseudo two-colour
diagram, or ``chromosome map'' (ChM) that maximises the separation
between the multiple 
populations. We have identified two main classes of GCs: Type~I,
including $\sim$83\% of the objects, and Type~II clusters. Both
classes host two main groups of stars, referred to in this series as
first (1G) and second generation (2G). Type~II clusters host
more-complex ChMs, exhibiting two or more parallel sequences of 1G and
2G stars. 
We exploit spectroscopic elemental abundances from literature to
assign the chemical composition to the distinct populations as
identified on the ChMs of 29 GCs. We find that stars in different
regions of the ChM have different composition: 1G stars share the same
light-element content as field stars, while 2G stars are enhanced in
N, Na and depleted in O. Stars with enhanced Al, as well as
stars with depleted Mg populate the extreme regions of the ChM. We
investigate the intriguing colour spread among 1G stars observed in
many Type~I GCs, and find no evidence for internal variations in light
elements among these stars, 
whereas either a $\sim 0.1$~dex iron spread or a variation in He among
1G stars remain to be verified. 
In the attempt of analysing the global properties of the multiple
populations phenomenon, we have constructed a universal ChM,
which highlights that, though very variegate, the phenomenon has
some common pattern among all the analysed GCs. The universal ChM
reveals a tight connection with Na abundances, for which we have
provided an empirical relation. 
The additional ChM sequences observed in Type~II GCs, are enhanced in
metallicity and, in some cases, $s$-process elements. Omega~Centauri can be
classified as an extreme Type~II GC, with a ChM displaying
three main extended ``streams'', each with its own variations in
chemical abundances. One of the most noticeable differences is found
between the lower and upper streams, with the latter, associated
with higher He, being also shifted towards higher Fe and lower Li
abundances. We publicly release the ChMs.
\end{abstract}


\section{Introduction}\label{sec:intro}

Multiple stellar populations in Milky Way globular clusters (GCs) are
now considered a rule. 
Spectroscopically, we know since a long time that GCs' stars are not
chemically homogeneous (e.g.\,Kraft 1994; Carretta et al.\,2009 and
references therein). 
The chemical abundances in light elements obey to specific patterns,
such as the O-Na and the C-N anticorrelations, and in some GCs a Mg-Al
anticorrelation is also observed (e.g.\,Ivans et al.\,1999; Yong et
al.\,2003; Marino et al.\,2008).  
These behaviors are interpreted as due to nucleosynthetic processes,
specifically by proton captures at high temperatures, occurring in a first
generation (1G) of stars (e.g. Ventura et al.\,2001; Decressin et
al.\,2007; Krause et al.\,2013; Denissenkov \& Hartwick\,2014), and
can be used as constraints to the mass of the polluters
(e.g.\,Prantzos, Charbonnel \& Iliadis\,2017). 
The most acknowledged scenarios predict that second generations
(2G) of stars form from material processed in 1G polluters, so filling
the Na-N-enhanced and O-C-depleted regions of the common
(anti)correlations abundance plots. 
However, the nature (e.g., mass range) of 1G stars where these
processes had taken place remains 
largely unsettled, as it is the sequence of events leading to the
formation of 2G stars (see e.g., Renzini et al.\,2015, hereafter
Paper~V, for a critical discussion).  

More recent observations have revealed that the multiple stellar
populations phenomenon in GCs is far more complex than previously
imagined. In particular, our {\it Hubble Space Telescope} ($HST$) UV
Legacy Survey of Galactic Globular Clusters, whose data are used in
the present work, has revealed and astonishing variety from cluster to
cluster of the multiple stellar populations  phenomenon, as documented
for 58 GCs (Piotto et al.\,2015, hereafter Paper~I; Milone et
al.\,2015a,b, 2017, 2018 hereafter Papers~II, III, IX, XVI).   

One very effective way of visualising the complexity of multiple
stellar populations is represented by a colour-pseudocolour plot that
we have dubbed {\it chromosome map} (ChM), with such maps having been
presented for 58 GCs in Paper~IX. 
The ChM plot, \x\ vs.\,\y\ (see Paper~II for a detailed discussion on
how to construct these maps),  owes its power to the capability of
maximising the photometric separation of different stellar populations
with even slightly different chemical abundances (Paper~II). 
In such plots red giant branch (RGB)  stars  are typically separated into two
distinct groups: one first-generation (1G) group and a
second-generation (2G) one (Paper~IX). Specifically, 1G stars are
located around the origin 
of the ChM (i.e., \x=\y=0), while 2G stars have large \y\ and low \x.  
Experiments based on synthetic spectra suggest that N abundance
variations, through the impact of molecular bands on the \y\ index,
are the main responsible of the ChM pattern. So far, this 
has been confirmed by spectroscopy in just a couple of GCs (Papers~III-IX).

The variety of the multiple populations phenomenon 
has prompted  efforts to classify GCs in different classes,
whose definition has changed from time to time, depending on the new features
being discovered. For example, in at least ten clusters over the
analysed 58, i.e.\,the 17 per cent of the 
entire sample, the 1G and/or the 2G sequences appear to be split,
indicating a much more complex distribution of chemical abundances  
compared to the majority of GCs. In Paper~IX these
clusters have been indicated as Type~II GCs, while the other, more
common, GCs were classified as Type~I.
All the GCs with known variations in heavy elements, including
iron, belong to this (photometrically defined) class of
objects. $\omega$~Centauri, with its 
well-documented large variations in the overall metallicity, is the
extreme example of a Type~II GC.
These GCs  were  early defined  as ``anomalous''
(Marino et al.\,2009, 2011a,b; 2015, see also Table~7 in Marino et
al.\,2018) as they display variations in the 
overall metallicity and/or in the  $slow$-neutron capture ($s$) elements 
(e.g.\,Marino et al.\,2009, 2015; Yong et al.\,2014; Johnson et
al.\,2015; Carretta et al.\,2011), suggesting that they
have experienced a much more
complex star formation history than the typical Milky Way GCs
(e.g.\,Bekki \& Tsujimoto\,2016; D'Antona et al.\,2016). 
The reason why this different behavior occurs remains to be
established, but, intriguingly,  
might be linked to a different origin with respect to the more common
GCs in the Galaxy (see discussion in Marino et al.\,2015).

Another surprising finding in Paper~IX is that the 1G group itself of
many Type~I GCs displays a spread of \x\ in the ChM that is not
consistent with 1G stars being a simple stellar population, i.e., they
don't make a chemical homogeneous sample.  
The 1G stars span a quite large range in \x, while sharing
approximately the same \y. The cause of this odd behavior has not 
been understood yet. Given that the \y\ axis is constant among these 1G
stars, they likely share the same content in N (and likely in
other light elements). Their different \col\ colour suggests a
difference in the stellar structure itself, i.e., in effective
temperature, rather than in 
the atmospheric abundances. Star-to-star variations in helium within
the 1G would be capable of producing such a \x\ spread without
affecting much \y\ and in Paper~XVI we consider various scenarios that
could have led to such helium enrichment, but none appears to work.   

Clearly, two major challenges in the understanding of the
multiple stellar populations pattern as displayed on the ChMs is the
presence of additional sequences at redder \x, in Type~II GCs, and
the large spread of \x\ towards bluer colours exhibited by 1G stars
in many GCs, both 
Type~I and Type~II. To start attacking these questions we here 
correlate the detailed chemical composition of stars (as from
high-resolution spectroscopy)  
with their location on the various sequences in the ChMs.
Relating spectroscopic abundances to photometric properties of GC
multiple populations was first pioneered  in Marino et al.\,(2008), 
where it was shown that the light element patterns, such as the O-Na and C-N
anticorrelations, are linked to different UV colours, due to the
CH/CN/NH/OH molecules in the UV spectrum (see also Paper~III). 

In this paper we exploit the  photometric database from the
$HST$ Legacy Survey supplemented by ground based photometry, and
couple it to  chemical 
abundances of the same stars as mined from the literature.  Thus, we
explore, for the first time for a large sample of GCs, the 
chemical properties of their stellar multiple populations as
photometrically revealed by ChMs. 
The paper is organised as follows.
In Section~\ref{sec:data} we describe the photometric and
spectroscopic dataset and use ChMs of GCs to identify the
distinct stellar populations; the chemical composition of the stellar
populations is derived and discussed in Section~\ref{sec:chemI},
separately for Type I and Type II GCs; 
Section~\ref{sec:omegacen} is specifically devoted to $\omega$~Centauri;
Section~\ref{sec:conclusions} is a final discussion and summary of the
results.  

\section{Data and data analysis}\label{sec:data}

In order to investigate the chemical composition of multiple stellar
populations in GCs, we combined multi-wavelength $HST$
photometry with spectroscopy. In addition, we used wide-field
ground-based photometry of four GCs, namely NGC\,1851, NGC\,5286,
NGC\,6656, NGC\,7089. 
The spectroscopic dataset is described in Section~\ref{sub:spec},
while Sections~\ref{sub:photHST} and ~\ref{sub:photGB} are devoted
to $HST$ and ground-based photometry, respectively. 

%%%%%%%%%%%%%%%%%%%%%%%%%%%%%%%%%%%%%%%%%%%%%%%%%%%%%%%%%%%%%%%%%%%%%%%%%%%%%%%
\begin{centering}
\begin{figure*}
\includegraphics[width=18cm]{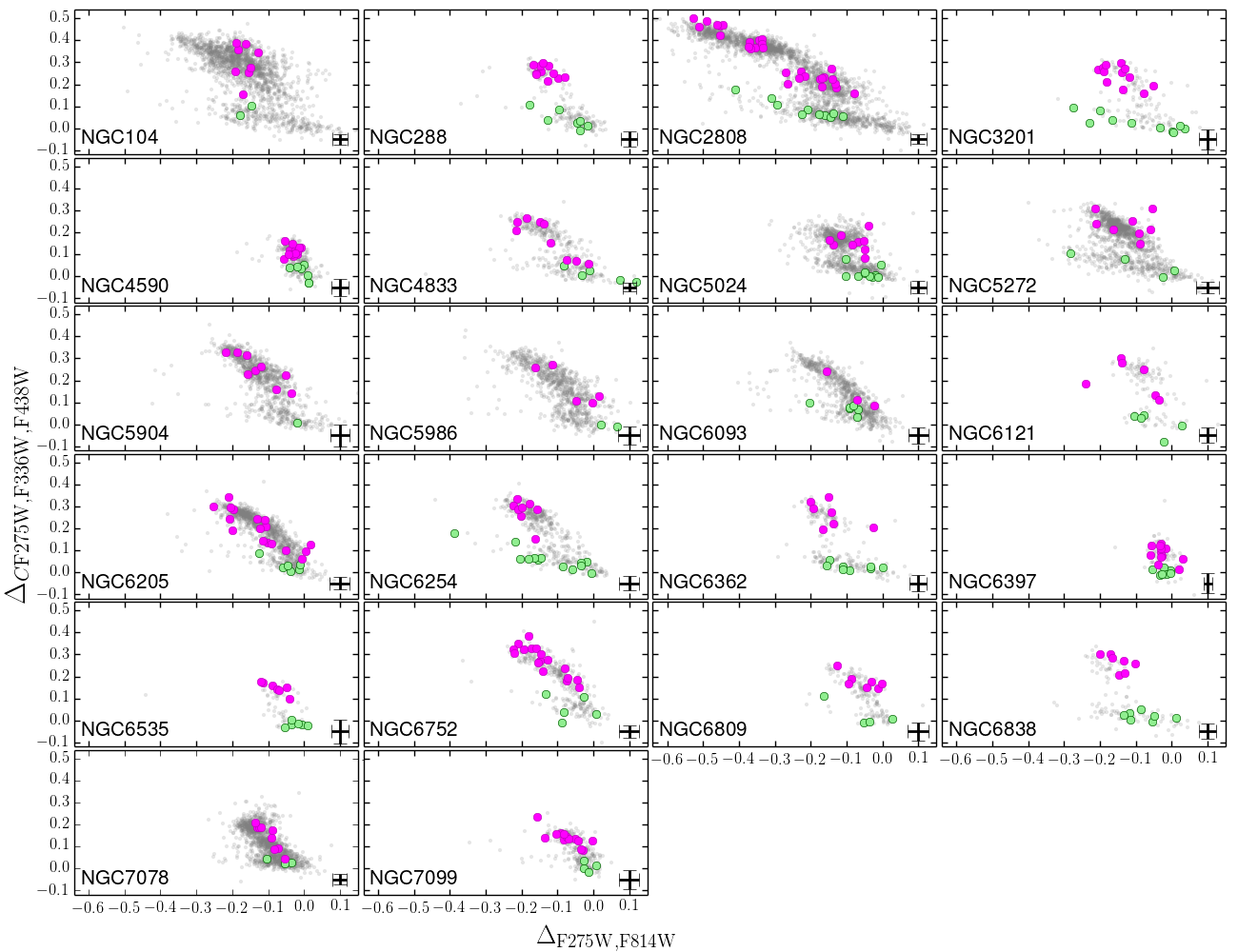}
\caption{Reproduction of the ChM diagrams from Paper~IX of the 22
    analysed Type~I GCs. 1G and 2G stars with available spectroscopy
    are marked with large green and magenta dots, respectively. Error bars
    are from Paper~IX.} 
\label{fig:maps}
\end{figure*}
\end{centering}
%%%%%%%%%%%%%%%%%%%%%%%%%%%%%%%%%%%%%%%%%%%%%%%%%%%%%%%%%%%%%%%%%%%%%%%%%%%%%%%

%%%%%%%%%%%%%%%%%%%%%%%%%%%%%%%%%%%%%%%%%%%%%%%%%%%%%%%%%%%%%%%%%%%%%%%%%%%%%
\begin{centering}
\begin{figure*}
 \includegraphics[width=17cm]{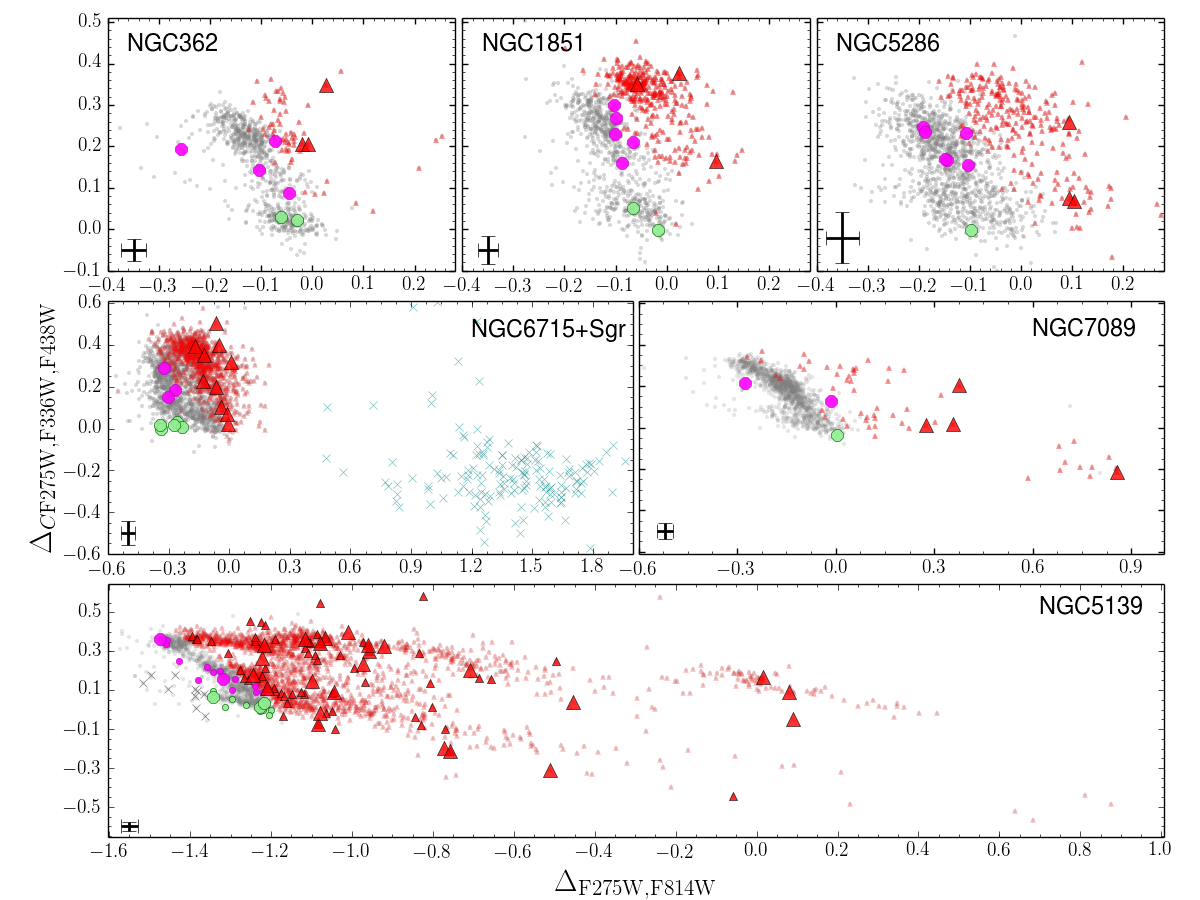}
   \caption{ChMs of the Type~II GCs NGC\,362, NGC\,1851, NGC\,5286,
     NGC\,6715 (M\,54), NGC\,7089 (M\,2) and NGC\,5139 ($\omega$\,Centauri)
     from Paper~IX. Red-RGB stars have been represented as red
     triangles, while green and magenta indicate the blue-RGB 1G and
     2G stars, respectively. In the ChM plane of M\,54, field stars in
     the Sagittarius dward (Sgr) are plotted as azure crosses. 
     For $\omega$~Centauri we have plotted both the targets 
     from Marino et al.\,(2011, big symbols) and Johnson \&
     Pilachowski\,(2010, small symbols); and, in addition of blue-RGB 1G
     and 2G stars, it is represented a third group of blue 1G stars with
     more negative \x\ that we have represented with grey crosses. The
     stars represented in grey crosses have not been considered in the
     abundance analysis of Section~\ref{sec:chemI}, and will be analysed in
     Section~\ref{sec:omegacen}. Error bars are from Paper~IX.} 
 \label{fig:mapA}
\end{figure*}
\end{centering}
%%%%%%%%%%%%%%%%%%%%%%%%%%%%%%%%%%%%%%%%%%%%%%%%%%%%%%%%%%%%%%%%%%%%%%%%%%%%%%%

\subsection{The spectroscopic dataset}\label{sub:spec}

In this work we investigate eleven chemical species, namely Li, N,
O, Na, Mg, Al, Si, K, Ca, Fe, and Ba, that are among the most-commonly
analysed in GCs studies. Each element has been taken into account only
when both photometry and high-resolution spectroscopy is available for
at least five stars in a given cluster.  Elemental abundances have
been taken from literature as  listed in Table~\ref{tab:data} which
provides, for each cluster, the number of stars for which the chemical
abundances of the various species were measured. 

\subsection{The $HST$ photometric dataset}\label{sub:photHST}

The multiple stellar populations along the RGB have been identified by
using the photometric catalogs published in Papers~I and IX as part of
the $HST$ UVIS Survey of Galactic GCs.  
Photometry has been derived from $HST$ images collected with the
Wide-Field Channel of the Advanced Camera for Surveys (WFC/ACS) and
the Ultraviolet and Visual Channel of the Wide-Field Camera 3
(UVIS/WFC3) as part of programs GO-11233, GO-12605, and GO-13297
(PI.\,G.\,Piotto, Paper\,I), and GO-10775 (PI.\, A.\,Sarajedini,
Sarajedini et al.\,2007; Anderson et al.\,2008), and from additional
archive data (Paper\,I and IX). 
Photometry has been corrected for differential reddening as in Milone
et al.\,(2012) and only stars that, according to their proper
motions, are cluster members have been included in our study.  
The analysed $HST$ images cover 
$\sim$2.7$\times$2.7 square arcmin over the clusters central
regions (see Papers~I and IX for details).  

For sake of clarity, we list here the definitions as from Paper~IX
that will be used along this paper to refer to the various stellar
populations as appearing on the ChMs:
\begin{itemize}
 \item{1G stars and 2G stars are separated by a line cutting through
     the minimum stellar density on the ChM. Thus, 1G stars are those
     located at lower \y and  will be represented by green dots if
     spectroscopic abundances are available;}  
\item{2G stars are located on bluer \x\ and higher \y, and will be
    represented by magenta dots if spectroscopic abundances are
    available.} 
\end{itemize} 
These definitions will be used both for Type~I and Type~II GCs.
In addition, for Type~II GCs only, we also define:
\begin{itemize}
\item{blue-RGB stars, those defining the main ChM 1G and 2G sequences,
    as observed in Type~I GCs; }  
\item{red-RGB stars are the objects located on redder ChM sequences,
    only found in Type~II GCs. Red-RGB stars for which spectroscopy is
    available will be represented as red triangles along the paper. }   
\end{itemize}

A note on the adopted naming scheme. We decided to use here the {\it
  pure} photometric classification of above. This is because of the
variety of the multiple stellar populations phenomenon, as seen in the ChMs.
As an example, from the chemical point of view, the presence of stellar populations
with different metallicity and $s$-process elements content is
the main distinctive feature of the Type~II GCs, with available
spectroscopy. However, this phenomenon looks variegate as well, as some
``red-RGB'' sub-populations have been observed not to be enhanced in
$s$-elements (e.g.\ in M\,2, Yong et al.\,2014, and 
NGC\,6934, Marino et al.\,2018). 
Furthermore, many Type~II GCs have a very tiny red component,
that has not yet been investigated in terms of chemical
abundances. So, at this stage, we prefer to keep our general naming
scheme introduced in Paper~IX for
ChMs, Type~I and II GCs, rather than adopting a more informative
nomenclature based e.g.\, on chemical abundances. 
This is to prevent assigning to all Type~II GCs properties that they might not have.
In the future, we may not exclude to adopt a more informative naming scheme. 

Figure~\ref{fig:maps} reproduces the ChMs of 22 Type\,I
GCs from Paper\,IX, where we have represented with green and magenta
dots all the 1G and 2G stars, respectively,  for which spectroscopy is
available.  
A collection of ChMs in six Type\,II GCs, namely NGC\,362,
NGC\,1851, NGC\,5286, NGC\,6715, NGC\,7089, and NGC\,5139
($\omega$~Centauri), is shown in Figure~\ref{fig:mapA}.
In these clusters, besides the common 1G and 2G groups (coloured green
and magenta in analogy with what done in Type~I GCs of
Figure~\ref{fig:maps}), we have represented in red the 
RGB stars located on the additional sequences, redder than the common
sequence formed by 1G and 2G stars.

The distinction between blue- and red-RGB
stars in Type~II GCs has also been made on a {\it classic}
colour-magnitude diagram (CMD), either the $m_{\rm F336W}$
vs.\,$m_{\rm F336W}-m_{\rm F814W}$ plot or using the $U$ vs.\,$U-I$
plot from ground-based photometry, as we will discuss in detail in
Section~\ref{sub:photGB}. 
The colours and symbols used to distinguish 1G, 2G, and red-RGB stars
introduced in Figures~\ref{fig:maps} and \ref{fig:mapA} will be used
consistently hereafter.  

As shown in the lower panel of Figure~\ref{fig:mapA},
$\omega$\,Centauri exhibits the most-complex ChM among Type~II GCs. In
particular, we note two main streams of red-RGB stars with small and
high values of \y\ that define the lower and upper envelope of the
map. These features will be discussed in detail in
Section~\ref{sec:omegacen}. 

%%%%%%%%%%%%%%%%%%%%%%%%%%%%%%%%%%%%% FIG 3 %%%%%%%%%%%%%%%%%%%%%%%%%%%%%%%%%%%
\begin{centering}
\begin{figure*}
 \includegraphics[height=7cm]{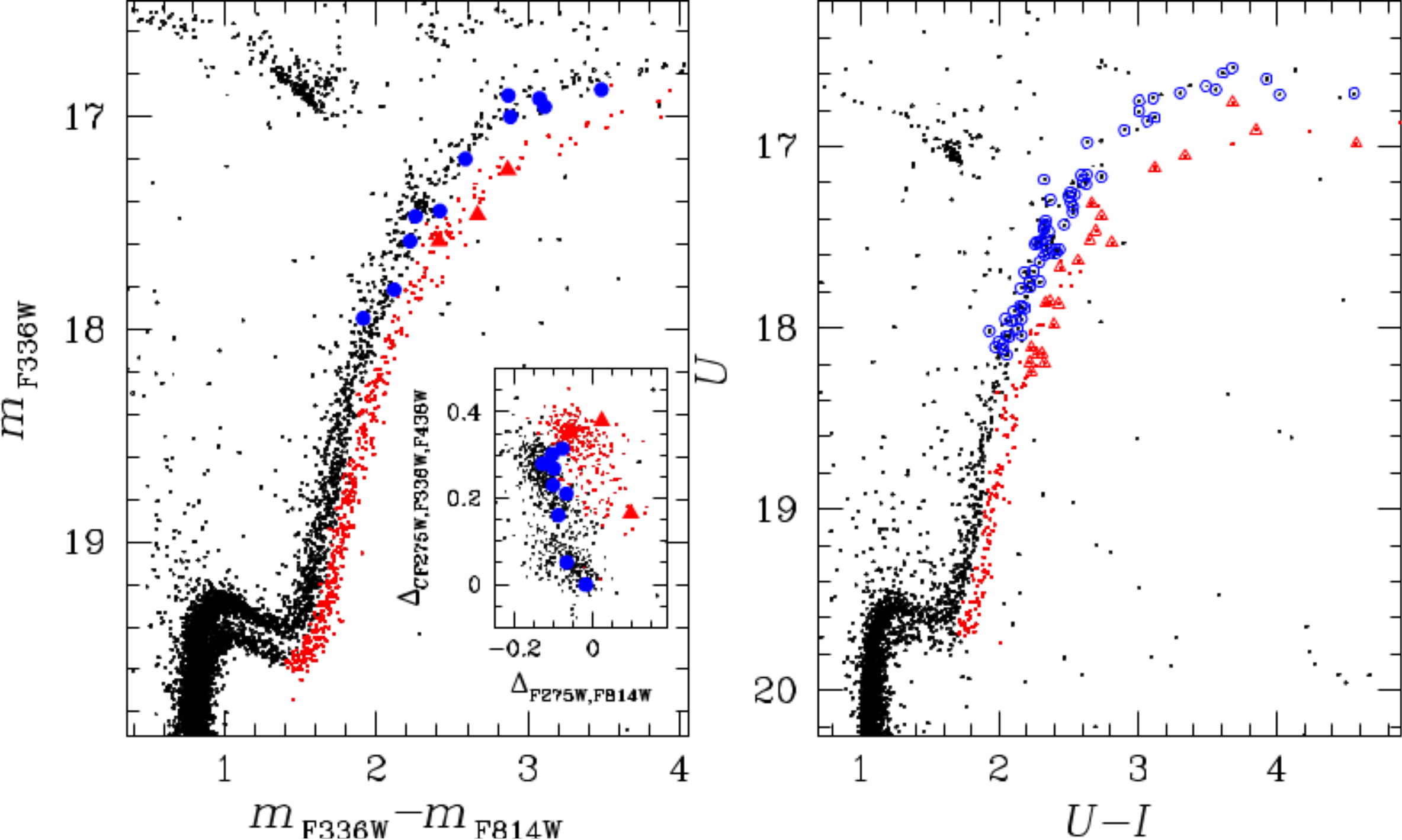}
 \includegraphics[height=7cm]{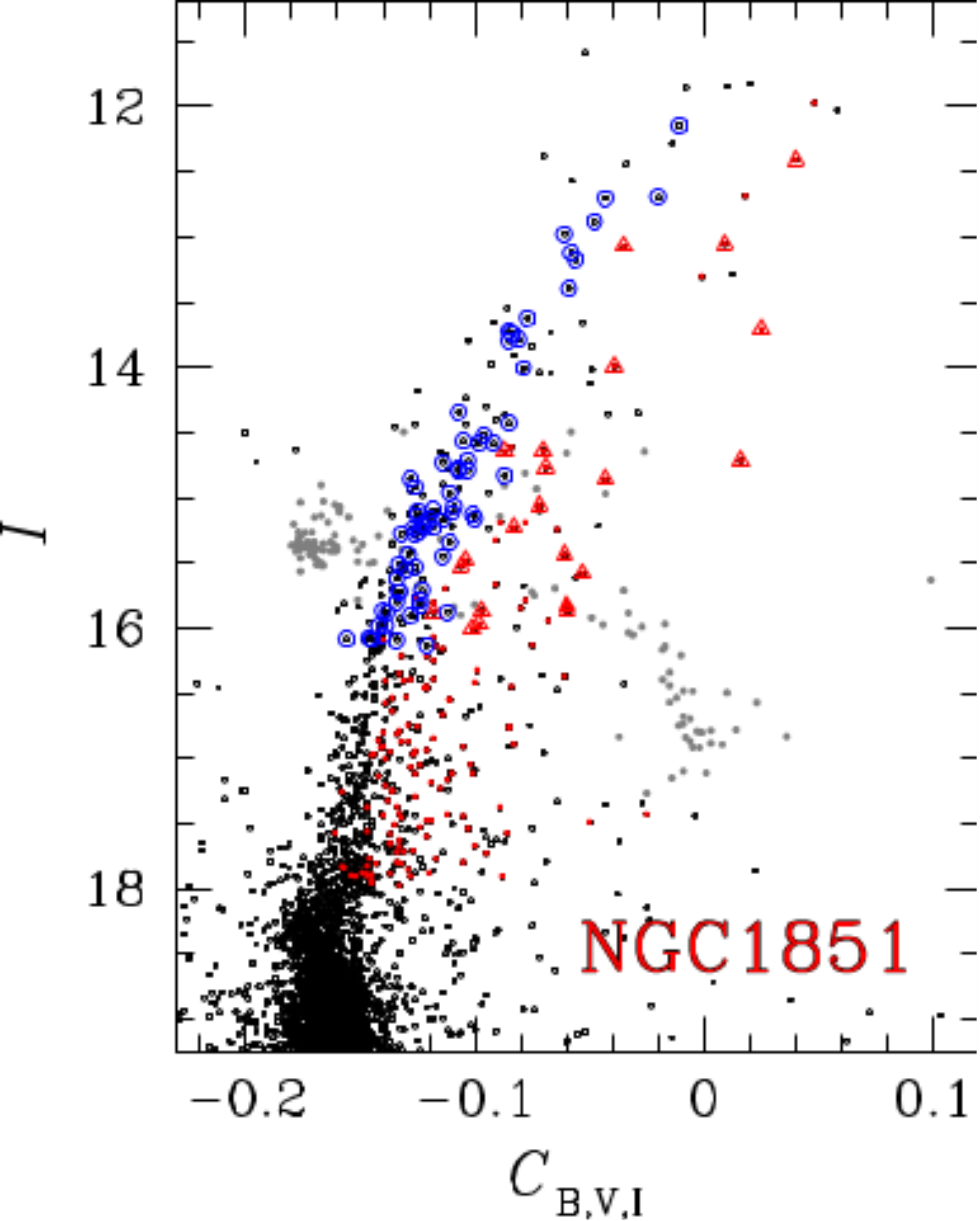}
 \caption{$m_{\rm F336W}$ vs.\,$m_{\rm F336W}-m_{\rm F814W}$ CMD of
   NGC\,1851 from $HST$ photometry (left panel). The ChM is shown in
   the left-panel inset. Middle and right panels show the $U$
   vs.\,$U-I$ CMD, and $I$ vs.\,$C_{\rm B,V,I}$ pseudo-CMD of
   NGC\,1851 from ground-based photometry, respectively. 
   Stars for which chemical abundances and $HST$ photometry with ChMs are
   available have been plotted with filled symbols: blue dots for
   blue-RGB stars and red triangles for the red-RGB ones. Stars with
   available abundances in the ground-based photometry are hereafter
   plotted with open symbols: blue open circles and red open triangles
   for blue- and red-RGBs, respectively.  
 } 
\label{fig:1851}
\end{figure*}
\end{centering}
%%%%%%%%%%%%%%%%%%%%%%%%%%%%%%%%%%%%%%%%%%%%%%%%%%%%%%%%%%%%%%%%%%%%%%%%%%%%%%%

%%%%%%%%%%%%%%%%%%%%%%%%%%%%%%%%%%%%% FIG 4 %%%%%%%%%%%%%%%%%%%%%%%%%%%%%%%%%%%
\begin{centering}
\begin{figure*}
  \includegraphics[width=5.6cm]{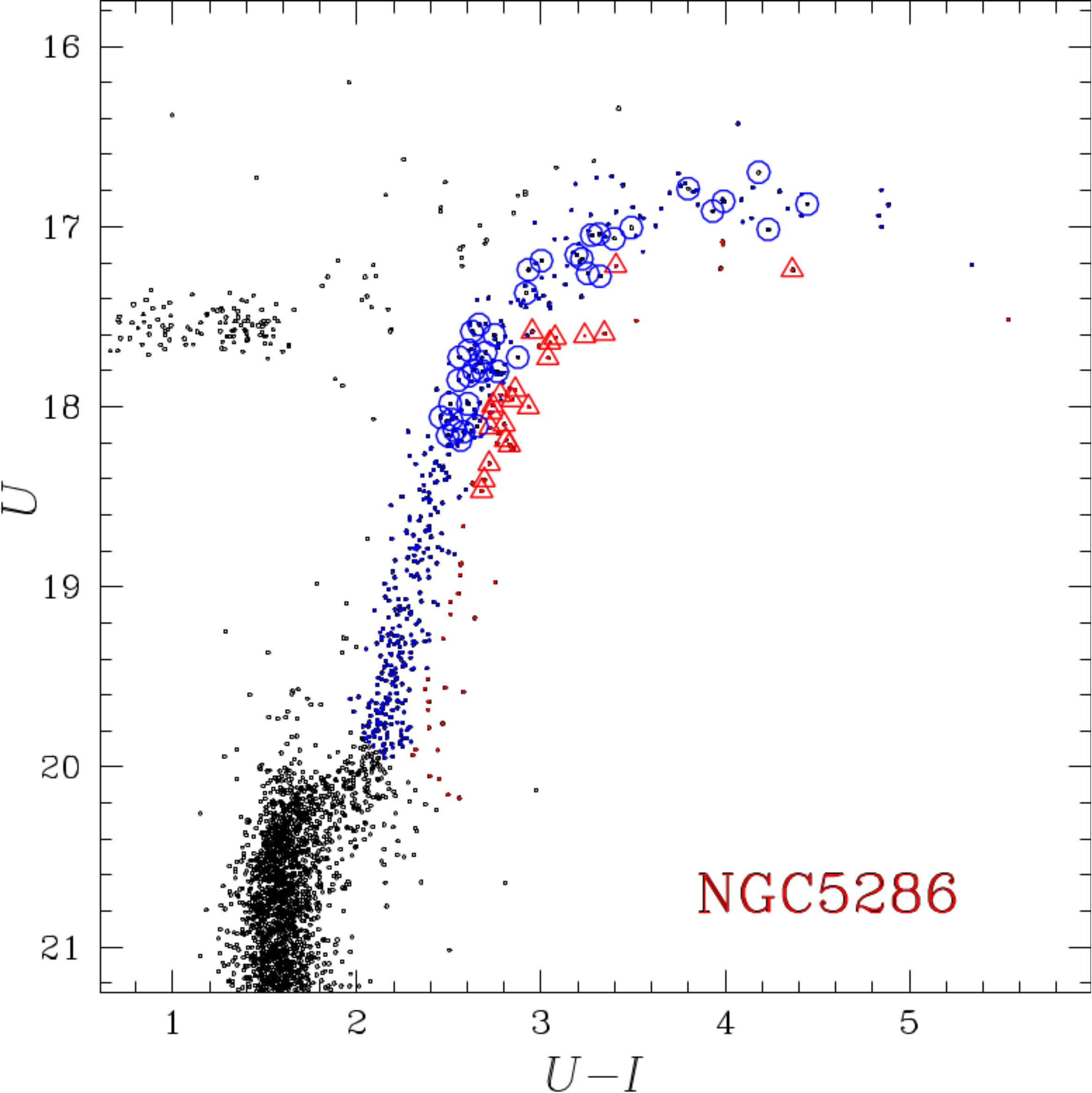}
  \includegraphics[width=5.6cm]{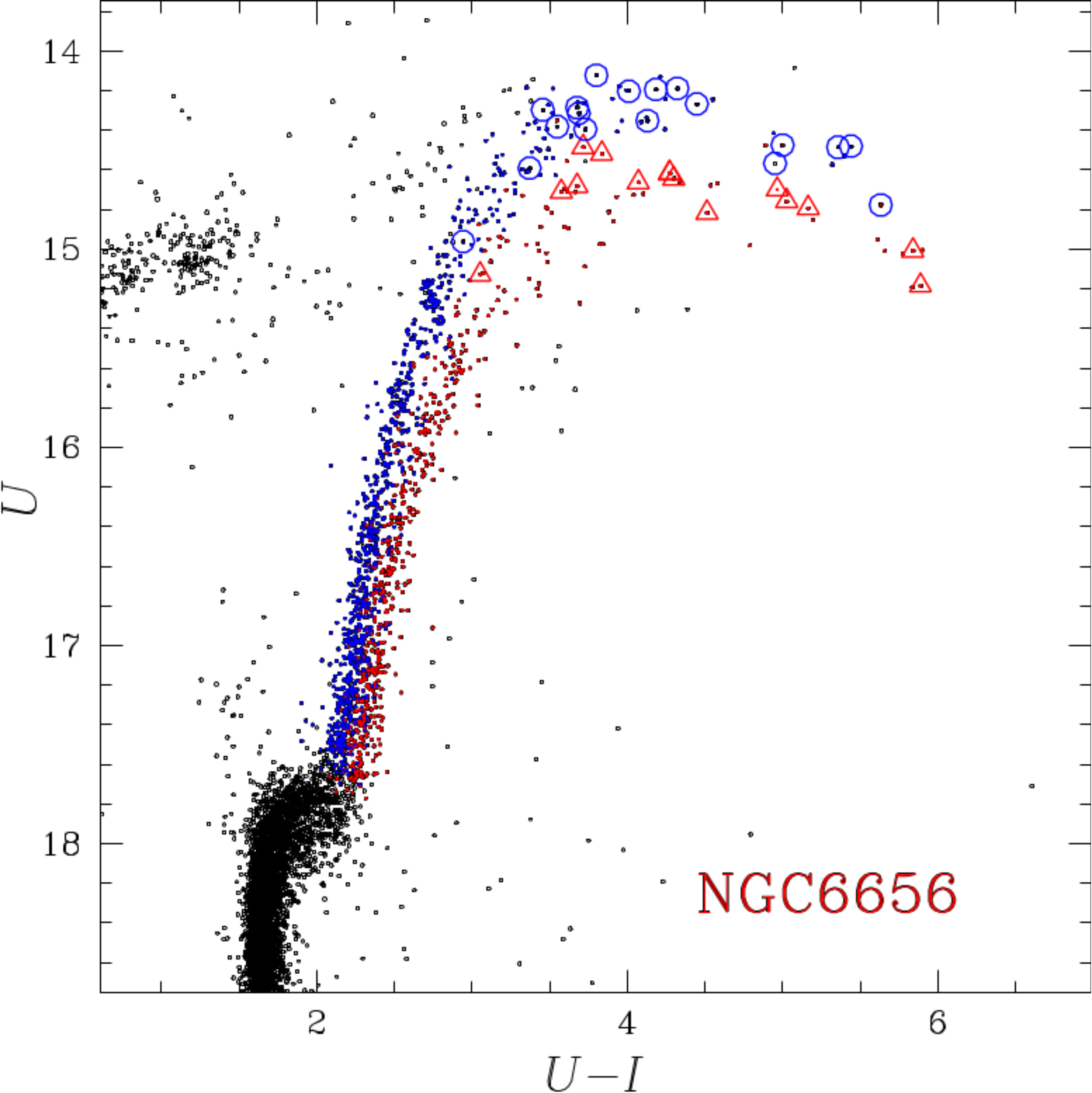}
  \includegraphics[width=5.6cm]{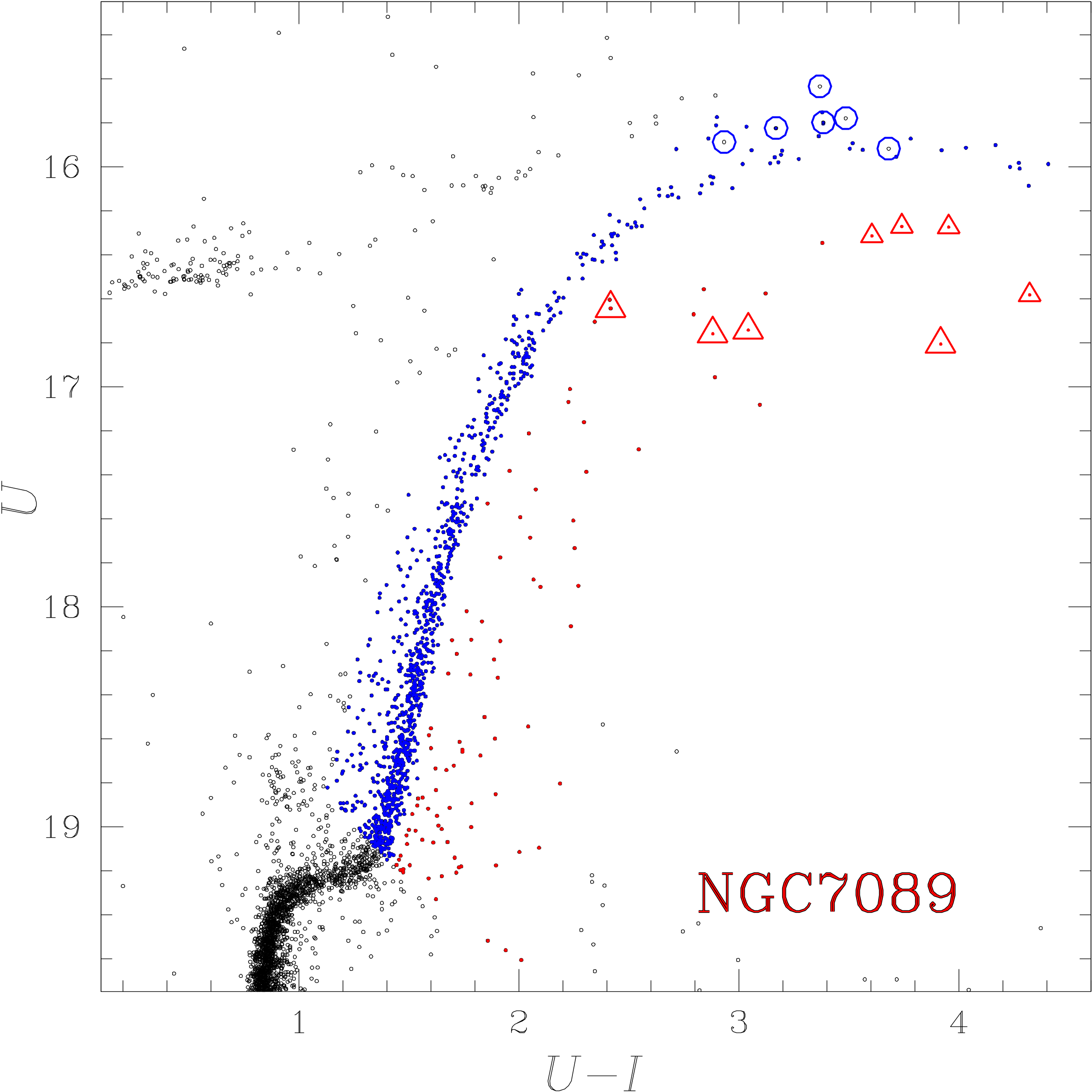}

  \includegraphics[width=5.6cm]{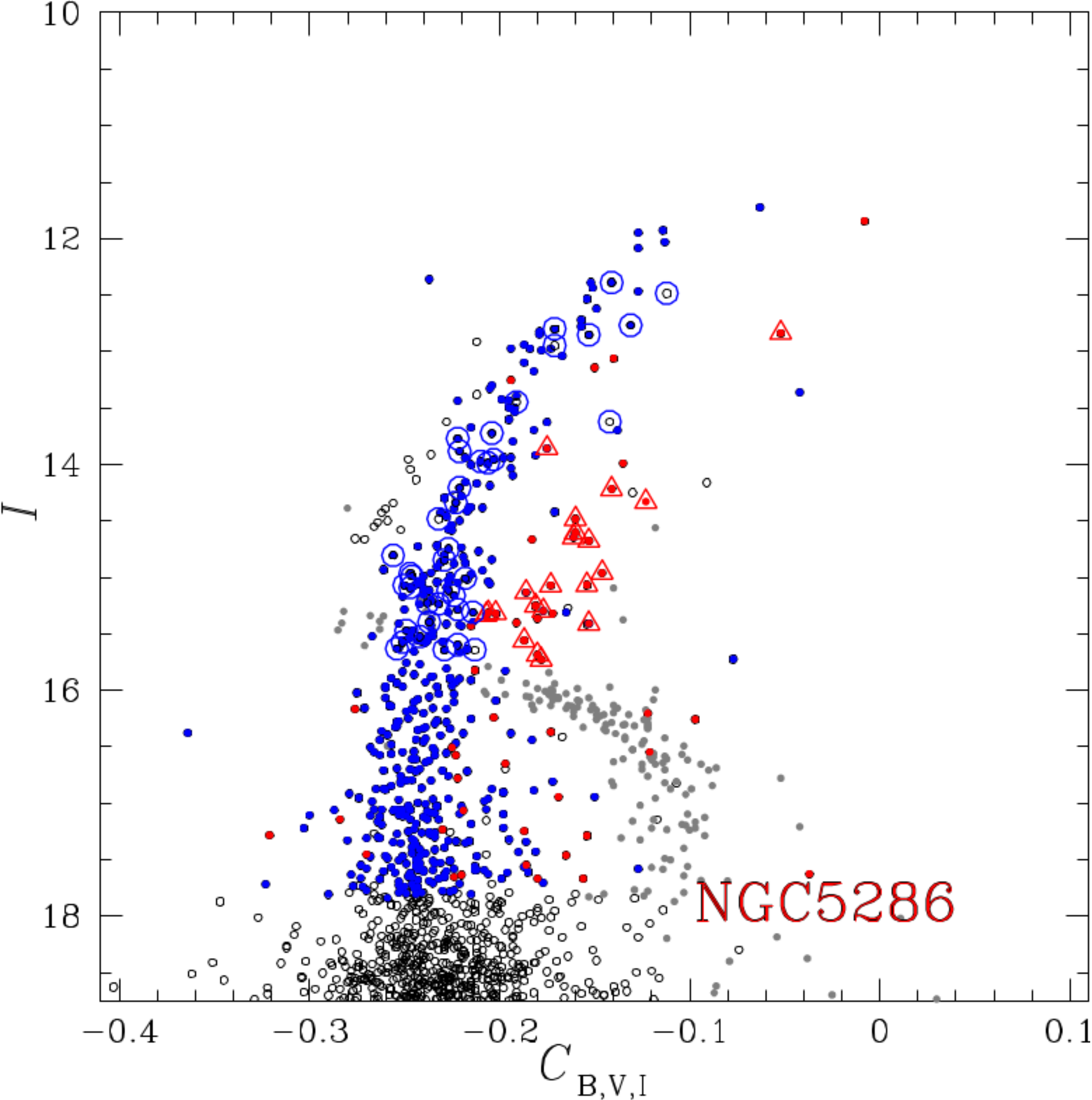}
  \includegraphics[width=5.6cm]{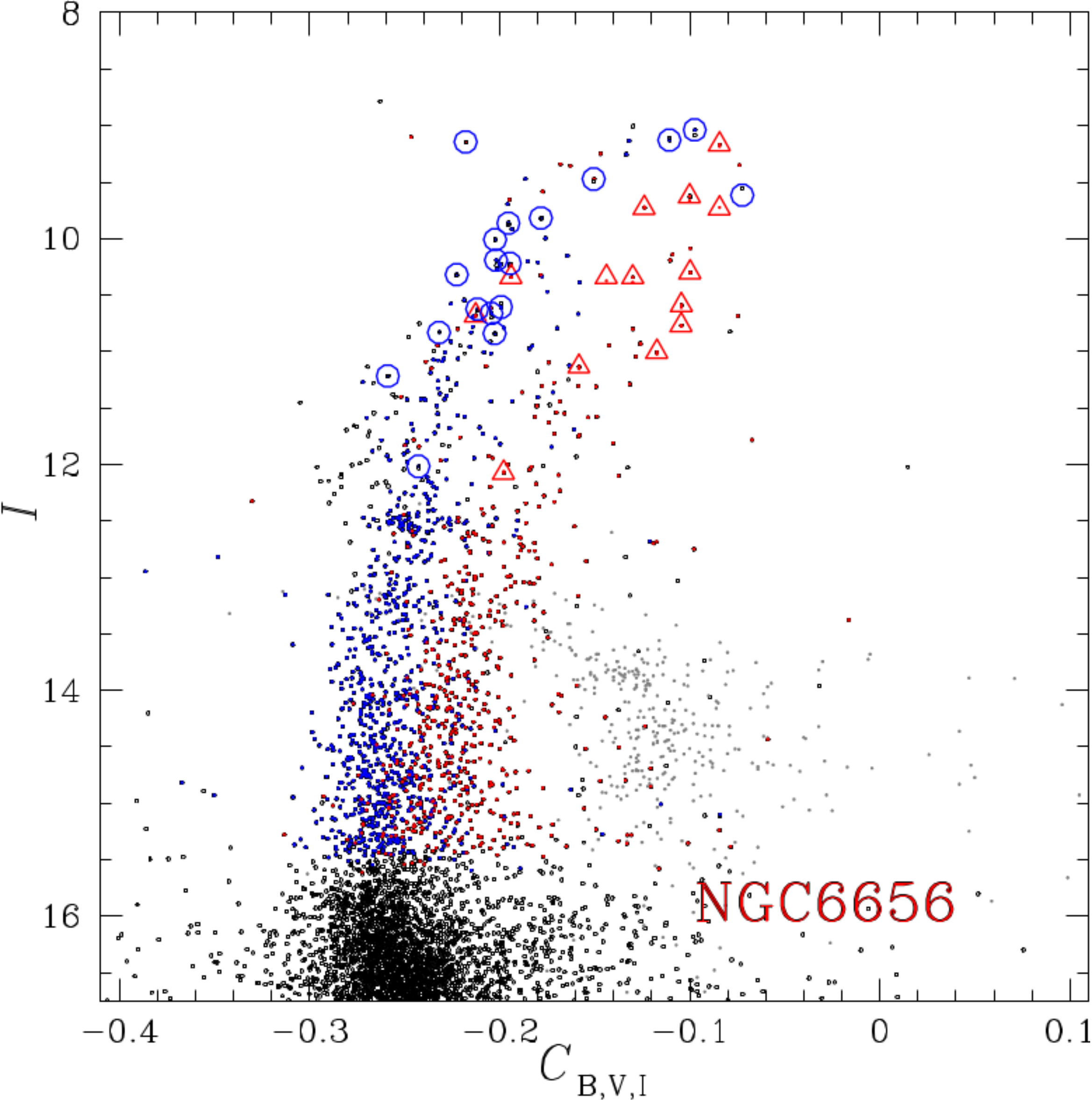}
  \includegraphics[width=5.6cm]{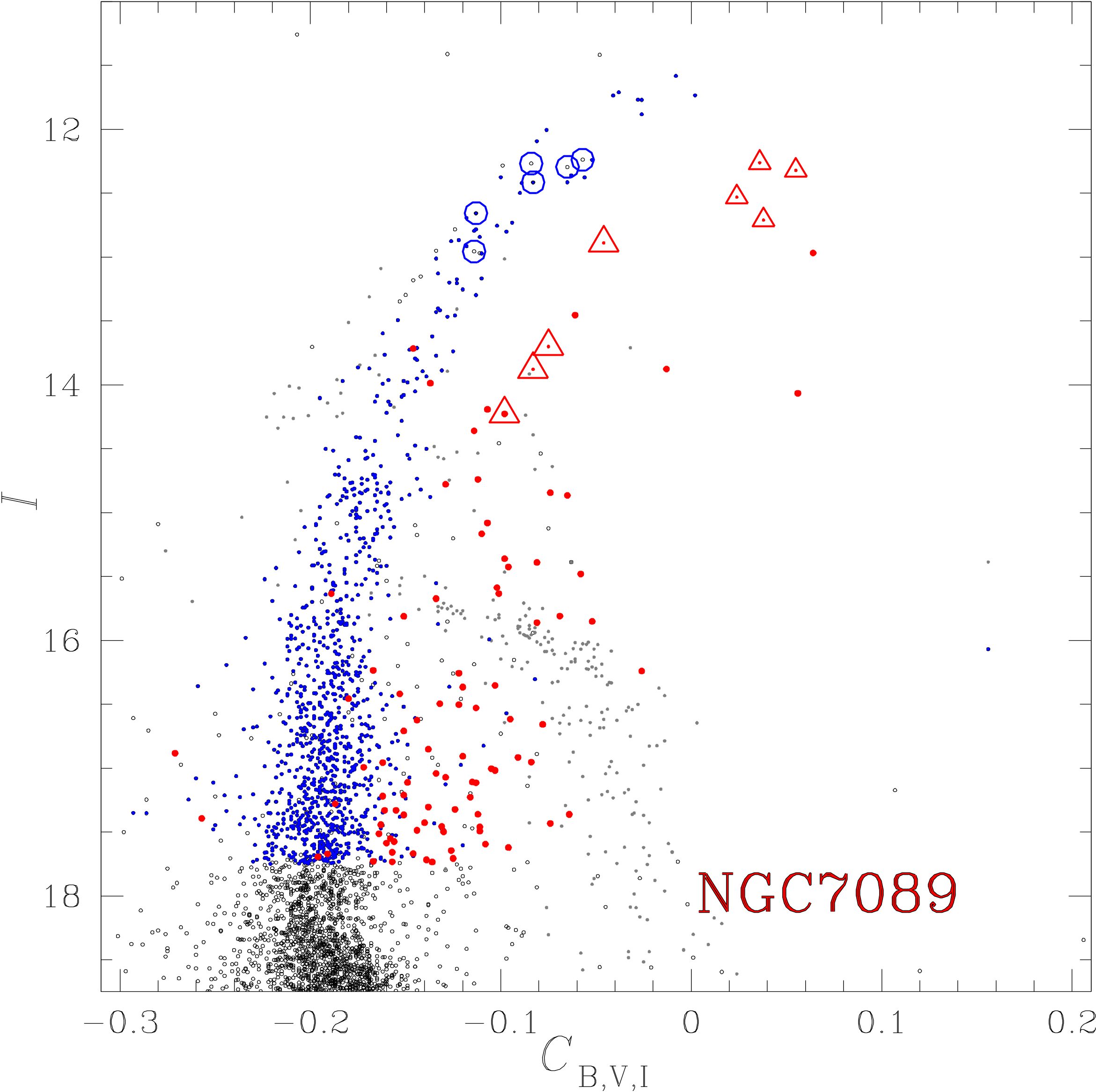}
  \caption{$U$ vs.\,$U-I$ CMD (upper panels) and $I$ vs.\,$C_{\rm
      B,V,I}$ pseudo CMD (bottom panels) of the Type~II GCs NGC\,5286,
    NGC\,6656 (M\,22) and NGC\,7089 (M\,2) from ground-based photometry. Red RGB
    stars are coloured red, while the blue points represent blue RGB
    stars. Spectroscopic targets are represented with red open triangles
    (red-RGB stars) and blue open circles (blue-RGB stars). 
} 
 \label{fig:GB}
\end{figure*}
\end{centering}
%%%%%%%%%%%%%%%%%%%%%%%%%%%%%%%%%%%%%%%%%%%%%%%%%%%%%%%%%%%%%%%%%%%%%%%%%%%%%%%

%%%%%%%%%%%%%%%%%%%%%%%%%%%%%%%%%%%%% FIG 2 %%%%%%%%%%%%%%%%%%%%%%%%%%%%%%%%%%%
\begin{centering}
\begin{figure*}
   \includegraphics[height=11.15cm]{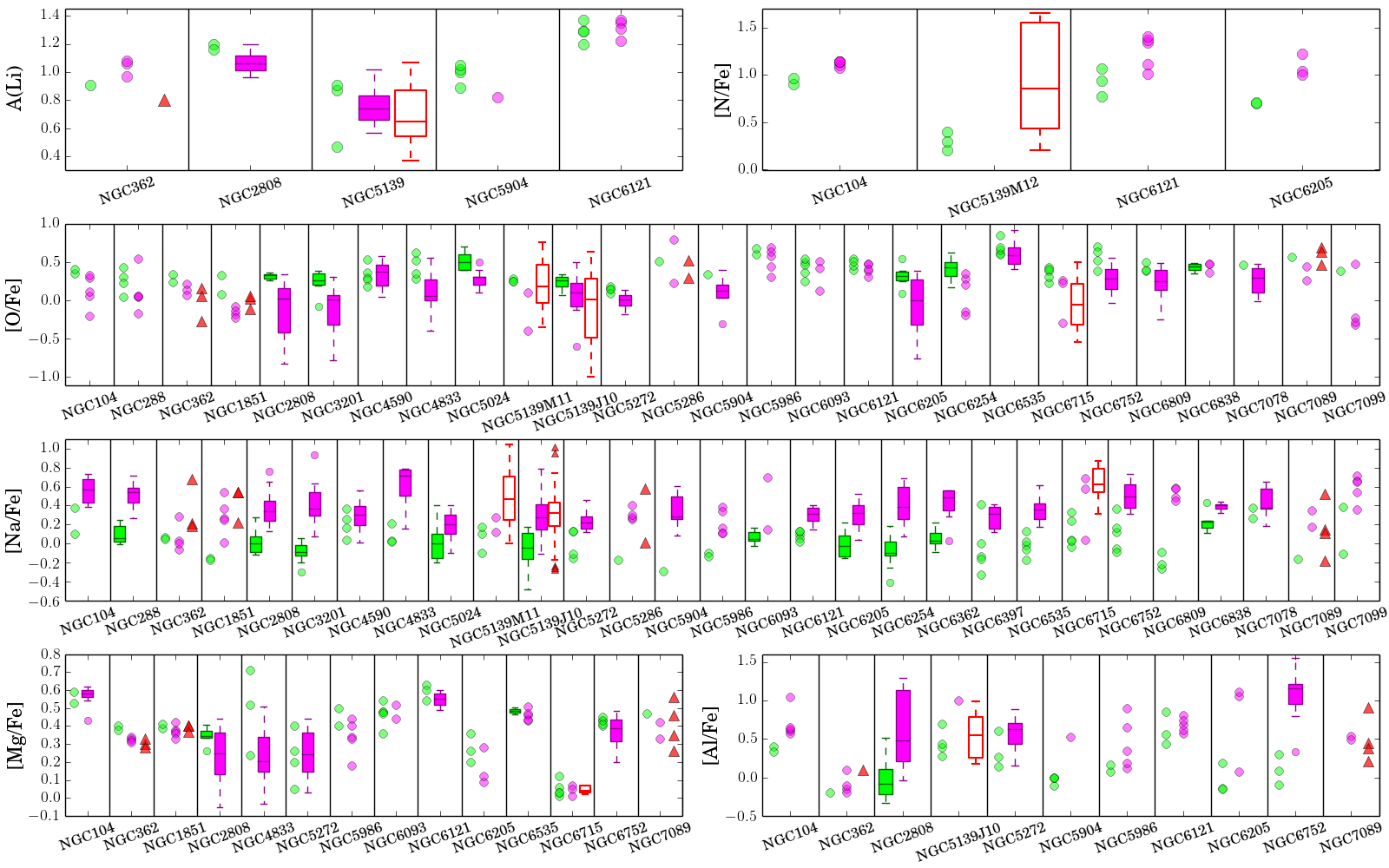}
   \caption{
Box-and-whisker plot for the chemical abundances of light elements
A(Li), [N/Fe], [O/Fe], [Na/Fe], [Mg/Fe], and [Al/Fe] for all the
clusters analysed here. 
Stars selected as 1G and 2G from the ChMs are
represented in green and magenta filled boxes, while stars on the red-RGB in
Type~II GCs have been represented in red empty boxes.
Each box represents the interquartile range (IQR) of the distribution,
with the median abundance marked by an horizontal line. 
The whiskers include observations that fall below the first quartile
minus 1.5$\times$IQR or above the third quartile plus 1.5$\times$IQR.
Small filled circles and triangles represent outliers or the
data for 1G/2G and red-RGB stars, respectively.
When less than five measurements are available for a given population,
data points are represented without box-and-whisker plot.
} 
\label{fig:chartNaO}
\end{figure*}
\end{centering}
%%%%%%%%%%%%%%%%%%%%%%%%%%%%%%%%%%%%%%%%%%%%%%%%%%%%%%%%%%%%%%%%%%%%%%%%%%%%%%%

%%%%%%%%%%%%%%%%%%%%%%%%%%%%%%%%%%%%%%%%%%%%%%%%%%%%%%%%%%%%%%%%%%%%%%%%%%%%%%%
\begin{centering}
\begin{figure*}
\includegraphics[height=16cm]{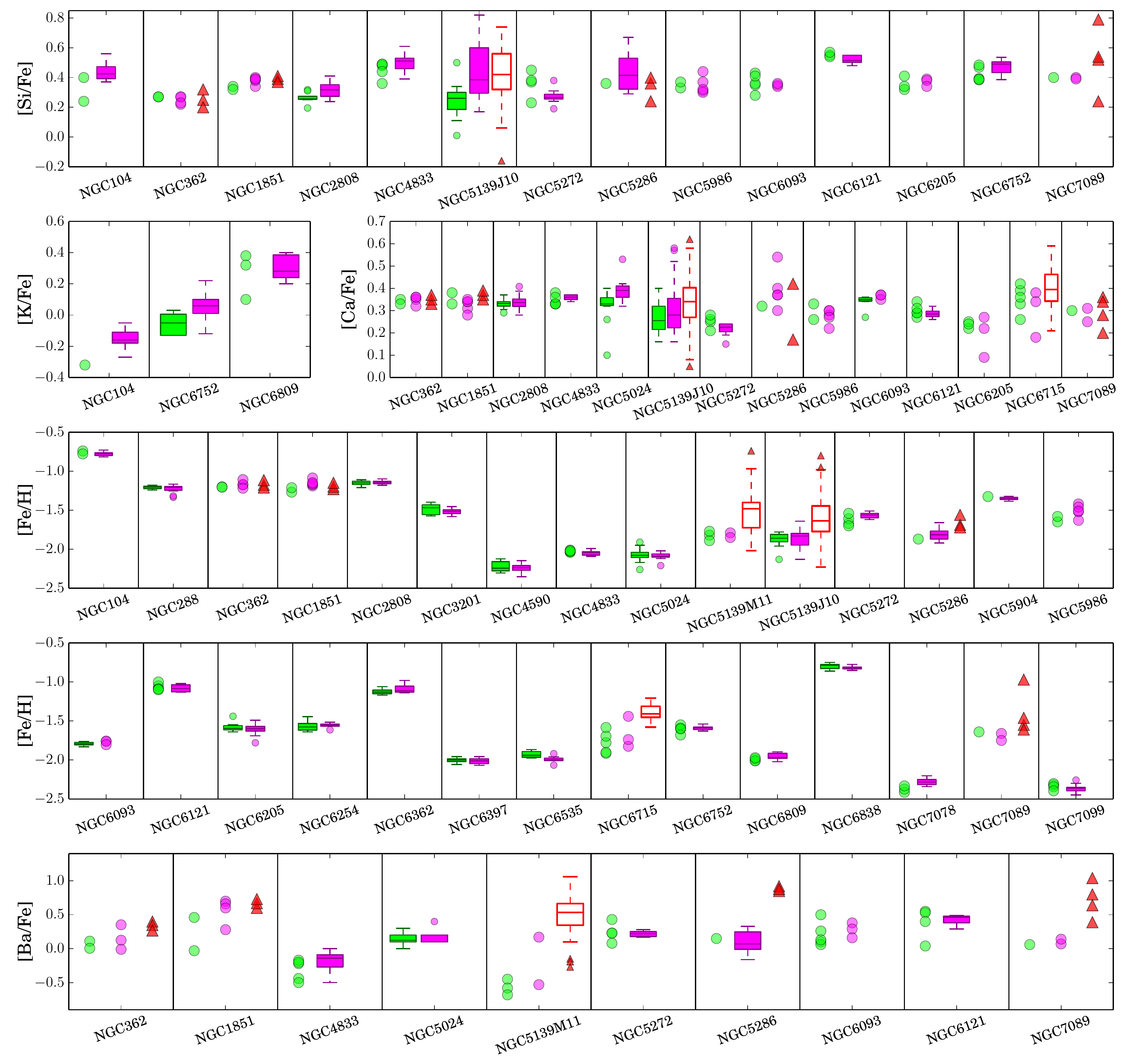}
\caption{Same as Figure~\ref{fig:chartNaO} for elements Si, K, Ca, Fe, and Ba.}
\label{fig:all}
\end{figure*}
\end{centering}
%%%%%%%%%%%%%%%%%%%%%%%%%%%%%%%%%%%%%%%%%%%%%%%%%%%%%%%%%%%%%%%%%%%%%%%%%%%%%%%

\subsection{The ground-based photometric dataset}\label{sub:photGB}

Ground-based photometry is used here for a sub-sample of Type~II GCs
only, because spectroscopic chemical abundances are available for many
stars lacking $HST$ observations. 
As discussed in Paper\,IX, Type\,II GCs are characterized by a 
bimodal RGB in the $m_{\rm F336W}$ vs.\,$m_{\rm F336W}-m_{\rm F814W}$
CMD, where the redder RGB is clearly connected with a faint sub-giant
branch (SGB).  
As an example, the redder RGB is evident in the $m_{\rm F336W}$
vs.\,$m_{\rm F336W}-m_{\rm F814W}$ CMD of NGC\,1851 plotted in the
left panel of Figure~\ref{fig:1851} where we indicate with small red dots
the red-RGB stars. 
Blue and red-RGB stars with available spectroscopy have been displayed
with large-filled blue dots and red triangles, respectively.
Clearly, stars in the blue RGB are located in the main ChM sequence,
which includes 1G and 2G stars (see the inset in
Figure~\ref{fig:1851}), while stars in the redder RGB correspond to
red-RGB stars, as defined in Section~\ref{sub:photHST}.

The middle panel of Figure~\ref{fig:1851} shows the $U$ vs.\,$U-I$ CMD of
NGC\,1851 from wide-field ground-based photometry. The split SGB and
RGB are clearly visible in both CMDs, i.e., from either ground-based
and $HST$ photometry. This fact is quite expected 
due to the similarity between $m_{\rm F336W}$ and $m_{\rm F814W}$
filters of UVIS/WFC3 and WFC/ACS and the Johnson $U$ and $I$ bands. 
The two main RGBs of NGC\,1851 are clearly visible also in the $I$
vs.\,$C_{\rm B,V,I}$ pseudo-CMD, introduced by Marino et al.\,(2015) and
shown in the right panel of Figure~\ref{fig:1851}.

In the following we will thus exploit the $U$ vs.\,$U-I$ CMD and the
$I$ vs.\,$C_{\rm B,V,I}$ pseudo-CMD from wide-field ground-based
photometry in order to increase the sample of stars for NGC\,1851,
NGC\,5286, and NGC\,7089 for which both photometry and spectroscopy
are available. Moreover, wide-field ground based photometry will allow
us to extend the analysis to NGC\,6656 (M\,22), for which no stars with spectroscopy are
available in the $HST$ field of view.
Specifically, for NGC\,1851, NGC\,5286, and NGC\,6656 we have used
the photometric data collected through $U$ filter of the Wide-Field
Imager (WFI) at the Max Planck 2.2m telescope at La Silla as part of
the SUrvey of Multiple pOpulations in GCs (SUMO; programme
088.A-9012-A, PI. A.\,F.\,Marino). Photometry and astrometry of the
WFI images have been carried out by using the method described by
Anderson \& King\,(2006). In addition, we have used $B$, $V$, $I$
photometry from the archive maintained by P.\,.B.\,Stetson
(Stetson\,2000).  
 
The CMDs of NGC\,5286 and NGC\,6656 are strongly contaminated by
field stars on the same line of sight of these two clusters. In order
to select a sample of probable cluster members we have  used stellar
proper motions from Gaia data release 2 (DR2, Gaia collaboration et al.\,2018).
The $U$ vs.\,$U-I$ CMDs and $I$ vs.\,$C_{\rm B,V,I}$ pseudo-CMDs for
NGC\,5286, NGC\,6656, and NGC\,7089 are shown in Figure~\ref{fig:GB}. 
All the stars in the ground-based CMDs of NGC\,1851, NGC\,5286, NGC\,6656 and
NGC\,7089 selected as red-RGB and blue-RGB stars are coloured red and
blue, respectively, with blue open circles and red open triangles
representing stars with available spectroscopy.

\section{The chemical composition of multiple populations over the
  Chromosome Maps}\label{sec:chemI}

Figures~\ref{fig:chartNaO} and \ref{fig:all} show the
A(Li), [N/Fe], [O/Fe], [Na/Fe], [Mg/Fe], [Al/Fe], [Si/Fe], [K/Fe],
[Ca/Fe], [Fe/H], and [Ba/Fe] abundances\footnote{Chemical
  abundances are expressed in the standard notation, as the
  logarithmic ratios with respect to solar values, [X/Y]=$\mathrm
  {log(\frac{N_{X}}{N_{Y}})_{star}-log(\frac{N_{X}}{N_{Y}})_{\odot}}$. For
lithium, abundances are reported as A(Li)=$\mathrm {log(\frac{N_{Li}}{N_{H}})_{star}}+12$.}
for all the stars in each 
cluster for which both spectroscopy and photometry are available. As
in the previous figures, we have used green and magenta to show 1G
stars and 2G stars, while red-RGB stars in Type~II GCs are coloured in
red. When more than five stars are available in each group, we show
the box-and-whisker plot, and the median abundance. The boxes in this
plot include the interquartile range (IQR) of the data, and the
whiskers extend from the first quartile minus 1.5$\times$IQR to the
third quartile plus 1.5$\times$IQR. 
The average abundances derived for each group of analysed stars, the
corresponding dispersion (r.m.s.) and the number of stars that we have
used are listed in Table~\ref{tab:abb}. 

In the following we will discuss in detail the chemical pattern
observed on the various portions of the ChMs by discussing Type~I and
Type~II GCs separately. Note that all the stars analysed here are
RGB stars, as, by construction, ChMs do not include asymptotic giant
branch stars (AGBs; see Paper~III for a detailed description on the
construction of ChMs)\footnote{So far, a study of the AGB ChM has
    been performed only for NGC\,2808 (Marino et al.\,2017) suggesting
    that the AGB of this GC does not host the counterpart of the most
    He-rich stars observed on the RGB ChM (Paper~III).}.   
Our goal is to investigate the chemical elements governing the shape
and variety observed on the maps for RGB stars.

%%%%%%%%%%%%%%%%%%%%%%%%%%%%%%%%%%%%% FIG 2 %%%%%%%%%%%%%%%%%%%%%%%%%%%%%%%%%%%
\begin{centering}
\begin{figure*}
\includegraphics[width=17.8cm]{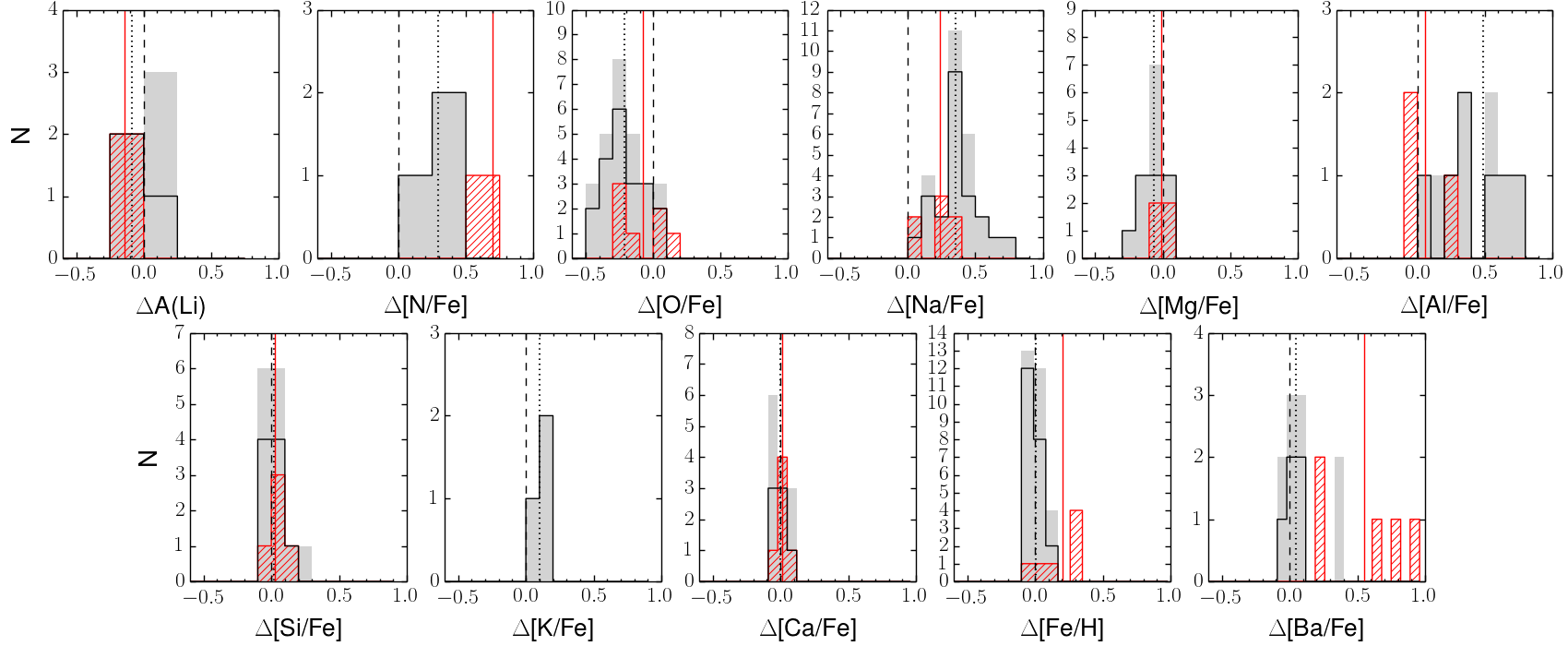}
  \caption{Distribution of the average abundance difference between 2G
    and 1G stars for all the analysed clusters (grey-shaded histogram)
    and for the Type~I GCs only (black histogram). 
    No difference is indicated by the black dashed line. Red histograms
    correspond to the distribution of the average abundance difference
    between red-RGB and blue-RGB stars in Type~II
    GCs. The dotted black vertical line and the continuous red line
    indicate the mean of the difference
    distributions between 2G and 1G stars and
    red-RGB and blue-RGB stars, respectively. To better visualise the
    different size of variations for distinct chemical species, the range in the {\it x}
    axis has been kept the same for all the elements.
  } 
 \label{fig:istogrammi}
\end{figure*}
\end{centering}
%%%%%%%%%%%%%%%%%%%%%%%%%%%%%%%%%%%%%%%%%%%%%%%%%%%%%%%%%%%%%%%%%%%%%%%%%%%%%%%

%%%%%%%%%%%%%%%%%%%%%%%%%%%%%%%%%%%%%%%%%%%%%%%%%%%%%%%%%%%%%%%%%%%%%%%%%%%%%%%
\begin{centering}
\begin{figure*}
 \includegraphics[width=17.5cm]{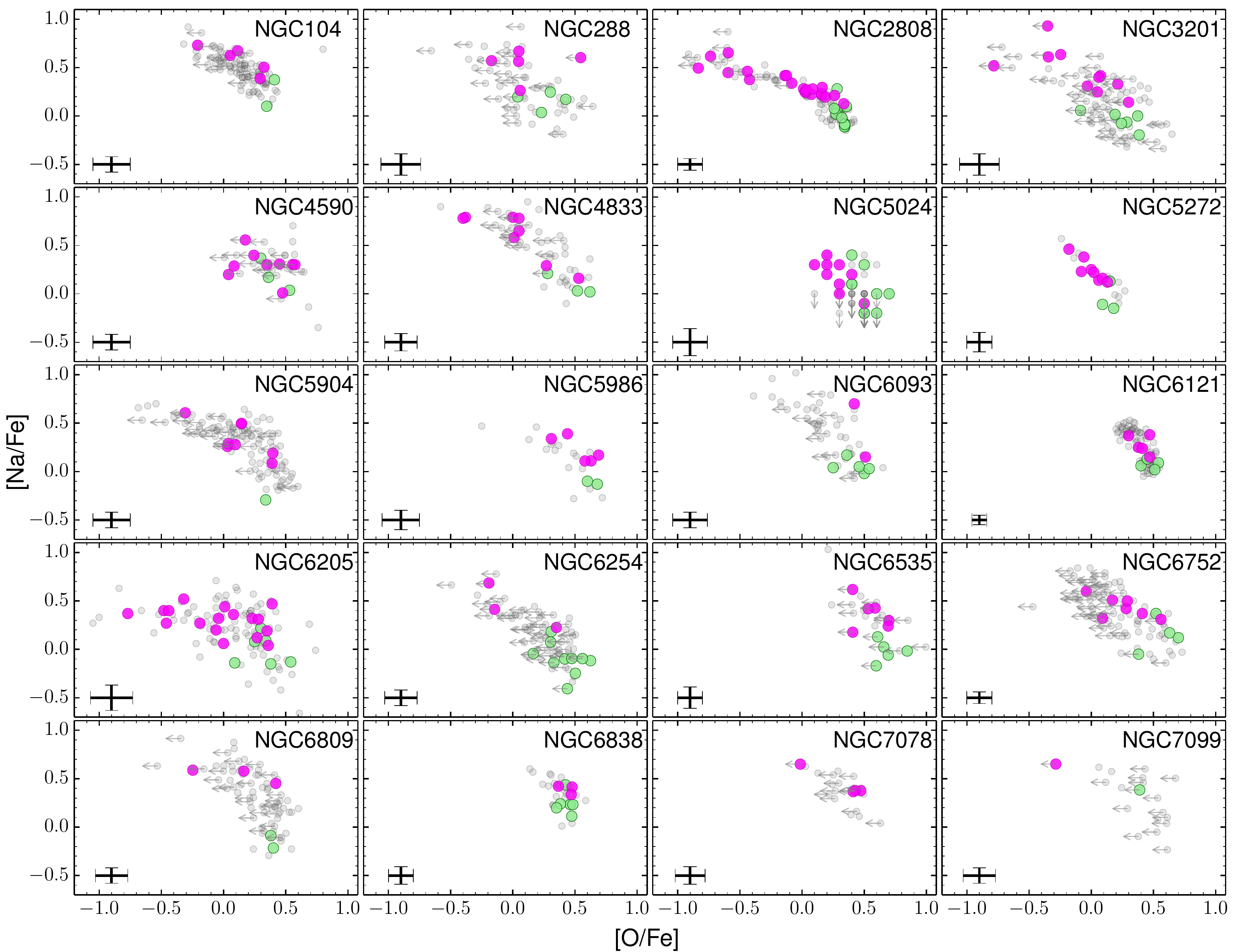}
  \caption{[Na/Fe] vs.\,[O/Fe] for 20 Type~I clusters for which both sodium
    and oxygen abundances are available from literature. 1G and 2G
    stars are represented with large green and magenta dots,
    respectively. The entire sample with spectroscopic abundances is
    represented with grey dots. The typical error bars are taken from
    the reference papers listed in Table~\ref{tab:data}. } 
 \label{fig:nao}
\end{figure*}
\end{centering}
%%%%%%%%%%%%%%%%%%%%%%%%%%%%%%%%%%%%%%%%%%%%%%%%%%%%%%%%%%%%%%%%%%%%%%%%%%%%%%%

%%%%%%%%%%%%%%%%%%%%%%%%%%%%%%%%%%%%% FIG 2 %%%%%%%%%%%%%%%%%%%%%%%%%%%%%%%%%%%
\begin{centering}
\begin{figure*}
   \includegraphics[width=17.5cm]{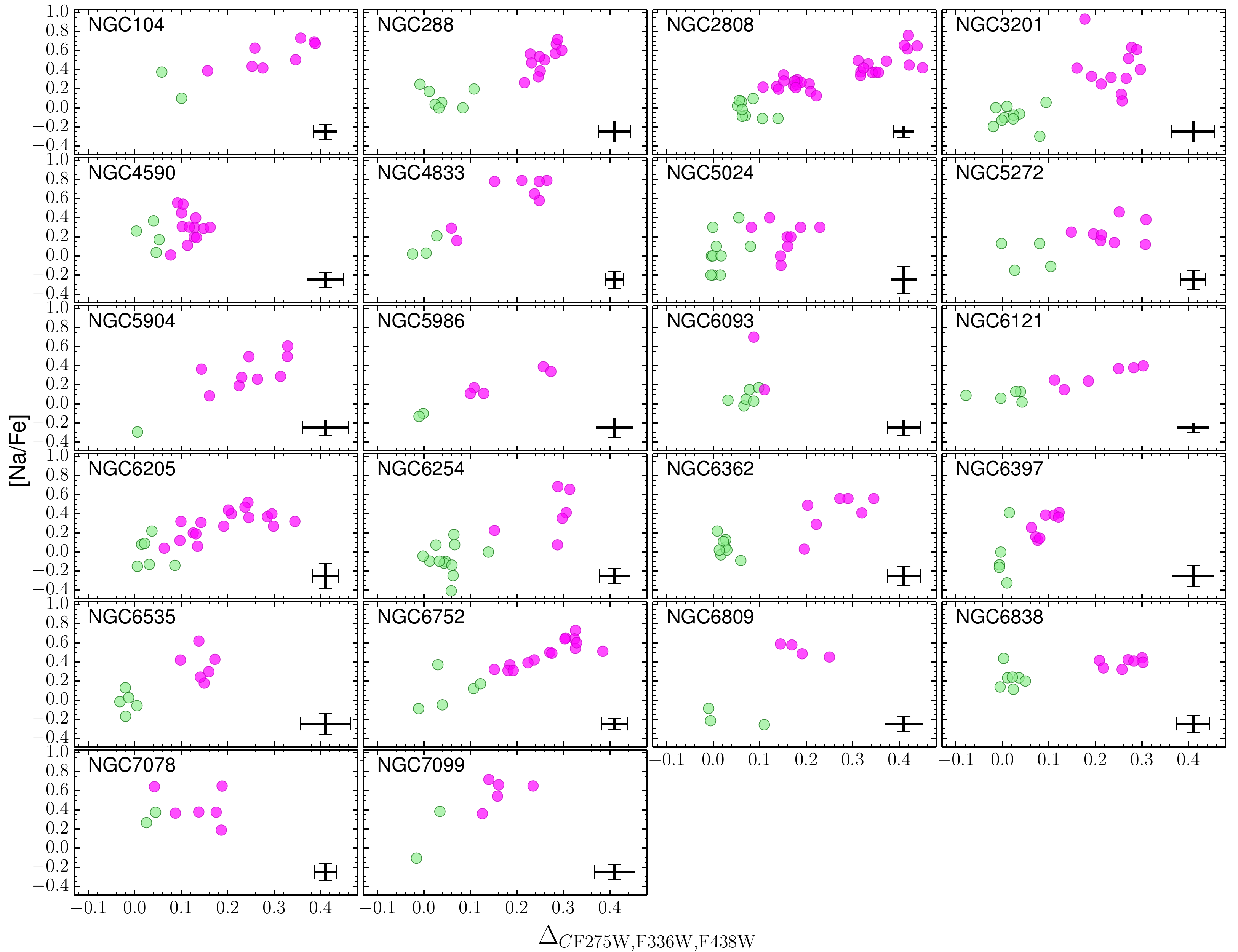}
  \caption{Sodium abundance as a function of \y. 1G and 2G stars are
    coloured green and magenta. The typical error bars for Na are taken from
    the reference papers listed in Table~\ref{tab:data}. Error bars
    for \y\ are from Paper~IX.}  
 \label{fig:nay}
\end{figure*}
\end{centering}
%%%%%%%%%%%%%%%%%%%%%%%%%%%%%%%%%%%%%%%%%%%%%%%%%%%%%%%%%%%%%%%%%%%%%%%%%%%%%%%

%%%%%%%%%%%%%%%%%%%%%%%%%%%%%%%%%%%%% FIG 2 %%%%%%%%%%%%%%%%%%%%%%%%%%%%%%%%%%%
\begin{centering}
\begin{figure*}
 \includegraphics[width=0.78\textwidth]{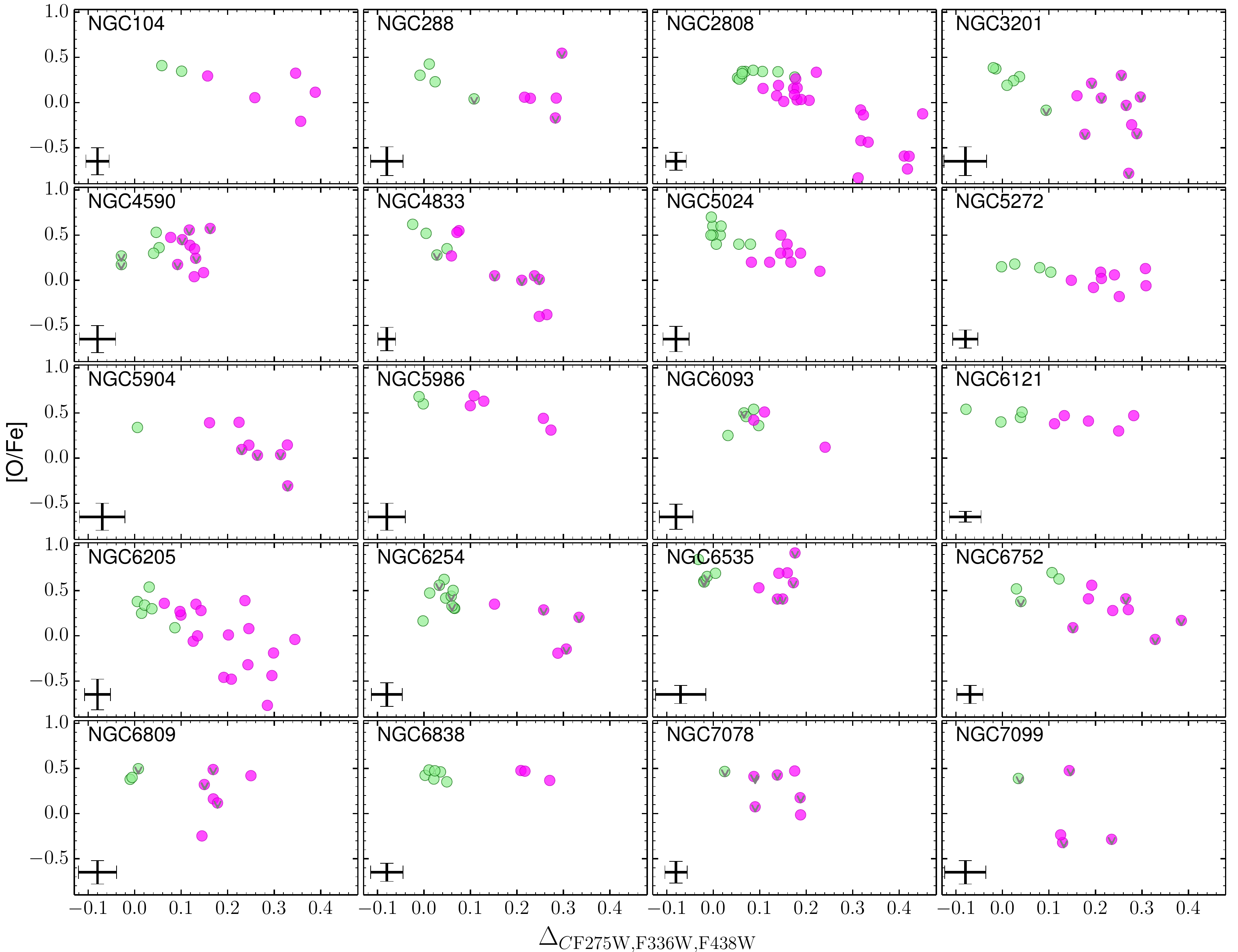}
  \caption{Oxygen abundance as a function of \y. 1G and 2G stars are
    coloured green and magenta. Upper limits are indicated with grey
    downward symbols. The typical error bars for O are taken from
    the reference papers listed in Table~\ref{tab:data}. Error bars
    for \y\ are from Paper~IX.}  
 \label{fig:oy}
\end{figure*}
\end{centering}
%%%%%%%%%%%%%%%%%%%%%%%%%%%%%%%%%%%%%%%%%%%%%%%%%%%%%%%%%%%%%%%%%%%%%%%%%%%%%%%

%%%%%%%%%%%%%%%%%%%%%%%%%%%%%%%%%%%%% FIG 2 %%%%%%%%%%%%%%%%%%%%%%%%%%%%%%%%%%%
\begin{centering}
\begin{figure*}
 \includegraphics[width=0.78\textwidth]{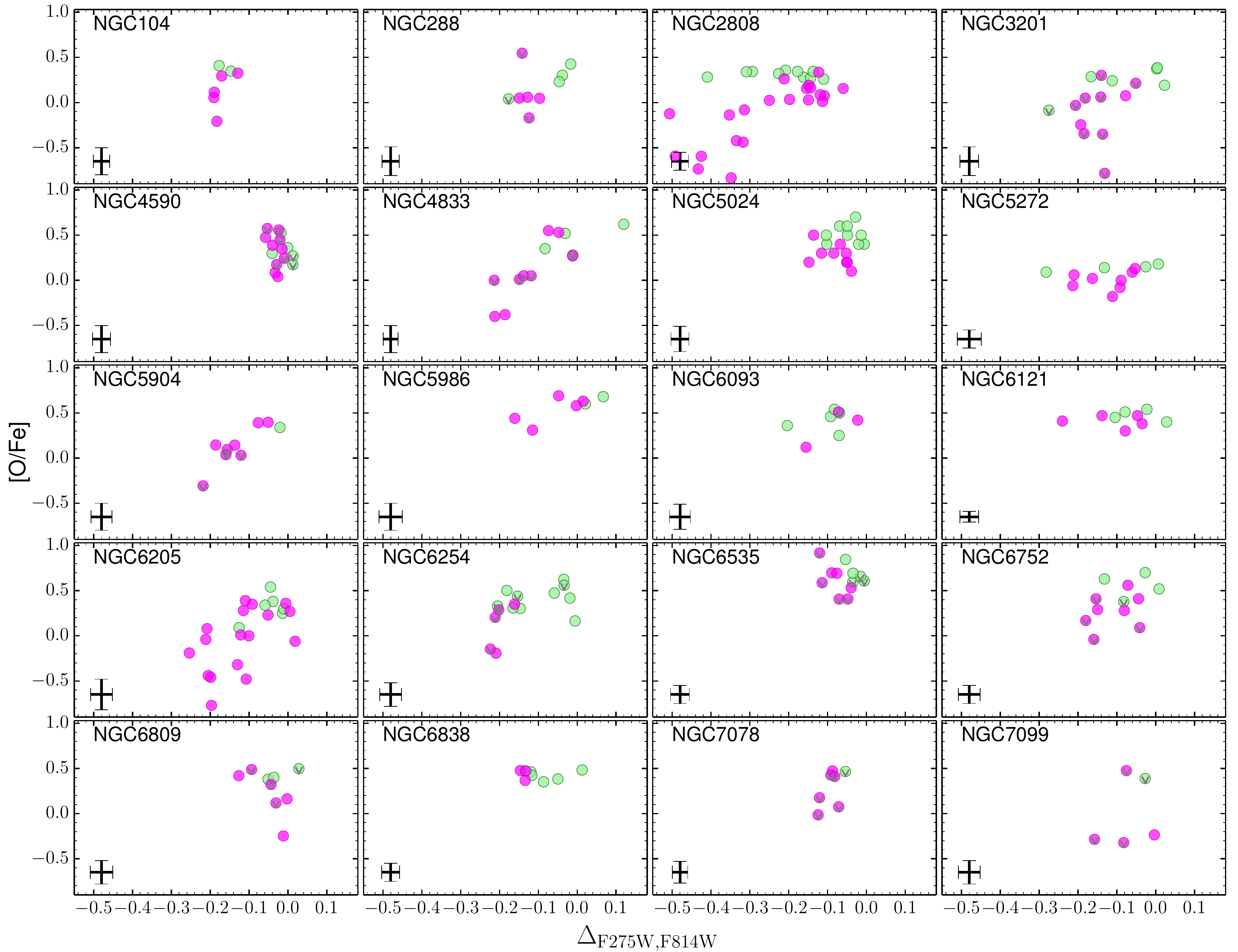}
  \caption{Oxygen abundance as a function of \x. Symbols are as in
    Figure~\ref{fig:oy}. The typical error bars for O are taken from
    the reference papers listed in Table~\ref{tab:data}. Error bars
    for \x\ are from Paper~IX.}
 \label{fig:ox}
\end{figure*}
\end{centering}
%%%%%%%%%%%%%%%%%%%%%%%%%%%%%%%%%%%%%%%%%%%%%%%%%%%%%%%%%%%%%%%%%%%%%%%%%%%%%%

%%%%%%%%%%%%%%%%%%%%%%%%%%%%%%%%%%%%% FIG 2 %%%%%%%%%%%%%%%%%%%%%%%%%%%%%%%%%%%
\begin{centering}
\begin{figure*}
   \includegraphics[width=0.78\textwidth]{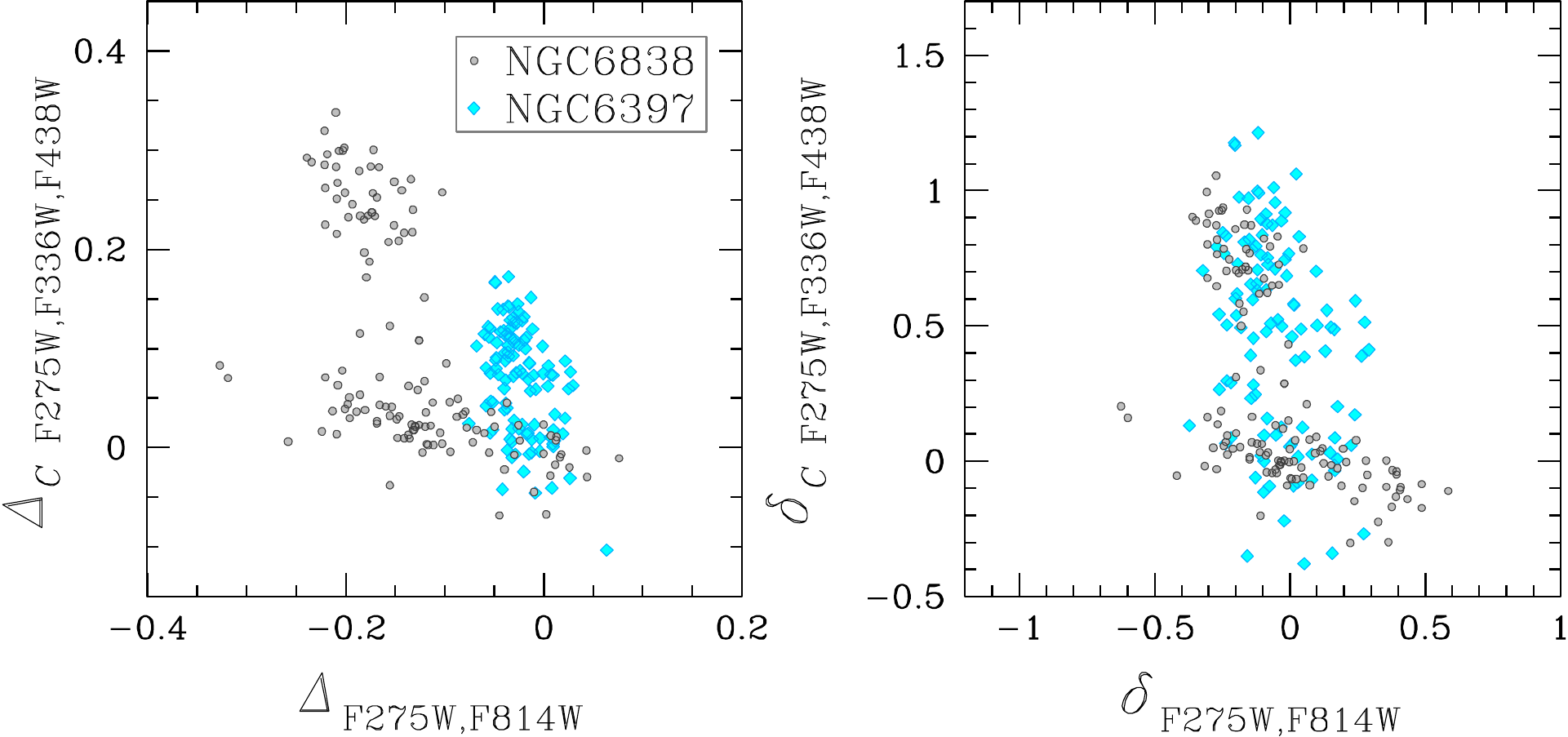}
  \caption{{\it Left panel}: Comparison of the ChMs of the two
      low-mass GCs NGC\,6838 ([Fe/H]=$-$0.78) 
      and NGC\,6397 ([Fe/H]=$-$2.02), represented in gray dots, and azure
      diamonds, respectively. {\it Right panel}: Comparison of the two
      universal ChMs for the same clusters (see text for details). We note
      that the two universal ChMs almost overlap each other suggesting that
      the metallicity dependence of the maps has been significantly reduced
      in the \dx-\dy\ plane.}  
 \label{fig:n6838n6397}
\end{figure*}
\end{centering}
%%%%%%%%%%%%%%%%%%%%%%%%%%%%%%%%%%%%%%%%%%%%%%%%%%%%%%%%%%%%%%%%%%%%%%%%%%%%%%%

%%%%%%%%%%%%%%%%%%%%%%%%%%%%%%%%%%%%% FIG 3 %%%%%%%%%%%%%%%%%%%%%%%%%%%%%%%%%%%
\begin{figure*}
    \centering
    \rotatebox[origin=c]{-90}{%
    \begin{minipage}{9in}
    \centering
        \includegraphics[width=\linewidth]{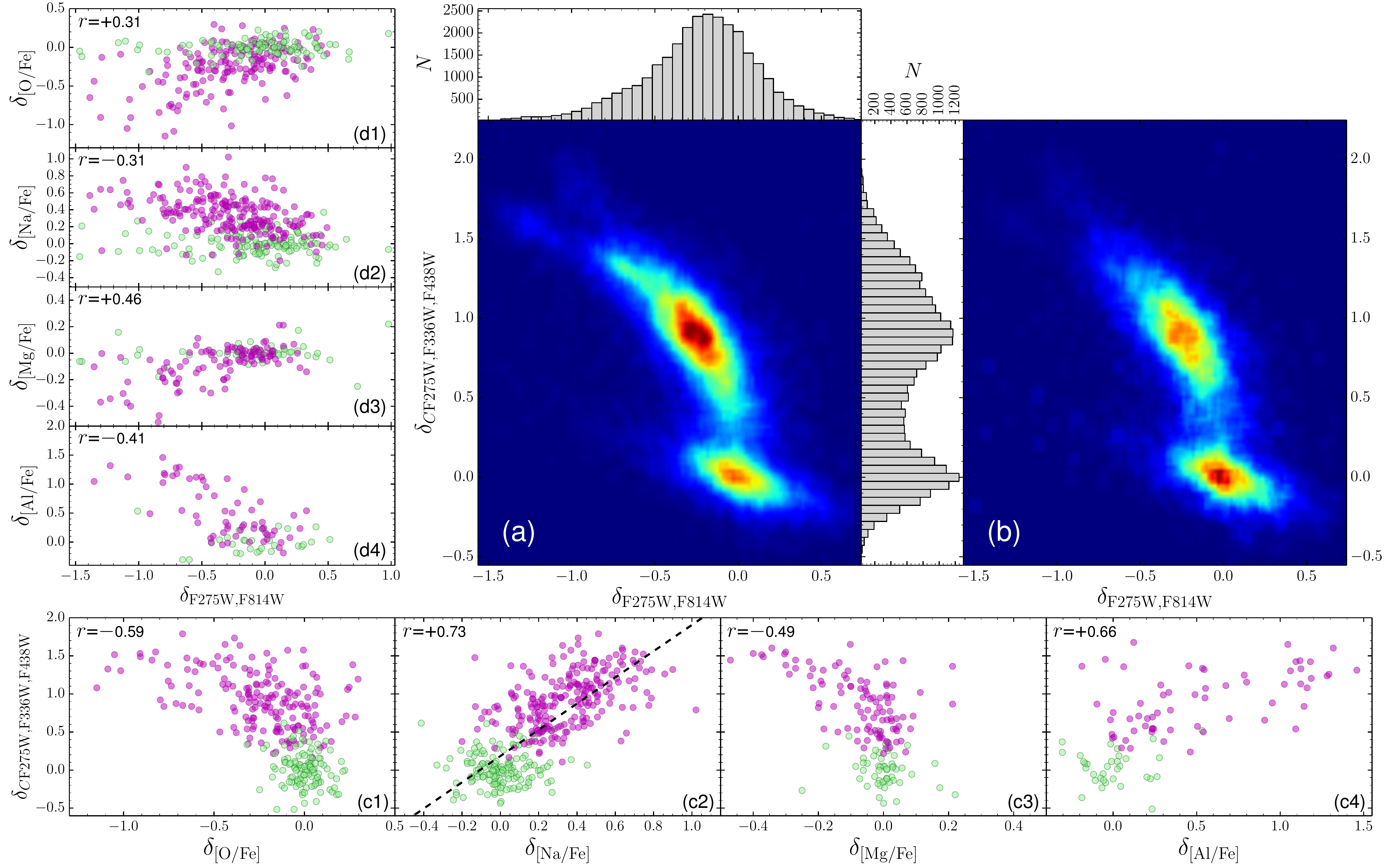}
    \caption{{\it Panels a, b}: Hess diagrams of the universal ChM for all the
        analysed GCs with [Fe/H]$<-0.6$. The color is indicative of
        the density of all the stars (panel (a)), and the stars
        normalised to the total number
        of stars in the parent cluster (panel (b)).
        The \dy\ and \dx\ histogram distributions for all the analyzed
        stars are shown as side plots of panel (a). 
        {\it Panels c1--c4}: \dy\ vs.\,the abundance ratios of O, Na, Mg
        and Al relative to the average abundances of 1G stars. 
        {\it Panels d1--d4}: the abundance ratios of O, Na, Mg
        and Al relative to the average abundances of 1G stars as a
        function of \dx. In all
        the panels with chemical abundances 1G and 2G stars, as
        selected from the ChMs in Paper~IX, are represented in green and magenta,
        respectively. For each element plot, we report the Spearman's
        correlation coefficient $r$. In panel (c2) we plot the
        least-square fit with data (see Section~\ref{sec:supermap} for details).}
    \label{fig:SuperMap}
    \end{minipage}
    }
\end{figure*}

%%%%%%%%%%%%%%%%%%%%%%%%%%%%%%%%%%%%%%%%%%%%%%%%%%%%%%%%%%%%%%%%%%%%%%%%%%%%%%%

\subsection{Type~I GCs}\label{sec:typeI}

As discussed in Section~\ref{sub:photHST} the ChM of Type~I GCs is
composed of two main sub-structures that we have named 1G and 2G. In this
section we investigate the difference in chemical content between
these two stellar populations only.
We start inspecting the boxplots displayed in Figures~\ref{fig:chartNaO} and
\ref{fig:all} from the lightest (Li) to heaviest (Ba) elements, for 1G
(green) and 2G stars (magenta).

Lithium abundances plotted in Figure~\ref{fig:chartNaO} have been
corrected for departures from the LTE assumption in the cases
NGC\,362, NGC\,2808, and $\omega$~Centauri; while for NGC\,5904 and NGC\,6121
(M\,4) the available abundances have been derived in LTE. 
Note that Li has been analysed here only for stars fainter than the
RGB bump, to minimise the effects of strong Li depletions that
  occur at this luminosity. 
With the exception of NGC\,2808 and $\omega$~Centauri, that will be
discussed in more details in Sections~\ref{sub:Li} and
\ref{sec:omegacen}, all the A(Li) abundances plotted in
Figure~\ref{fig:chartNaO} are measurements, with no upper limits.  
No obvious difference is seen in the box-and-whisker plot of
Figure~\ref{fig:chartNaO} between the A(Li) of 1G and 2G, though 
very few stars are available and only in a few clusters. 
In the case of NGC\,2808, for which we have two 1G stars and six
  2G stars, the difference is $\Delta$A(Li)$_{\rm
    {2G-1G}}$=$-$0.11$\pm$0.07, a $\sim$1.5~$\sigma$
  difference, but a clearer difference can be seen when plotting Li
  abundances with other elements (see Section~\ref{sub:Li}).
Given the nuclear {\it fragility} of lithium one would have expected
it to be strongly depleted in 2G stars, unless stars making the material for
the formation of 2G stars were also making some fresh lithium (e.g.,
Ventura, D'Antona \& Mazzitelli\,2002).  
We note however that the lithium abundance in 2G stars depends
also on the Li content of the pristine gas, with which the material
ejected from whatever polluter is likely diluted (D'Antona et al.\,2012).

Nitrogen abundances are available only for a few 1G and 2G stars in
NGC\,104 (47~Tucanae), M\,4 and NGC\,6205, suggesting higher N 
for 2G stars, as expected. 
In the case of $\omega$~Centauri we find that red-RGB stars span a
wide range of N of more than 1~dex, and have on average higher N than
1G blue-RGB stars, but this case will be discussed more in detail in
Section~\ref{sec:typeII}.

In the past years, sodium and oxygen abundances have been widely used
to investigate the multiple stellar populations in several of GCs
(e.g.\,Carretta et al.\,2009).    
Hence, among stars studied in the Legacy Survey of GCs, sodium and oxygen
abundances are available for a relatively high number of stars in 
22 and 20 Type~I GCs, respectively.  
As illustrated in Figure~\ref{fig:chartNaO}, on average 2G stars have
depleted O with respect to the 1G stars in most clusters. 
However, the most evident differences
between 1G and 2G are in the Na abundances, with the 2G having
systematically higher Na, 
then strengthening the notion of the ChM being an optimal tool to separate
GC stellar populations with different light elements chemical content.

Remarkably, the [O/Fe] distribution of 1G stars is generally
consistent with typical observational errors for these spectroscopic
measurements, $\sim$0.10-0.20~dex. On the other hand, when more
measurements are available, 2G stars display wider oxygen spreads.
Some GCs might have 1G Na distributions somewhat wider than 
expected from observational errors alone, e.g.\,NGC\,5024, NGC\,6752, but
it is difficult to  make definitive conclusions given the very small number
of observed 1G objects.   

No large difference is seen between the Mg abundances of 1G and 2G
stars in our dataset, as plotted in Figure~\ref{fig:chartNaO}. The
average Mg values of 1G and 2G stars, listed in Table~\ref{tab:abb},
however suggest that there is some hint (in most cases a
$\sim$1~$\sigma$ differene) for 2G stars to be Mg-depleted with
respect to the 1G ones.  
Note indeed that for many GCs considered for Mg abundances here, Mg-Al
anticorrelations have been reported in the literature (see Table~\ref{tab:data}). 
The GCs with the most clear enhancement in Al in 2G stars are
47~Tucanae, NGC\,2808, NGC\,5986, NGC\,6205 and
NGC\,6752. Stars in the 2G of NGC\,2808 show 
a broad Al distribution, consistent with the very extended 2G on the ChM
of this GC.
Only one 2G star has Al measurements available for
$\omega$~Centauri and NGC\,5904, and in both cases the abundance is
higher than in 1G stars. Lower degree of Al enhancement is observed
for NGC\,5272.

Enhancements in silicon in 2G stars, linked to depletions in Mg and
enrichments in Al, suggest that the temperature of the H burning
generating this pattern exceeded ${T}_{6}\sim 65$~K (Arnould et
al.\,1999), because at this temperature the reaction
{\phantom {}}$^{27}$Al($p$,$\gamma$)$^{28}$Si becomes dominant over
$^{27}$Al($p$,$\alpha$)$^{24}$Mg. 
The upper panel of Figure~\ref{fig:all} suggests that in most GCs
there is no obvious evidence
for Si abundance variations between 1G and 2G stars, indicating that,
if present, they should be small. The
exceptions are NGC\,2808, where the 2G stars enriched in Al
(Figure~\ref{fig:chartNaO}) have on average higher Si, and
$\omega$~Centauri. Possible small Si enrichment in 2G is likely
present in NGC\,4833 and NGC\,6752.

From Figure~\ref{fig:all}, no remarkable difference is generally observed in 
the elements K, Ca, Fe, and Ba between 1G and 2G.
The fact that the elements shown in Figure~\ref{fig:all} do not display
any strong variation between 1G and 2G, suggests that the chemical
abundances for heavier elements 
are not significantly involved in shaping the ChMs of Type~I GCs. 
In general, we note that, as the 1G stars are typically a lower
fraction than the 2G stars (see Paper~IX), in most cases
only few stars are available in this population.  

To further compare the chemical composition of the different stellar
populations, we have calculated for each element the difference
between the average abundance of 2G and 1G stars (Table~\ref{tab:abb}). 
The histogram distribution of such differences are represented with grey-shaded
histograms for all the clusters (Type~I and Type~II together) in
Figure~\ref{fig:istogrammi}, where
we have marked the mean difference value with a black-dotted line. The
corresponding distribution for Type~I clusters only, is represented by the
black histograms. 
These histograms well illustrate the chemical elements that are
mostly involved in determining the distributions of stars along the
ChMs, namely N, O and Na. Indeed, among the inspected species, these
three elements have the highest mean difference in the
chemical abundances between  1G and 2G stars, purely selected on the ChM.
Still, it would be important to have far more stars with measured
nitrogen, as this element is expected to show the largest differences
between 1G and 2G stars and to drive much of the 1G-2G difference in the ChMs.  

Thus, the ChM is a very effective tool in separating stellar
populations with different light elements, as typically observed in
Milky Way GCs (Papers III and XIV).
The capability of ChMs to isolate stars with the typical chemical
composition of different stellar populations in GCs is also
illustrated in Figure~\ref{fig:nao},
where we plot the 1G and 2G stars on the Na-O plane. 
This figure represents the Na-O anticorrelation for the 20 Type~I GCs where
both Na and O abundances are available on the ChMs. 
The grey dots show the full spectroscopic samples, whereas the large green and magenta dots  
represent 1G stars and 2G stars, respectively, as identified on the
ChMs. As expected, 1G stars are clustered in the region of the Na-O
plane with high oxygen and low 
sodium, while 2G stars have, on average, lower oxygen and higher
sodium abundance. 

The role of each element in separating stars along the \x\ and \y\
axis of the ChMs has been investigated by exploring possible
correlations between the stellar abundance and the
pseudo-colours \x\ and \y. 
For each combination of element and pseudo-colour we have determined
the statistical correlation between the two quantities by using the
Spearman's rank correlation coefficient, $r$. The corresponding
uncertainty has been estimated as in Milone et al.\,(2014) by means of
bootstrapping statistics. To do this we have generated 1,000
equal-size resamples of the original dataset by randomly sampling with
replacement from the observed dataset. For each $i$-th resample, we
have determined $r_{\rm i}$ and considered the 68.27$^{\rm th}$
percentile of the $r_{\rm i}$ measurements ($\sigma_{\rm r}$) as
indicative of the robustness of $r$. 

In Table~\ref{tab:cor} we provide the derived $r$ values for the
entire sample of analysed stars and for the individual groups of
1G and 2G stars. These groups have
been analysed separately only when both photometry and spectroscopy are
available for five stars or more. 
An inspection of the results listed in Table~\ref{tab:cor} suggests
that there is no straightforward correlation for any element with
\x\ and \y\, with the exception of Na. 
Figure~\ref{fig:nay} reveals that in most clusters there is a strong
correlation between [Na/Fe] and \y\ as confirmed by the Spearman's
correlation coefficients listed in Table~\ref{tab:cor}. 
Among the clusters with the clearest Na-\y\ correlations, 
NGC\,2808, NGC\,5904, NGC\,6205, and NGC\,6752 exhibit a significant
[Na/Fe]-$\Delta_{\rm F275W, F814W}$ anticorrelation with 2G stars spanning a
wide range of Na. 
Oxygen anti-correlates with \y\ and correlates with \x\ in most
GCs (see Figs.~\ref{fig:oy} and ~\ref{fig:ox}).
Clusters with
extended ChMs, and larger internal variations in He, e.g.\, NGC\,2808,
show the most significant trends between the ChM and O abundances.

Although sodium is the element that best correlate with the ChM
pattern, this does not mean that Na is the driver of 2G differences from 1G in the ChMs.
Indeed, the sodium abundance itself does not directly affect any of
fluxes in the passbands used to construct the maps, which instead are
affected by nitrogen coming from the destruction of carbon and
oxygen. So, there must be a N-Na correlations and a N-O 
anticorrelation, but given the sparse determinations of N, they remain
to be adequately documented by spectroscopic observations for 
large samples of stars in each population observed in different GCs.

\subsection{A universal chromosome map and relation with chemical abundances}\label{sec:supermap}

In this Section we analyse the ChMs and chemical abundances of 1G and
2G stars, and try to envisage general patterns (if present) in the variegate zoo of
ChMs, that can be valid for all the GCs.
To this aim we examine here both Type~I and Type~II GCs, but for the
latter we consider only 1G and 2G stars (i.e. only blue-RGB).

The \x\ and \y\ width of the ChM dramatically changes from one cluster to another and
mainly correlates with the cluster metallicity. In particular, low-mass GCs
define a narrow correlation between the ChM width and [Fe/H] (e.g.
Figures\,20 and 21 from Paper~IX). This observational evidence is
consistent with the fact that a fixed variation in helium, nitrogen, and
oxygen provide smaller $m_{\rm F275W}-m_{\rm F814W}$ and $C_{\rm
F275W,F336W,F438W}$ variations in metal-poor GCs than in the metal-rich ones.
As an example, in the left panel of Figure\,\ref{fig:n6838n6397} we compare the ChMs
of NGC\,6397 (azure diamonds) and NGC\,6838 (gray dots), which are
two low-mass GCs with metallicities [Fe/H]=$-2.02$ and [Fe/H]=$-0.78$
(Harris\,1996, 2010 version). While both clusters exhibit a quite
simple ChM, the \x\ and \y\ extension of stars in NGC\,6838 is significantly
wider than that of NGC\,6397.

In an attempt to compare the ChMs of GCs with different metallicities
we defined for each cluster the quantities:
\begin{equation}
  \dx=\frac{\x-\x^{\rm 1G,0}}{D1}
\end{equation}
where \x$^{\rm 1G,0}$ is the median value of \x\ for 1G stars, and $D1=$0.14$\times$[Fe/H]$+$0.44
is the straight line that provides the best fit of GCs with $M_{\rm
V}<-7.3$ in the \x\ vs.\,[Fe/H] plane, and

\begin{equation}
\dy =\frac{\y-\y^{\rm 1G,0}}{D2}
\end{equation}
where \y$^{\rm 1G,0}$ is the median value of \y\ of 1G stars, and
$D2=$0.03$\times$[Fe/H]$^{2}+$0.21$\times$[Fe/H]$+$0.43
is the square linear function that provides the best fit of GCs with
$M_{\rm V}<-7.3$ in the $C_{\rm F275W,F336W,F438W}$ vs.\,[Fe/H] plane.

The resulting \dy\ vs.\,\dx\ plots of NGC\,6397 and NGC\,6838 RGB stars in shown in
the right panel of Figure\,\ref{fig:n6838n6397} and reveals that these
{\it normalised} ChMs almost overlap
each other. This fact suggests that the dependence on metallicity of
the classical ChM is significantly reduced in this plane.

Now, that the metallicity-dependance has been reduced, we can compare
the ChMs for all the clusters in our sample on the 
\dx-\dy\ plane, and obtain an universal ChM. 
The result of this comparison is shown in Figure~\ref{fig:SuperMap}.
The \dy\ vs.\,\dx\ Hess diagram for all GCs with
[Fe/H]$<-0.6$ is plotted in panel (a), where we also show
the \dy\ and \dx\ histogram distributions for all the analyzed
stars. The Hess diagram plotted in the panel (b) is derived in such a
way that the stars of each GC have been normalized to the total number of stars in that cluster. 

These universal maps are useful tools to investigate the global
properties of the multiple stellar populations phenomenon in GCs.
Noticeably:
\begin{itemize}
\item{two major overdensities are observed on the \dy-\dx\ plane;}
\item{a clear separation between 1G and 2G stars exists that corresponds to \dy$\sim 0.25$;}
\item{as expected the first overdensity corresponds to the 1G population, which
appears to occupy a relatively narrow range in \dy. The extension in
\dx\ is larger, but most stars are located within
$-$0.3$\lesssim$\dx$\lesssim +$0.3;} 
\item{the bulk of 2G stars are clustered around \dy$\sim 0.9$ with a poorly-populated tail 
of stars extended towards larger values of \dy. This suggests that
even if the ChMs is variegate, the dominant 2G in GCs has similar properties.}
\end{itemize}

The universal ChM allows us to investigate the global variations in
light elements in the overall sample of analysed GCs, in the \dx-\dy\ plane.
Panels (c1)--(c4) of Figure~\ref{fig:SuperMap} represent \dy\ as a function of the O, Na, Mg, and
Al abundance ratios relative to the average abundances of 1G stars. 
We find significant correlation with $\delta$[Al/Fe] and
$\delta$[Na/Fe]. The \dy--$\delta$[Na/Fe] relation is well 
reproduced by the straight line \dy=1.72$\times \delta$[Na/Fe]$+$0.18 that is
obtained by the least-squares fit.  
The tight relation and the high value of the Spearman's rank
correlation coefficient ($r=+$0.73) suggests that such relation could be
exploited for empirical determination of the relative sodium abundance
in GCs. 
We also find significant anticorrelations between \dy\ and
$\delta$[O/Fe] and $\delta$[Mg/Fe]. In the latter case, we note that
only stars with \dy$\gtrsim 0.8$ exhibit magnesium variations, meaning
that Mg-depleted stars are located in more extreme regions of ChMs.

Panels (d1)--(d4) show the relation between \dx\ and $\delta$[O/Fe],
$\delta$[Na/Fe], $\delta$[Mg/Fe] and $\delta$[Al/Fe]. 
In these cases the significance of the correlations is lower than for
\dy\ as indicated by the Spearman's rank correlation coefficients
quoted in the figure.

\subsection{Search for intrinsic abundance variations among 1G stars}

Already in Paper~III it was shown that 1G stars in NGC\,2808 exhibit a spread
in \x\ with two major clumps that we have named A and B. The chemical
composition of the sub-population B is constrained by the abundances
derived by Carretta et al.\,(2006) showing that these stars have the
same composition of halo field stars with similar
metallicities. Unfortunately, no spectroscopic measurements are available for
sub-population A stars. A comparison of the appropriate synthetic spectra and the observed
colours then revealed that population B is consistent with having an
helium abundance $\sim$0.03 higher in mass fraction with respect to population-A stars
(Paper\,III). Alternatively, the 1G spread in \x\ may be ascribed  
to population-A stars being enhanced in [Fe/H] and [O/Fe] by $\sim$0.1~dex
with respect to population-B stars (Paper\,III and D'Antona et
al.\,2016), but having the same helium.  

The wide spread in \x\ among Type~I clusters was fully documented in
Paper~IX, measuring the full \x\ width $W^{\rm 1G}$ for all the program
clusters and showing that it mildly correlates with the cluster mass,
and can reach up to $\sim 0.3$ mag in some clusters being negligibly
small in others. Moreover, the 1G width $W^{\rm 1G}$ mildly correlates
with the 2G width $W^{\rm 2G}$ as well, and in several cases it even
exceeds it. 

The discovery that even 1G stars do not represent a chemically
homogeneous population has further complicated, if possible, the
already puzzling phenomenon of multiple stellar populations in
GCs. The helium abundance differences among both 1G and 2G stars has
then been thoroughly investigated in Paper~XVI, where it was shown
that, if the 1G width is entirely due to a helium spread then the
helium variations among 1G stars changes dramatically from one cluster
to another, ranging from  
$\delta Y^{\rm 1G}\simeq 0$ to $\sim 0.12$, with an average $<\!\delta
Y^{\rm 1G}\!>\simeq 0.05$. However, 
we have been unable to find a process that would enrich material in
helium without a concomitant  production of nitrogen at the expenses
of carbon and oxygen (Paper~XIV). Thus, the alternative option of a metallicity
spread among 1G stars is still on the table. 

We note that the presence of a spread in iron, or oxygen or any other
chemical species among 1G stars would result in a positive correlation
between the \x\ and the abundance
of these elements.  Therefore, the Spearman's rank correlation
coefficients between these 
quantities listed in Table~\ref{tab:cor} may provide further insights on
the origin of the 1G spread. However, from Table~\ref{tab:cor} we do
not find strong evidence for positive a correlation between \x\ and any element
abundance in the analysed GCs, with few exceptions.

%%%%%%%%%%%%%%%%%%%%%%%%%%%%%%%%%%%%% FIG 2 %%%%%%%%%%%%%%%%%%%%%%%%%%%%%%%%%%%
\begin{centering}
\begin{figure*}
\   \includegraphics[width=12cm]{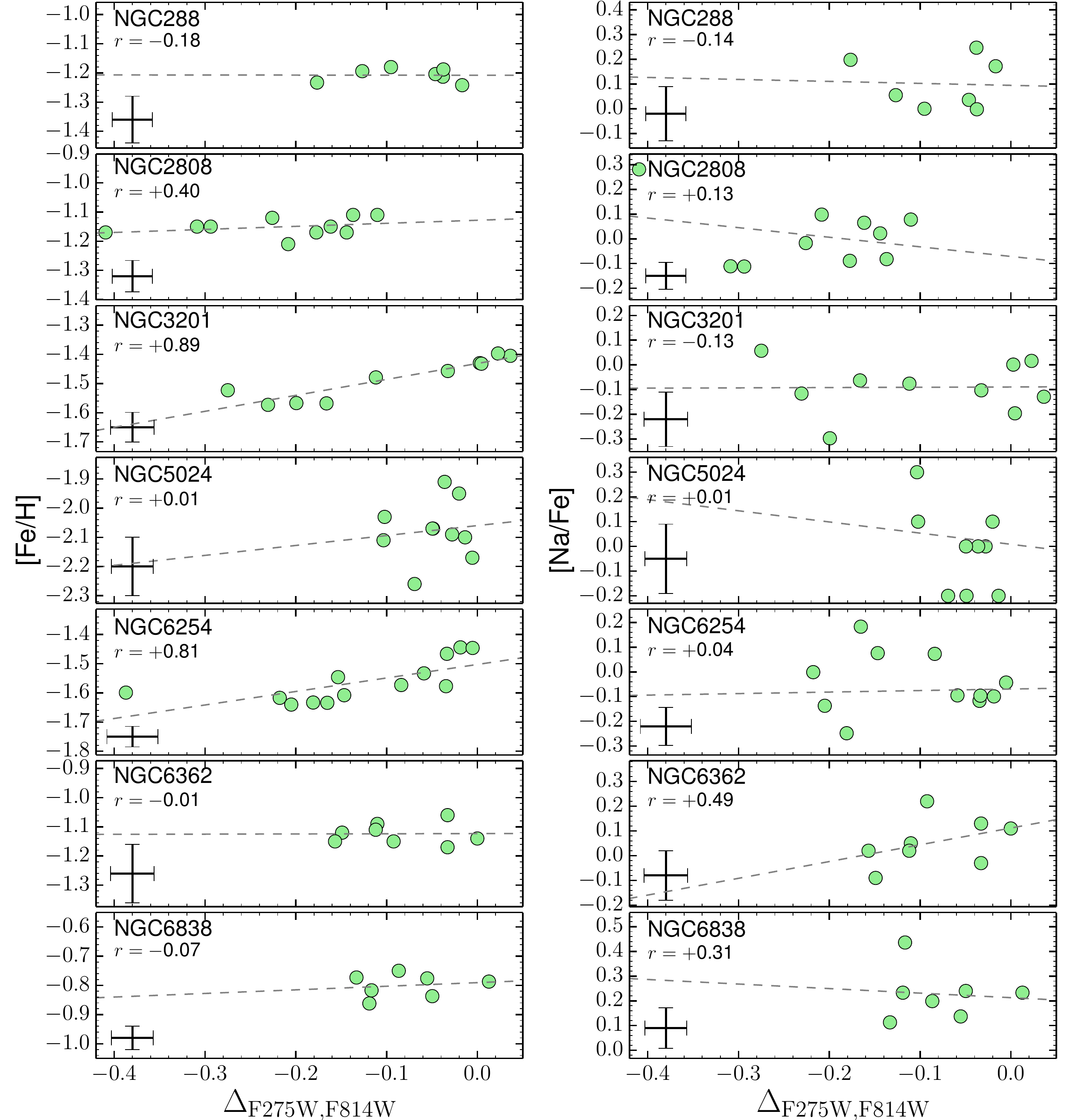}
  \caption{[Fe/H] (left panels) and sodium abundances relative to
    Fe (right panels) as a function of \x\ for the 1G stars. Only the GCs for
    which at least seven  stars are available on the 1G region of the
    ChMs have been considered. In each panel, we report the Spearman
    correlation coefficient. The typical error bars for Na and Fe are taken from
    the reference papers listed in Table~\ref{tab:data}. Error bars
    for \x\ are from Paper~IX.}
 \label{fig:Fe1G}
\end{figure*}
\end{centering}
%%%%%%%%%%%%%%%%%%%%%%%%%%%%%%%%%%%%%%%%%%%%%%%%%%%%%%%%%%%%%%%%%%%%%%%%%%%%%%%

In Figure~\ref{fig:Fe1G} we show the [Fe/H] and [Na/Fe] abundances as
a function of \x\ for the clusters for which data for at least seven stars are
available.  
The \x-[Fe/H] plots suggest that there is no significant correlation
in five out of the seven clusters. Only in the case of NGC\,3201 and
NGC\,6254 we have positive and significant correlations ($r \gtrsim 0.8$). 
The fact that NGC\,3201 and NGC\,6254 also exhibit a very-extended
sequence of 1G makes it tempting to speculate that small internal iron
variations among the 1G stars in these clusters may be responsible for 
the 1G spread in \x. 
We note here that for NGC\,3201 the presence of intrinsic
metallicity spread has been proposed by Simmerer et al.\,(2013; see
also Kravtsov et al.\,2017). However, Mucciarelli et al.\,(2015)
found that the RGB sample analysed in this GC does not show any
evidence for intrinsic variations in Fe. An investigation of
connection between a possible Fe variation and the 1G extension in
\x\ is still undergone. 

The lack of a significant trend in the other GCs suggests that iron
variations, if present, are smaller and not detectable, which might
also be due to the smaller range in \x\ (explored)\footnote{As an
  example in NGC\,2808 the full range in \x\ for 1G stars has not been
properly analysed in terms of chemical abundances.} in these
clusters.   
Still, the small number of analysed 1G stars prevents
us from any definitive conclusion even in the more promising  cases of
NGC\,3201  
and NGC\,6254. It is also worth noting  that small spurious
metallicity variations can be actually be due to systematic errors in the
determination of the stellar effective temperature. Indeed Carretta et
al.\,(2009) have determined the effective temperature by assuming that
all the stars have the same helium and metallicity, and projecting
their stars on one single fiducial, which may not be realistic for such
large variations in \x.  

Remarkably, the lack of any significant correlation between \x\ and Na
abundances (right panels in Figure~\ref{fig:Fe1G}) suggests that light
element abundances, likely including C, N and O, can be ruled out as
the responsible for the high spread in 1G stars observed in some GCs.
We conclude this section by emphasising the need to further investigate
this phenomenon with chemical abundances obtained from high-resolution
spectroscopy for many more stars than done so far.
  
%%%%%%%%%%%%%%%%%%%%%%%%%%%%%%%%%%%%%%%%%%%%%%%%%%%%%%%%%%%%%%%%%%%%%%%%%%%%%%%
\begin{centering}
\begin{figure*}
\includegraphics[width=17cm]{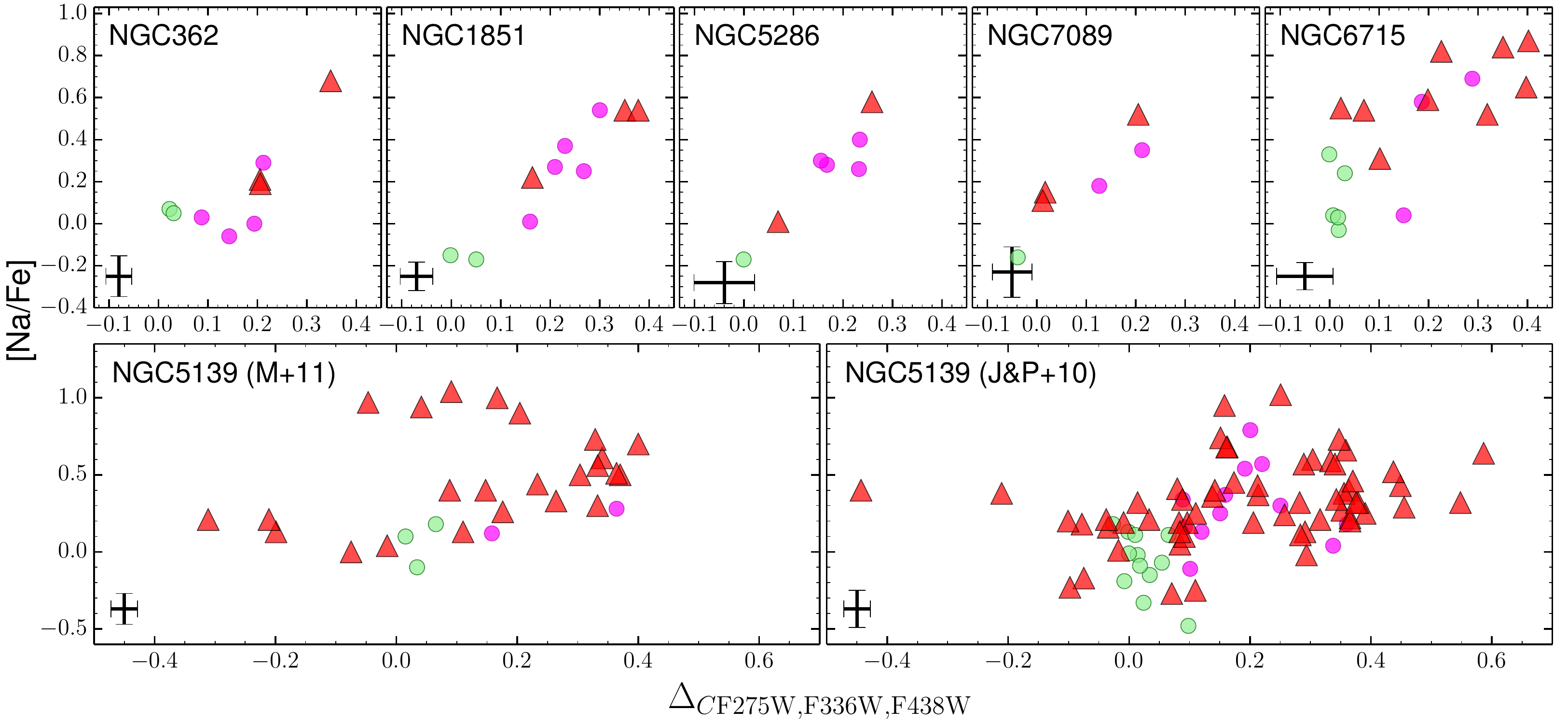}
\caption{Sodium abundance as a function of \y\ for a sample of Type~II
  clusters. Blue-RGB 1G and 2G stars are represented with green and
  magenta dots, respectively, while red-RGB stars are plotted with
  red triangles, without distinguishing between the corresponding 1G
  and 2G. The typical error bars for Na are taken from the reference
  papers listed in Table~\ref{tab:data}. Error bars for \y\ are from
  Paper~IX.}    
 \label{fig:naANO}
\end{figure*}
\end{centering}
%%%%%%%%%%%%%%%%%%%%%%%%%%%%%%%%%%%%%%%%%%%%%%%%%%%%%%%%%%%%%%%%%%%%%%%%%%%%%%%

%%%%%%%%%%%%%%%%%%%%%%%%%%%%%%%%%%%%%%%%%%%%%%%%%%%%%%%%%%%%%%%%%%%%%%%%%%%%%%%
\begin{centering}
\begin{figure*}
\includegraphics[width=16.5cm]{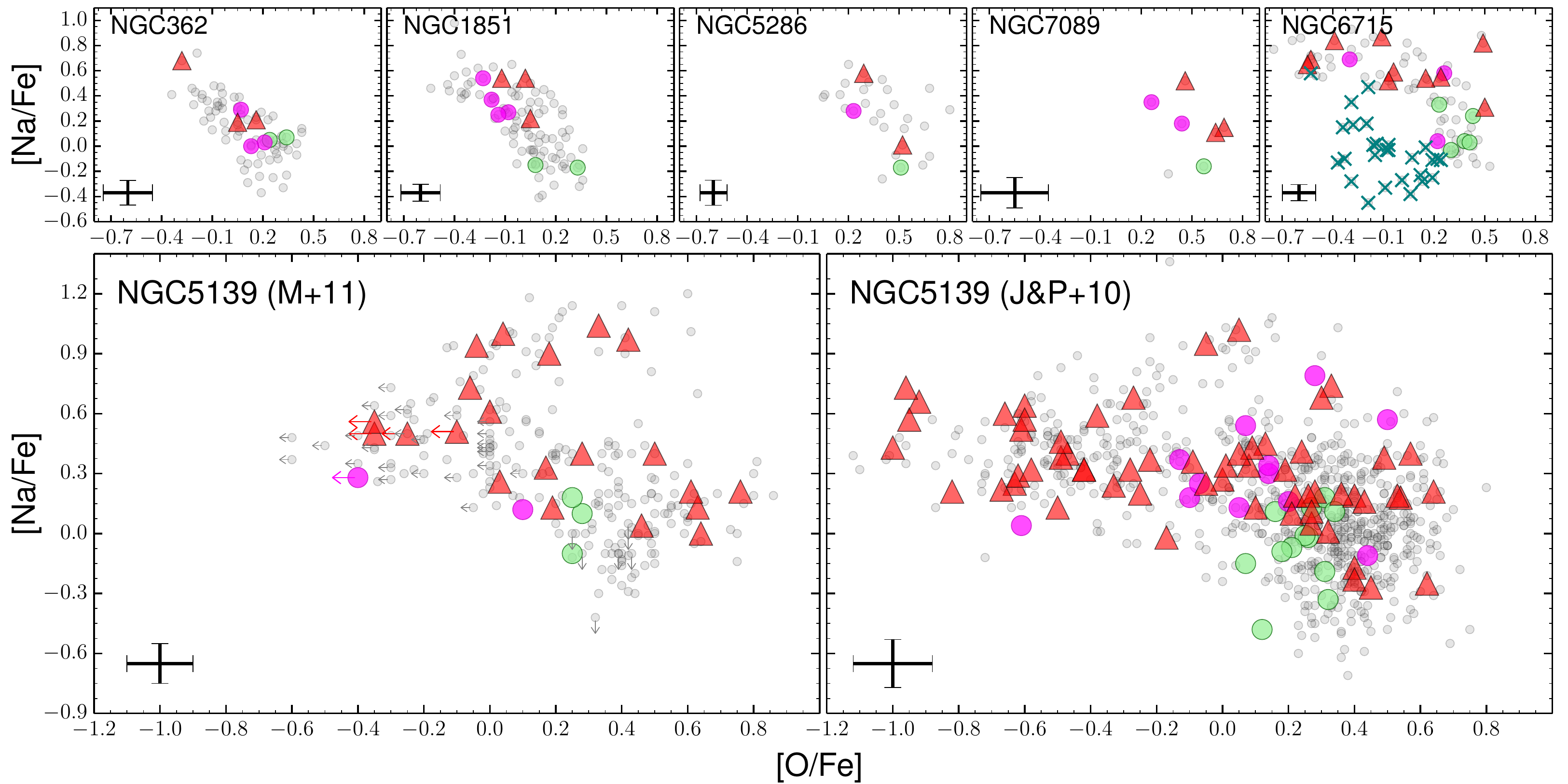}
\caption{[Na/Fe] vs.\,[O/Fe] for seven Type~II clusters with available
  abundances. The symbols are the same as in Figure~\ref{fig:naANO} with
  blue crosses in the NGC\,6715 (M\,54) panel indicating stars of the
  Sagittarius dwarf galaxy. The typical error bars for O and Na are taken
  from the reference papers listed in Table~\ref{tab:data}. } 
\label{fig:NaOano}
\end{figure*}
\end{centering}
%%%%%%%%%%%%%%%%%%%%%%%%%%%%%%%%%%%%%%%%%%%%%%%%%%%%%%%%%%%%%%%%%%%%%%%%%%%%%%%

\subsection{Type~II GCs}\label{sec:typeII}

In addition to the 1G and 2G observed in Type~I GCs, the ChMs of
Type~II clusters  
display stellar populations extending on redder colours, and
distributing on a distinct red-RGB on the $m_{\rm F336W}$ vs.\,$m_{\rm
  F336W}-m_{\rm F814W}$, or $U$ vs. $(U-I)$ CMDs
(Sections~\ref{sub:photHST} and \ref{sub:photGB}). 
On the chemical side, the presence of stellar populations with
different metallicity and content of $s$-process elements is the
main distinctive features of the ``anomalous'' GCs as were defined in Marino et
al.\,(2015) from pure chemical abundance evidence. 
The fact that all the Type~II GCs analysed spectroscopically exhibit
star-to-star variations in heavy elements suggested that the class of
Type~II GCs, identified from photometry, corresponds to the ``anomalous''
GCs previously identified from spectroscopy. 

In this section we explore the chemical abundances-ChMs connections for 
all the stellar populations observed in Type~II GCs:
the 1G and 2G components of the blue-RGB, corresponding to those discussed in
Section~\ref{sec:typeI} for Type~I GCs, and the red-RGB populations.
ChMs in Figure~\ref{fig:mapA} immediately suggest that the red-RGB
component includes stars at different \y\ implying that, similarly to
the blue-RGB, it hosts both 1G and 2G stars.

From the ChMs presented in Paper~IX, as well as here from
Figure~\ref{fig:mapA}, one can see that among Type~II GCs the
red-RGB/blue-RGB number ratio varies considerably from cluster to
cluster, whereas  
for most GCs the 2G stars (with higher \y\ values) are the dominating
stellar component in the red-RGB. 
This sets an important constraint when trying to imagine the sequence
of events that led to the two distinct (blue and red) populations in
these special clusters. 
We now pass to discuss the chemical tagging of the various
subpopulations of this group of clusters as identified on their ChMs. 

%%%%%%%%%%%%%%%%%%%%%%%%%%%%%%%%%%%%%%%%%%%%%%%%%%%%%%%%%%%%%%%%%%%%%%%%%%%%%%%
\begin{centering}
\begin{figure*}
\includegraphics[width=17cm]{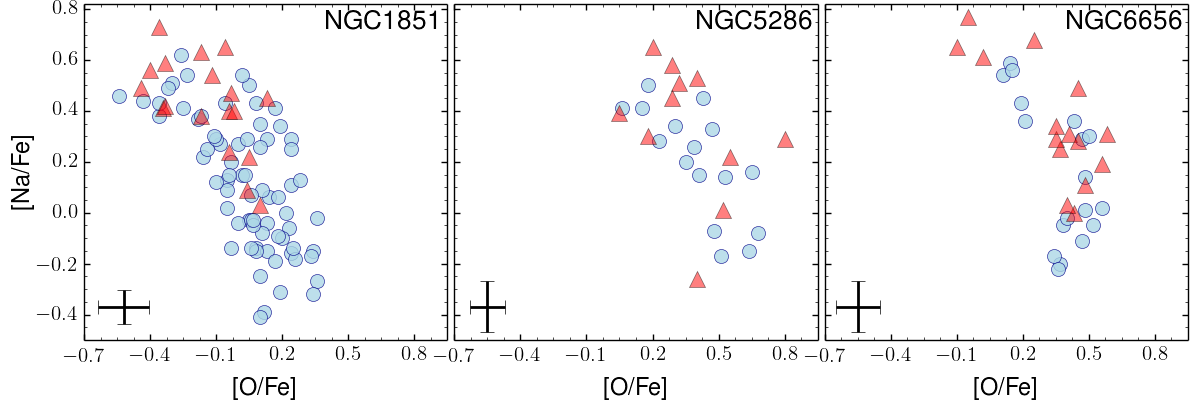}
 \caption{An expanded view of the [Na/Fe] vs.\,[O/Fe] anticorrelation
   for the clusters NGC\,1851, NGC\,5286 and NGC\,6656 (M\,22) with
   blue-RGB and red-RGB stars being shown as light-blue circles and
   red triangles, respectively. The typical error bars for O and Na are taken
  from the reference papers listed in Table~\ref{tab:data} and from
  Marino et al.\,(2009, 2011) for M\,22.} 
 \label{fig:NaOanoG}
\end{figure*}
\end{centering}
%%%%%%%%%%%%%%%%%%%%%%%%%%%%%%%%%%%%%%%%%%%%%%%%%%%%%%%%%%%%%%%%%%%%%%%%%%%%%%%

%%%%%%%%%%%%%%%%%%%%%%%%%%%%%%%%%%%%% FIG 2 %%%%%%%%%%%%%%%%%%%%%%%%%%%%%%%%%%%
\begin{centering}
\begin{figure*}
\includegraphics[width=17cm]{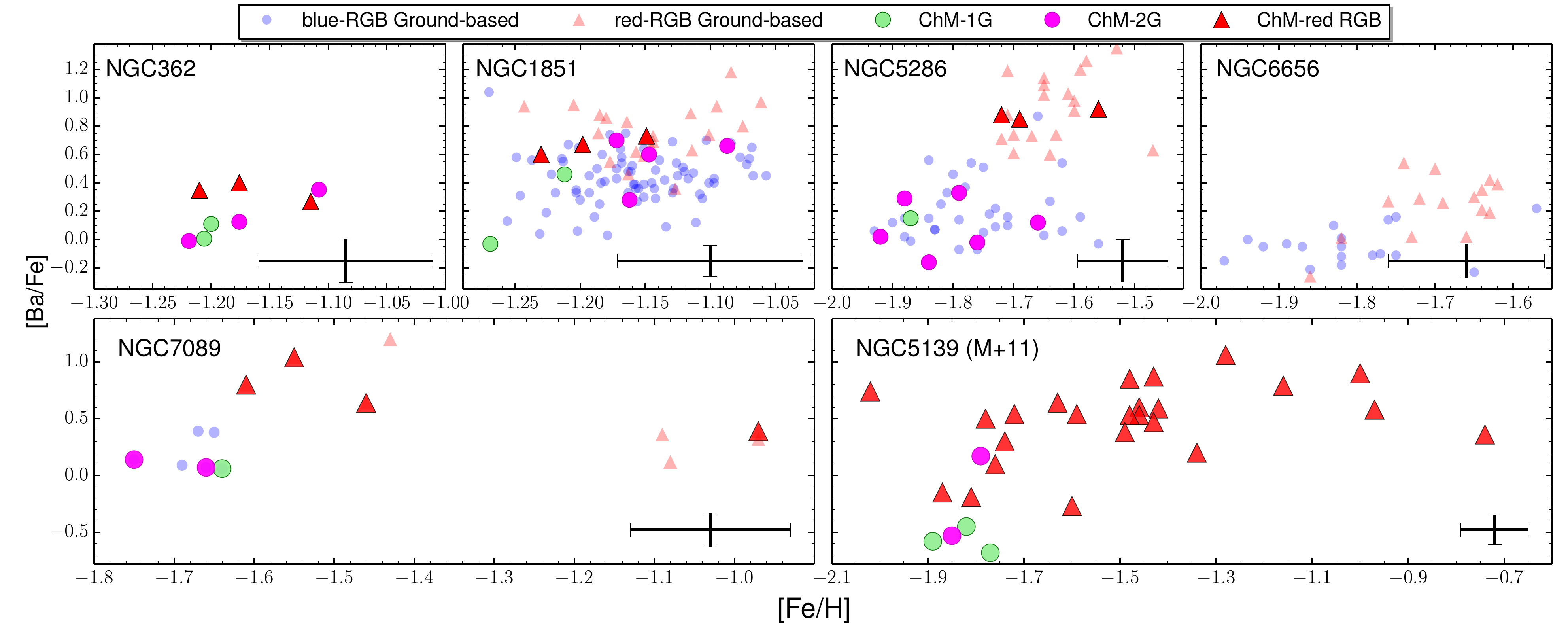} 
\caption{Barium as a function of iron abundance for Type\,II GCs.
Symbols are as in Figure~\ref{fig:NaOano}.
For NGC\,1851, NGC\,5286, NGC\,6656, and NGC\,7089 blue and red-RGB
stars with available ground-based photometry have been represented as
small blue dots and red triangles, respectively. A legend of the
symbols is shown on the top of the figure. The typical error bars for
Fe and Ba are taken from the reference papers listed in
Table~\ref{tab:data} and from Marino et al.\,(2009, 2011) for M\,22.}  
 \label{fig:FeBaANO}
\end{figure*}
\end{centering}
%%%%%%%%%%%%%%%%%%%%%%%%%%%%%%%%%%%%%%%%%%%%%%%%%%%%%%%%%%%%%%%%%%%%%%%%%%%%%%%

As shown in Figure~\ref{fig:all}, lithium for red-RGB stars is
available only for one star in NGC\,362, and 28 stars in
$\omega$~Centauri. Keeping in mind that our sample includes only one
star, we note that the red-RGB star in NGC\,362 has lower Li than
blue-RGBs. Red-RGB stars seem to have on average slightly lower Li
also in $\omega$~Centauri, which will be discussed in more details in
Sections~\ref{sub:Li} and \ref{sub:LiStreams}. 

Among Type~II GCs, N abundances are available on the ChM only for
three blue-RGB 1G stars and 13 red-RGB stars in $\omega$~Centauri. 
Figure~\ref{fig:chartNaO} shows that
the [N/Fe] abundances distribution in the red-RGB stars spans more than one dex,
in agreement with the presence of a strong C-N anticorrelation among the stars
in this cluster (e.g.\,Marino et al.\,2012).

The same figure shows also that in $\omega$~Centauri and
NGC\,6715 (the Type~II GCs with abundances available for a larger number of
stars) the red-RGB stars have larger variations in O, and  
extend to higher Na abundances compared to blue-RGB stars.  
Although for a small number of stars, these features are observed
also in the other Type~II GCs plotted in Figure~\ref{fig:chartNaO}. If
the large spreads in O and Na among red-RGB stars are consistent with
the presence of internal anticorrelations in this stellar group
(e.g. Marino et al.\,2009, 2011; Johnson \& Pilachowski\,2010), it is
noteworthy that there is a tendency of red-RGBs to have higher mean Na
than blue-RGBs. In our sample we do not detect any significant
difference in neither Mg nor Al between blue and red-RGB stars. We
note however that Yong et al.\,(2014) find small increase in these
elements in red-RGB stars in M\,2. 

From Figure~\ref{fig:all}, the typical $\alpha$ elements Si
and, on a less degree, Ca appear to increase going from the blue- to the red-RGB
stars in $\omega$~Centauri. 
This trend is seen for Si in M\,2 and Ca in M\,54.

Perhaps most importantly, differences are observed in the iron
abundance between blue- and red-RGB stars, no matter whether belonging
to the respective 1G or 2G. 
In $\omega$~Centauri, NGC\,5286, M\,54 and
M\,2 red-RGB stars have higher [Fe/H], as already documented in
the literature (e.g.\,Marino et al.\,2015; Carretta et al.\,2010a;
Yong et al.\,2014). Higher Ba abundances 
characterise the red-RGB stars of all the Type~II GCs in our sample. 
These results indicate that stars chemically-enriched in metallicity (Fe) 
and $s$-process elements have a specific location on the ChMs, which is
on redder-\x\ 1G and 2G sequences shown in
Figure~\ref{fig:mapA}{\footnote {By analysing four stars in the
    Type~II GC NGC\,6934, Marino et al.\,(2018) find no evidence for
    chemical enrichment in the $s$-elements among red RGB stars, that
    resulted instead to be enriched in Fe. Given the small sample size
  and the fact that the red RGB analysed are only at low \y, the
  authors did not exclude $s$-process elements enrichment among other
  red RGB stars in this cluster.}} .

%%%%%%%%%%%%%%%%%%%%%%%%%%%%%%%%%%%%% FIG 2 %%%%%%%%%%%%%%%%%%%%%%%%%%%%%%%%%%%
\begin{centering}
\begin{figure*}
   \includegraphics[width=17.4cm]{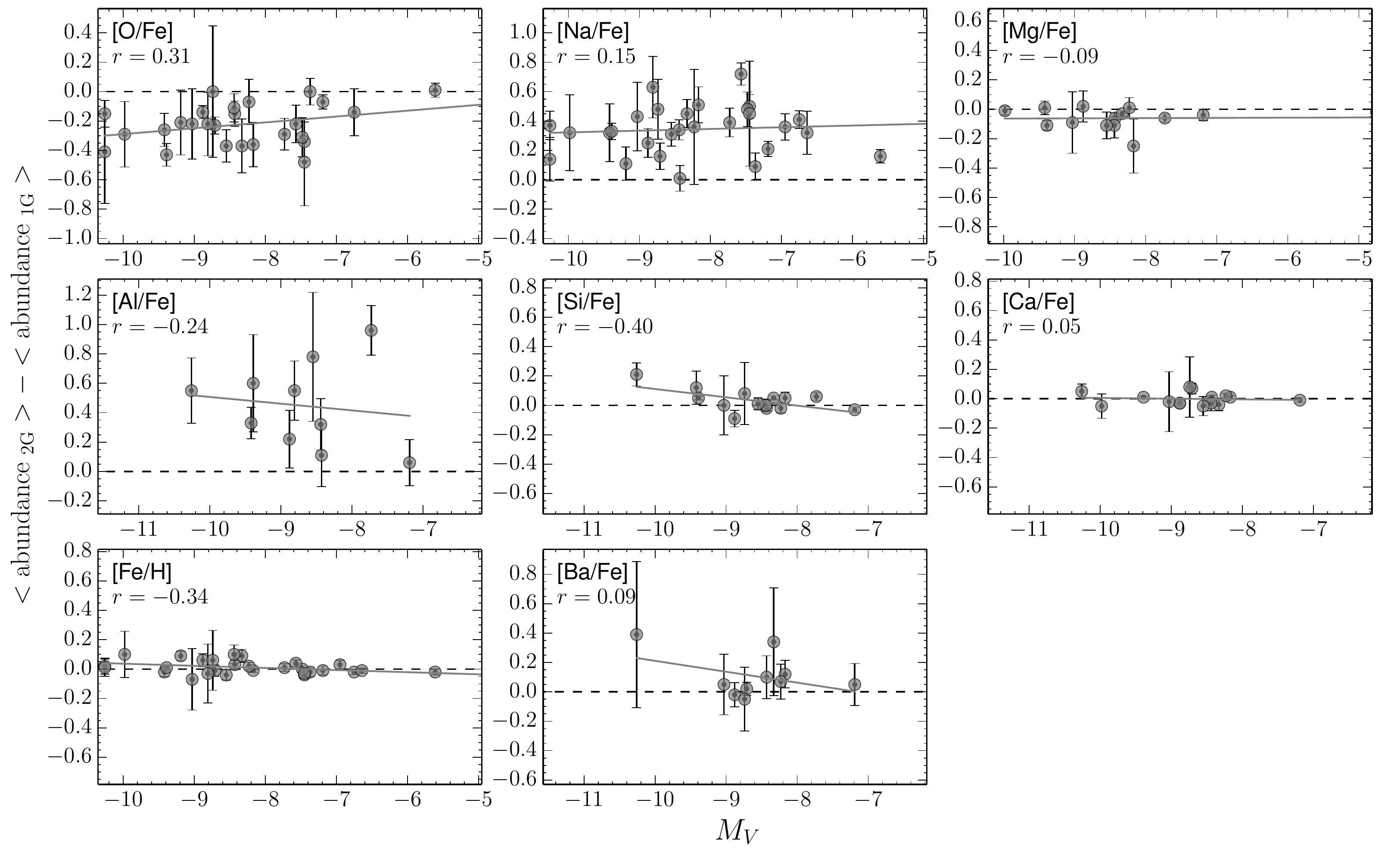}
  \caption{Abundance differences relative to Fe 
    $<\mathrm {abundance_{\ 2G}}> - <\mathrm {abundance_{\ 1G}}>$ (and relative to H in 
    the case of Fe) between 2G and 1G stars (from
    Table~\ref{tab:abb}) as a function of the absolute magnitude $M_{V}$. 
    Error bars are the square root sum of the statistical errors
    associated with the average values (r.m.s.$/\sqrt{(N-1)}$,
    with $N$ the number of measurements) of 1G and 2G stars; when only
    one measurement is available, the statistical error has been
    assumed equal to 0.20~dex. The dashed black line indicates no
    difference. For comparison purposes, the {\it x} and {\it y} axis
    have the same size in all the panels.  
    Each panel reports the Spearman correlation coefficient $r$.}
 \label{fig:2Gminus1G}
\end{figure*}
\end{centering}
%%%%%%%%%%%%%%%%%%%%%%%%%%%%%%%%%%%%%%%%%%%%%%%%%%%%%%%%%%%%%%%%%%%%%%%%%%%%%%%

%%%%%%%%%%%%%%%%%%%%%%%%%%%%%%%%%%%%% FIG 2 %%%%%%%%%%%%%%%%%%%%%%%%%%%%%%%%%%%
\begin{centering}
\begin{figure*}
\   \includegraphics[width=14cm]{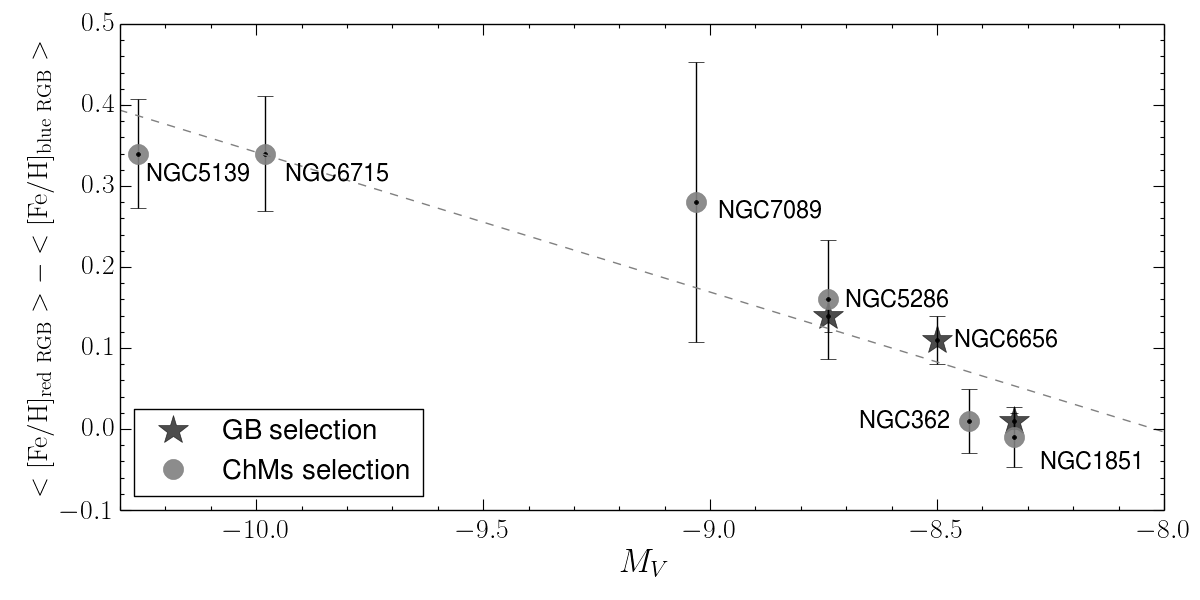}
  \caption{Differences in [Fe/H] between red and blue RGB stars as a
    function of the absolute magnitude for Type~II GCs. The differences
    obtained from the ChMs selection are represented as grey-filled
    circles; for NGC\,5286, NGC\,1851, and NGC\,6656 (M\,22) we plot the Fe
    differences between the red and blue RGBs as selected from the
    ground-based (GB) CMDs described in Section~\ref{sub:photGB}
    (star-like symbols). 
    Error bars are the square root sum of the statistical errors
    associated with the 
    average values (r.m.s.$/\sqrt{(N-1)}$, with $N$ the number of
    measurements) of blue and red RGB stars. The dashed grey line is
    the best-fit with the results obtained from the ChMs selection
    (grey dots), plus NGC\,6656. }
 \label{fig:FeMv_ANO}
\end{figure*}
\end{centering}
%%%%%%%%%%%%%%%%%%%%%%%%%%%%%%%%%%%%%%%%%%%%%%%%%%%%%%%%%%%%%%%%%%%%%%%%%%%%%%%

Similarly to what has been done for 1G and 2G components of Type~I
clusters (see Section~\ref{sec:typeI}), 
we have calculated for each element the difference
between the average abundance of the red-RGB and the blue-RGB stars
(lumping together the corresponding 1G and 2G) and we 
have plotted the distribution of such difference in
Figure~\ref{fig:istogrammi} by using red-shaded histograms.  
The corresponding mean differences are marked with red vertical lines. 
From these results, we see that blue and red-RGB stars abundance
differences in Type~II GCs are dominated by Fe and Ba
($s$-elements) variations.

The sodium abundances are nicely correlated with \y\, for both blue-
and red-RGB stars, as displayed in Figure~\ref{fig:naANO}, just like
in Type~I GCs (see Figure~\ref{fig:nay}). The red-RGB stars show their own
internal range in \y, as well as in Na, and tend to overlap with the
blue-RGB 2G stars in these plots. 
Data plotted in Figure~\ref{fig:naANO} for NGC\,362, NGC\,1851,
NGC\,5286, NGC\,7089, and NGC\,6715 suggest indeed that the red-RGB
stars are shifted towards both slightly higher \y\ and Na values 
than blue-RGB ones.

Sodium abundances are plotted against oxygen abundances in Figure~\ref{fig:NaOano}. 
Blue- and red-RGB stars do not appear to segregate from each others in
these plots, emphasising that these elements are not responsible for
the splitting of the ChMs of these clusters.
As a general trend, 1G and 2G stars (no matter whether blue or red)
are located on the Na-poor/O-rich and on the Na-rich/O-poor regions, respectively.
Similarly to blue-RGB stars, red-RGB ones exhibit their own Na-O anticorrelation.

For completeness, together with the Type~II GC M\,54 (NGC\,6715), we
show in Figure~\ref{fig:NaOano} the stars belonging to the
Sagittarius dwarf galaxy (the likely host of M\,54), as identified by
Carretta et al.\,(2010a) from their position in the CMD. While most of
these stars are clustered around ([O/Fe];[Na/Fe])$\sim$($-0.1$;$-0.2$)
we note a few stars with higher sodium and, possibly, lower oxygen may
be present.  
However, there is no strong evidence for the Sagittarius stars
exhibiting the 1G/2G dichotomy which is ubiquitous among GCs. 
Unfortunately the available data do not allow us to decide whether  the
few stars with high [Na/Fe] values are either contaminants  from
NGC\,6715 stars, or spectroscopic errors,  or if a small fraction of
Na-rich (2G-like) stars is also present in this dwarf galaxy. 

The Na-O anticorrelation of RGB stars in $\omega$~Centauri
is shown in the lower panels of Figure~\ref{fig:NaOano} (Marino et
al.\,2011b, left panel and Johnson \& Pilachowski 2010, right
panel).
Similar to Type~I GCs, blue-RGB 1G stars have high O and low Na
abundance and 2G stars are enhanced in Na and depleted in O; and, 
similarly to the other Type\,II GCs, red-RGB stars define their own
Na-O anticorrelation. 
Furthermore, as previously noted, it appears that red-RGB stars  
have, on average, slightly higher Na abundance than blue-RGB stars. 
Data are consistent with a Na-O anticorrelation for red-RGB stars
largely overlapping with that of blue-RGB stars, slightly extended
towards higher Na, a pattern that can be recognised also in some upper
panels of Figure~\ref{fig:NaOano}. 
The extreme ChM of this cluster displays the presence of well-extended
red streams which will be discussed in further details in
Section~\ref{sec:omegacen}.  

For NGC\,1851, NGC\,5286, and NGC\,6656 we took advantage of %our
ground-based photometry (Section~\ref{sub:photGB}) to increase
our sample of blue- and red-RGB stars from the CMDs of
Figure~\ref{fig:GB}, and plotted them in Figure~\ref{fig:NaOanoG}. 
The figure confirms that blue- and red-RGB stars follow basically the same 
Na-O anticorrelation, with a slight predominance of low-O and high-Na
in red RGB stars, in particular in NGC\,1851. 

The majority of Type~II GCs that have been properly studied 
spectroscopically exhibit internal metallicity variations, with NGC\,362
and NGC\,1851 being possible exceptions (Carretta et al.\,2013;
Villanova et al.\,2010). 
Iron abundance has been investigated in 26 clusters studied in Papers
I and IX, including six Type~II GCs. [Fe/H] is also available for
blue- and red-RGB stars identified by using ground-based photometry
in four Type~II GCs including NGC\,6656 for which there are no stars
with jointly available spectroscopy and ChM photometry. 
We find that red-RGB stars are significantly enhanced in iron with
respect to the remaining RGB stars in most Type~II GCs and the average
iron difference ranges from $\sim$0.15 dex for NGC\,5286 to $\sim 1$
dex for $\omega$\,Centauri. In NGC\,362 and NGC\,1851 there appears to
be no appreciable iron difference between red- and blue-RGB stars. 
However, we note that the absence of a significant
difference between the [Fe/H] of red and blue RGB stars in NGC\,1851
(found here and by Lardo et al.\,2012)  
appears to be at odd with the Gratton et al.\,(2012) finding that the faint
SGB stars of this cluster are more-metal rich than those on the bright SGB.
Clearly, whether a small metallicity difference exists in this cluster
(and in NGC\,362) between the blue- and the red-RGB stars is yet to be
firmly established. 

As barium is the most commonly studied $s$-process element in GCs, 
it has been taken as representative of this group of chemical species. 
Red-RGB stars are significantly enhanced in Ba with respect to blue-RGB stars. 
This is illustrated in Figure~\ref{fig:FeBaANO} where we plot [Ba/Fe]
as a function of [Fe/H] for six Type~II GCs. 
These results confirm that stars enhanced in both iron and barium
populate the red-RGB of NGC\,6656 and NGC\,5286 (Marino et al.\,2009,
2011a, 2015). Moreover, Figure~\ref{fig:FeBaANO} shows that stars of
NGC\,5286, NGC\,6656, NGC\,7089 and $\omega$\,Centauri follow similar
patters in the [Ba/Fe] vs.\,[Fe/H] plane as early suggested by Da
Costa \& Marino (2011) and Marino\,(2017) for the
NGC\,6656-$\omega$\,Centauri pair.    

The connection between iron and $s$-process elements seems more
controversial for NGC\,1851. Similarly,
NGC\,362 apparently does not follow the [Ba/Fe] vs.\,[Fe/H] trend
observed in other Type~II GCs. 

The evidence presented in this section definitively demonstrates that
Type~II GCs identified 
from photometry correspond to the class of ``anomalous'' GCs that were 
spectroscopically identified (Marino et al.\,2009; 2015). 
The evidence that blue- and red-RGB correspond to stars with different
abundance of heavy elements is consistent with previous 
findings that the $s$-rich and $s$-poor stars in ``anomalous'' GCs
host stars with different light-element abundance (Marino et
al.\,2009, 2011a, 2015; Carretta et al.\,2010b; Yong et al.\,2014). 

All in all, these  facts demonstrate that ChMs offer an efficient tool
to identify GCs with intrinsic heavy-elements variations. 

\subsection{Relations with the parameters of the host GC}\label{sec:correlations} 

Having measured the mean chemical abundances for 1G and 2G stars as
identified on the ChMs, we have investigated 
their relations with GCs' metallicity and absolute magnitude M$_{V}$.  
To this aim, we derived the Spearman's correlation coefficient between
the average difference in chemical 
abundances between 2G and 1G ($<\mathrm {abundance_{\ 2G}}> -
<\mathrm {abundance_{\ 1G}}>$) 
and the cluster [Fe/H] and luminosity (M$_{V}$, a proxy for the
cluster mass). The results indicate 
that no strong correlation exists neither with [Fe/H] nor with
M$_{V}$, as shown in Figure~\ref{fig:2Gminus1G} for M$_{V}$ and
the chemical species for which data exist for more than five clusters
(both Type~I and Type~II). 

To investigate possible correlations that might be present for Type~II
GCs, we have derived the  average metallicity difference
$\Delta$([Fe/H])$ = <$[Fe/H]$_{\rm blue-RGB}-$[Fe/H]$_{\rm red-RGB}>$
between blue and red RGB stars in each cluster.  
Again, this quantity does not distinguish between 1G and 2G stars, but
lumps together, separately, all blue- and red-RGB stars. 
 This selection is blind towards the
various stellar populations with different Fe that could be present on
distinct red RGBs. As an example, at least in $\omega$~Centauri and
NGC\,7089 there are more than one stellar population with enhanced Fe,
defining different red RGB sequences. 
This translates into larger statistical errors associated with the
$\Delta$([Fe/H]) of those GCs. 
Furthermore, our $\Delta$([Fe/H]) values hide the real range in Fe in Type~II
GCs, and can be observationally biased,
depending on how many stars are available in each population with
different Fe (see Table \ref{tab:data}).

Having in mind all these possible shortcomings, we plot in
Figure~\ref{fig:FeMv_ANO} the $\Delta$([Fe/H]) as a function of M$_{V}$.
Interestingly, there is a quite strong correlation, with  Spearman's
coefficient $r>$0.9.  
If confirmed, this relation would imply that more massive Type~II GCs
have been able to retain some  material ejected by supernovae and formed
stellar populations with enhanced Fe abundances.
The obvious caveat is that the samples of measured stars in each
cluster is rather small.  
For example, if we add to the current sample NGC\,6934 (Marino et
al.\,2018), which 
has not been included in this study because only four stars have been
observed spectroscopically, we get a point at
$\Delta$([Fe/H])$=$0.20~dex and M$_{V}=-$7.45, virtually wiping out
the correlation. 
Of course, the present M$_{V}$ of this cluster gives  just an estimate
of its current mass, which could be very different from the mass at birth.

%%%%%%%%%%%%%%%%%%%%%%%%%%%%%%%%%%%%%%%%%%%%%%%%%%%%%%%%%%%%%%%%%%%%%%%%%%%%%%%
\begin{centering}
\begin{figure*}
  \includegraphics[width=12cm]{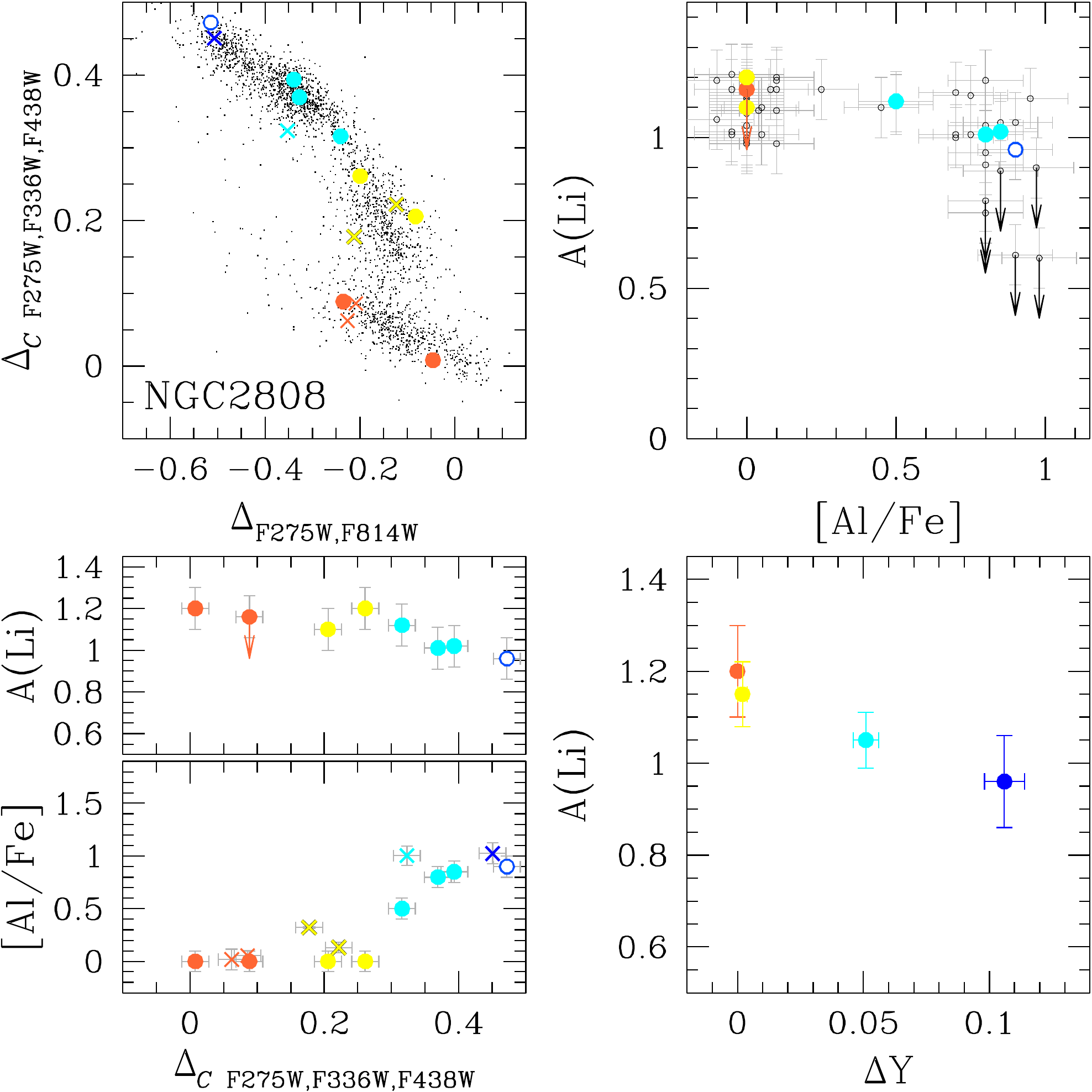}
  \caption{\textit{Upper-left panel}: Reproduction of the ChM of
    NGC\,2808 from Paper~IX. Spectroscopic targets studied by 
    D'Orazi et al.\,(2015) and Carretta\,(2014) of the populations B,
    C, D, and E defined in Paper~III have been marked with large
    orange, yellow, cyan, and blue dots and crosses,
    respectively. These colours have been used consistently in the
    other panels of this figure. \textit{Upper-right panel}: A(Li)
    vs.\,[Al/Fe] from D'Orazi et al.\,(2015). {\it Lower-left panel}:
    A(Li) and [Al/Fe] as a function of \y, while in the
    lower-right panel we plot the average lithium abundance as a
    function of the average relative helium content of populations
    B--E.   } 
 \label{fig:Li}
\end{figure*}
\end{centering}
%%%%%%%%%%%%%%%%%%%%%%%%%%%%%%%%%%%%%%%%%%%%%%%%%%%%%%%%%%%%%%%%%%%%%%%%%%%%%%%

%%%%%%%%%%%%%%%%%%%%%%%%%%%%%%%%%%%%%%%%%%%%%%%%%%%%%%%%%%%%%%%%%%%%%%%%%%%%%%%
\begin{centering}
\begin{figure*}
  \includegraphics[width=15cm]{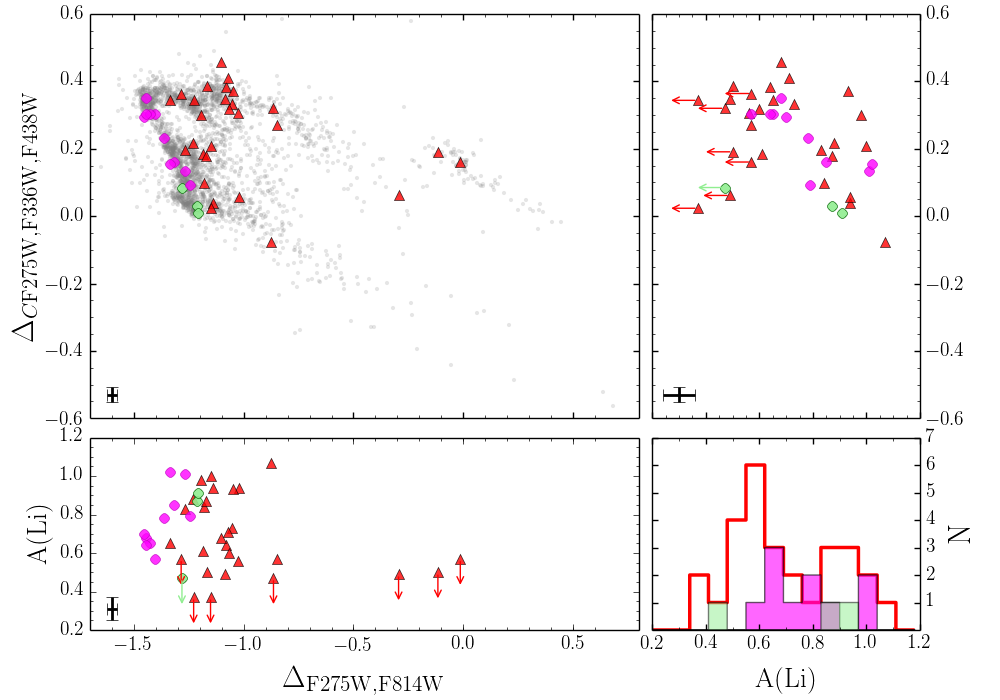}
  \caption{In the upper-left panel small grey dots represent the ChM of
      $\omega$~Centauri. Stars represented with large coloured symbols
      are those with available Li abundances (or upper limits) from
      Mucciarelli et al.\,(2018). As in Figure~\ref{fig:mapA}, red-RGB
      stars have been represented as red triangles, while green and
      magenta indicate the blue-RGB 1G and 2G stars. 
      Lithium abundances as a function of \x, and \y\ are represented
      on the bottom and right-side panels, respectively. 
      The bottom-right panel is the histogram distribution of the Li
      abundaces for 1G (green), 2G (magenta) and red-RGB stars (red),
      respectively.
      }
 \label{fig:LiOme}
\end{figure*}
\end{centering}
%%%%%%%%%%%%%%%%%%%%%%%%%%%%%%%%%%%%%%%%%%%%%%%%%%%%%%%%%%%%%%%%%%%%%%%%%%%%%%%

\subsection{Lithium}\label{sub:Li}

Although Li can provide important constraints to the formation
mechanisms of multiple stellar populations in GCs, as it is easily
destroyed by proton-capture in stellar environments, its observed
abundance depends on the interplay between several mechanisms
that can decrease the internal and surface Li content of low-mass stars
at different phases of their evolution. 
A drop in the surface Li abundance is observed in the subgiant branch,
as a consequence of the Li dilution due to the first dredge-up, and at
the luminosity of the RGB bump (Lind et al.\,2009) when thermohaline
mixing occurs (Charbonnel \& Zahn\,2007). 

All the stars discussed in this section are
RGBs less-luminous that the RGB bump, such that their abundances have not
been affected by the strong drop due to the occurrance of thermohaline
mixing. However, the Li content observed in these stars is depleted with
respect to their primordial abundances. 

The Li abundance in the five clusters with measurements
ranges from A(Li)$\sim 0.4$ to $\sim 1.4$ (see
Figure~\ref{fig:chartNaO}), with no obvious differences between 1G and
2G stars. This compares to a typical value $\sim 1$ for RGB stars
having experienced the full first dredge up but not 
having reached yet the RGB bump level (Lind et al.\,2009).

NGC\,2808 hosts stellar populations with extreme photometric and
spectroscopic properties (e.g.\,Piotto et al.\,2007; Carretta et
al.\,2006) and provides an ideal case to investigate the lithium
content of stellar populations with very different chemical
composition. 
In Paper~III we have analysed the ChM of NGC\,2808 and we have
identified at least five stellar populations, namely A--E with
different helium and light element abundances. Specifically,
population A and B have nitrogen and oxygen abundance consistent with
halo field stars with the same metallicity and correspond to 1G stars,
while populations C, D, and E are enhanced in sodium, depleted in
oxygen and correspond to the second generation. 

Lithium and aluminum abundances have been determined from spectroscopy
by D'Orazi et al.\,(2014) for stars in four of the stellar populations
identified in Paper~III. These stars are marked with coloured large
dots in the ChM plotted in the upper-left panel of
Figure~\ref{fig:Li} while in the upper-right panel we
show lithium as a function of the aluminum abundances. 
These plots reveal that  for the available stars the lithium abundance
 changes only slightly from one population to another, with the population-E
star being depleted in lithium by only $\sim$0.2~dex with respect to
population-B stars. 
Note that a few stars with high Al abundance, but without any ChM
information, have lower Li, including some upper limits (see D'Antona
et al.\,2019, for a discussion).

The slight descrease of A(Li) as a function of \y\ is  
evident from the lower-left panels of Figure~\ref{fig:Li}.
This figure also shows that population-D and -E stars with
large \y\ cluster around distinct high values of [Al/Fe].
Similarly, as shown in Figure~\ref{fig:nay}, 
population-E, which has the highest \y\ values, also has higher Na abundances. 
We note that the group of stars with high
aluminum abundance (populations-D and E) host
both stars with higher Li and stars with slightly lower abundances,
suggesting that these populations, at least population-D, might not be
chemically homogeneous. 
Unfortunately, the small sample with available ChM information does not
allow us to make stronger conclusions. 

Finally, in the lower-right panel of Figure~\ref{fig:Li} we have plotted the
average lithium abundance of population B, C, D and E stars against the
relative helium enhancement as derived in Paper~III. 
The presence of stars with extreme helium abundance but only slightly
Li-depleted (population~E) is a challenge for  scenarios that aim
to explain the formation of multiple stellar populations in GCs.  
Indeed, under normal circumstances, depleting O and enhancing Na and
Al requires very high temperatures, such that Li is totally destroyed,
unless the ``Cameron'' Li production process is in operation, such as
e.g., in massive AGB stars (Ventura et al.\,2002). 

In Figure~\ref{fig:LiOme} we report the stars with available Li
abundances (from Mucciarelli et al.\,2018) on the ChM of
$\omega$~Centauri. Similarly to previous figures, 1G and 2G stars in
the blue-RGB are plotted green and magenta dots, respectively; red-RGB
stars are plotted as red triangles. The Li abundance
distribution for the three stellar groups (lower-right panel)
suggests a wide range for 2G stars. Two out of three 1G stars have
among the highest Li in the sample, while the other one, which has
the highest \y\ and lowest \x\ among 1G stars, 
has low Li (A(Li)$\lesssim$0.5). Lithium for 2G stars decreases with
\x\ and with increasing \y; however 
Li has been measured for all the 2G stars in our sample, suggesting
that this element is detected even in the 
stellar atmospheres of stars with the highest \y\ (see the
bottom-left and upper-right panels). 

The red-RGB stars have a much broader distribution in Li, pointing to
the coexistence of stars with relatively-high abundances, comparable to
those observed in 1G stars, and stars with the lowest abundances in
the sample, with many objects having only upper limits.
Comparing the Li abundances for these stars with the location on the
ChM it is clear that most red-RGB stars with high \y\ have low Li
measurements or upper limits. This in turn suggests that these
stars, also enriched in Na (Figure~\ref{fig:naANO}), formed from a
highly-processed material enriched in Na, highly-depleted in O and Li,
and are likely the most He enhanced stars in $\omega$~Centauri. 
To the best of our knowledge, only in the AGB scenario
relatively high level of lithium can coexist with high oxygen-depletion
(D'Antona et al.\,2019).  
   
In general, the presence of Li in 2G stars requires mixing with pristine gas if
the polluters were massive stars, as such stars destroy lithium. As
mentioned above, AGB stars can both produce and destroy lithium and in
this regard the presence of an extreme 2G star in NGC\,2808, extremely
oxygen depleted but with fairly high lithium (A(Li)$\sim 1$) 
hints in favour of AGB polluters. Indeed, in
such case one cannot appeal to mixing of massive star ejecta with
pristine material, as this would have restored a high oxygen
abundance. On the other hand, the presence of a Li-poor 1G star in
$\omega$~Centauri (see Figure~\ref{fig:chartNaO}) remains quite puzzling.

%%%%%%%%%%%%%%%%%%%%%%%%%%%%%%%%%%%%%%%%%%%%%%%%%%%%%%%%%%%%%%%%%%%%%%%%%%%%%%%
\begin{centering}
\begin{figure*}
\includegraphics[width=18cm]{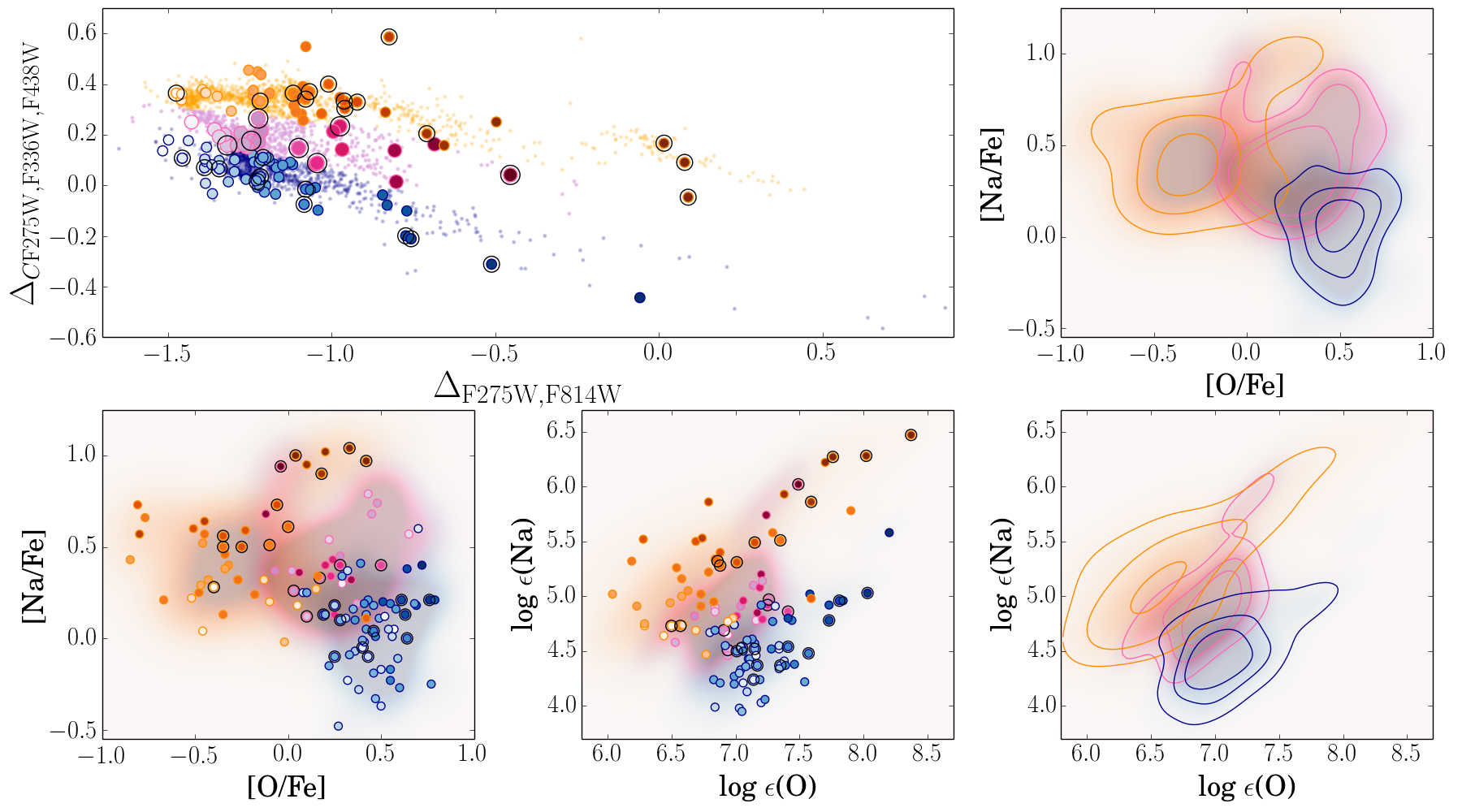}
\caption{Chromosome map of $\omega$~Centauri (top-left panel): three
  main streams of stars are evident at different levels of \y. The
  lower, middle, and upper streams have been coloured in light blue,
  pink and orange colours, respectively. Larger dots are stars with
  available chemical abundances from spectroscopy from Johnson \&
  Pilachowski\,(2010, dots without the black circles) and Marino et
  al.\,(2011, black-circled points). Oxygen abundances from Johnson \&
  Pilachowski\,(2010) have been shifted by $+$0.15~dex to account for
  the systematic difference with the O abundances inferred by Marino et
  al.\,(2011). To each star with spectroscopic
  data a gradient in colour has been assigned proportional to its \x\
  value. In the lower-left and -middle panels we represent the
  observed Na-O anticorrelation in abundances relative to Fe (left)
  and in absolute abundances (middle) for the stars represented on the
  ChM. The two right-side panels
  represent the same Na-O planes with contour density levels for stars
  belonging to each separate stream of the map. }
\label{fig:OmegaNaO_ChM}
\end{figure*}
\end{centering}
%%%%%%%%%%%%%%%%%%%%%%%%%%%%%%%%%%%%%%%%%%%%%%%%%%%%%%%%%%%%%%%%%%%%%%%%%%%%%%%%

%%%%%%%%%%%%%%%%%%%%%%%%%%%%%%%%%%%%% FIG 2 %%%%%%%%%%%%%%%%%%%%%%%%%%%%%%%%%%%
\begin{centering}
\begin{figure*}
  \includegraphics[width=18.cm]{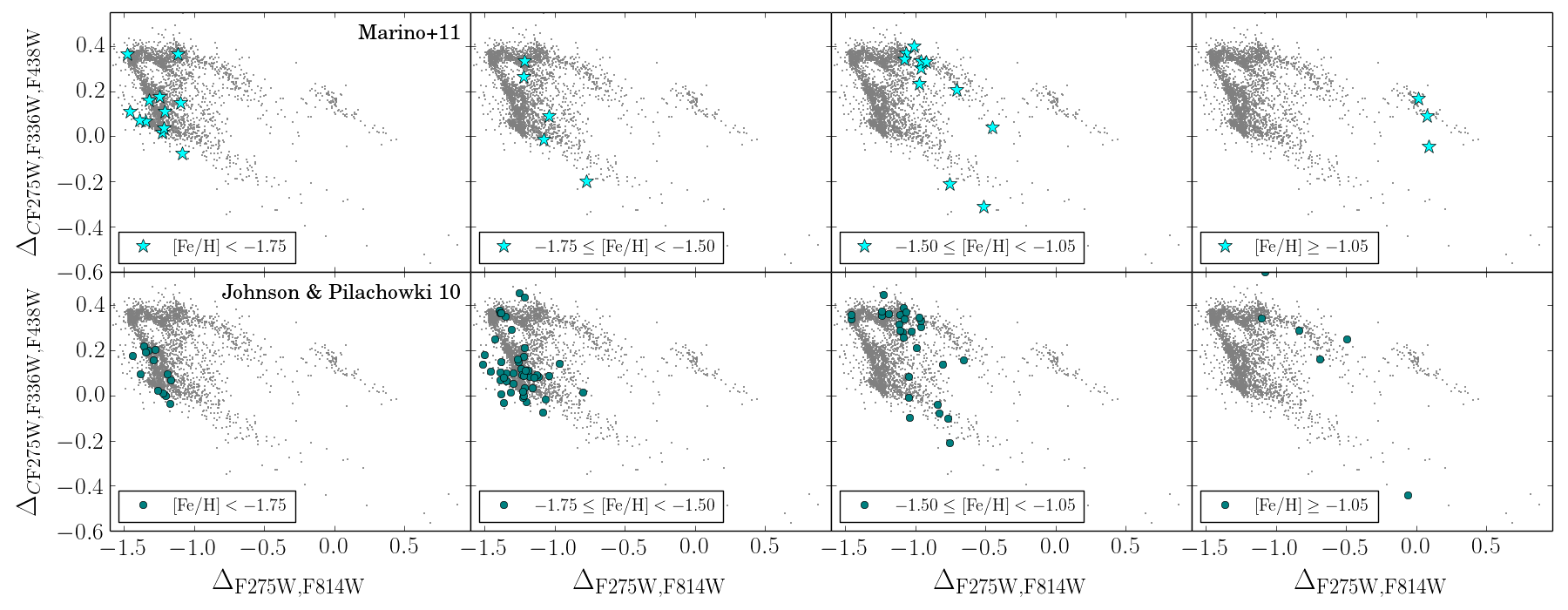}
   \caption{From the left to the right the position of stars with
     increasing [Fe/H] are plotted on the ChM of
     $\omega$~Centauri. The Marino et al.\,(2011) and Johnson \&
     Pilachowski\,(2010) samples are plotted on the top and bottom
     panels, respectively. Due to the systematic Fe difference between
     the two samples, the Johnson \& Pilachowski\,(2010) [Fe/H] values have been
     shifted by $+$0.15~dex.}
 \label{fig:OmegaFe_ChM_0}
\end{figure*}
\end{centering}
%%%%%%%%%%%%%%%%%%%%%%%%%%%%%%%%%%%%%%%%%%%%%%%%%%%%%%%%%%%%%%%%%%%%%%%%%%%%%%%

%%%%%%%%%%%%%%%%%%%%%%%%%%%%%%%%%%%%%% FIG 2 %%%%%%%%%%%%%%%%%%%%%%%%%%%%%%%%%%%
\begin{centering}
\begin{figure*}
\includegraphics[width=17.7cm]{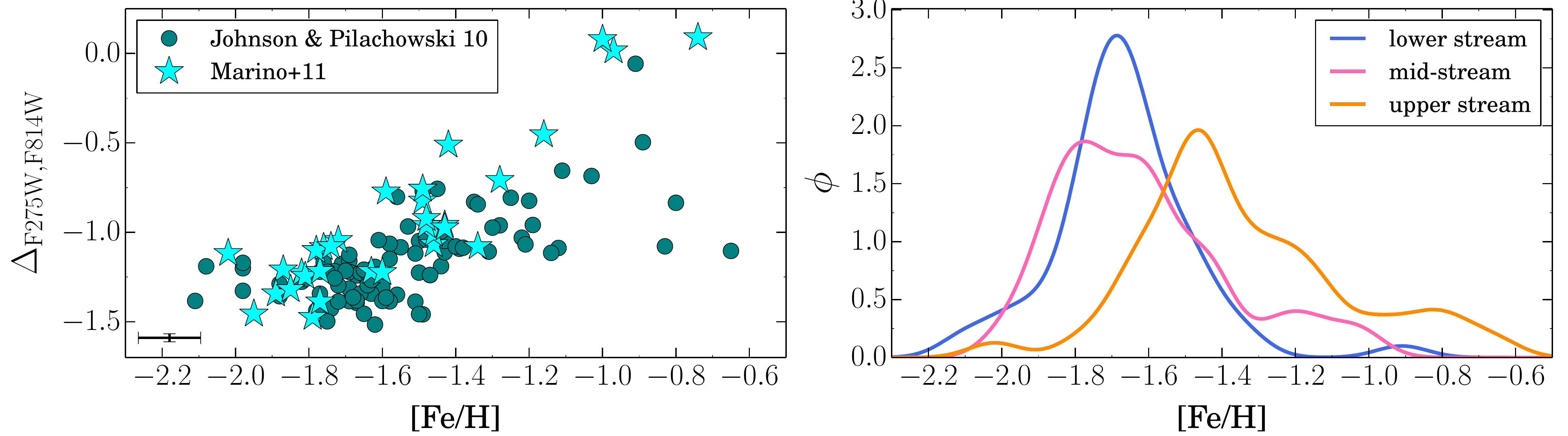}
\caption{{\it Left panel}: \x\ as a function of [Fe/H] for
  $\omega$~Centauri. Stars in Marino et al.\,(2011) and Johnson \&
     Pilachowski (2010) samples have been represented as aqua
     star-like symbols and teal-blue circles, respectively. {\it Right panel}: Gaussian kernel density
   distribution of [Fe/H] for the three streams by combining the
   Marino et al.\,(2011) and Johnson \& Pilachowski (2010)
   samples, with the latter values increased by $+0.15$~dex to account
 for the systematic difference between the two datasets (see Marino et
 al.\,2011). }
\label{fig:OmegaFe_ChM}
\end{figure*}
\end{centering}
%%%%%%%%%%%%%%%%%%%%%%%%%%%%%%%%%%%%%%%%%%%%%%%%%%%%%%%%%%%%%%%%%%%%%%%%%%%%%%%%

%%%%%%%%%%%%%%%%%%%%%%%%%%%%%%%%%%%%%%%%%%%%%%%%%%%%%%%%%%%%%%%%%%%%%%%%%%%%%%%
\begin{centering}
\begin{figure*}
  \includegraphics[width=17cm]{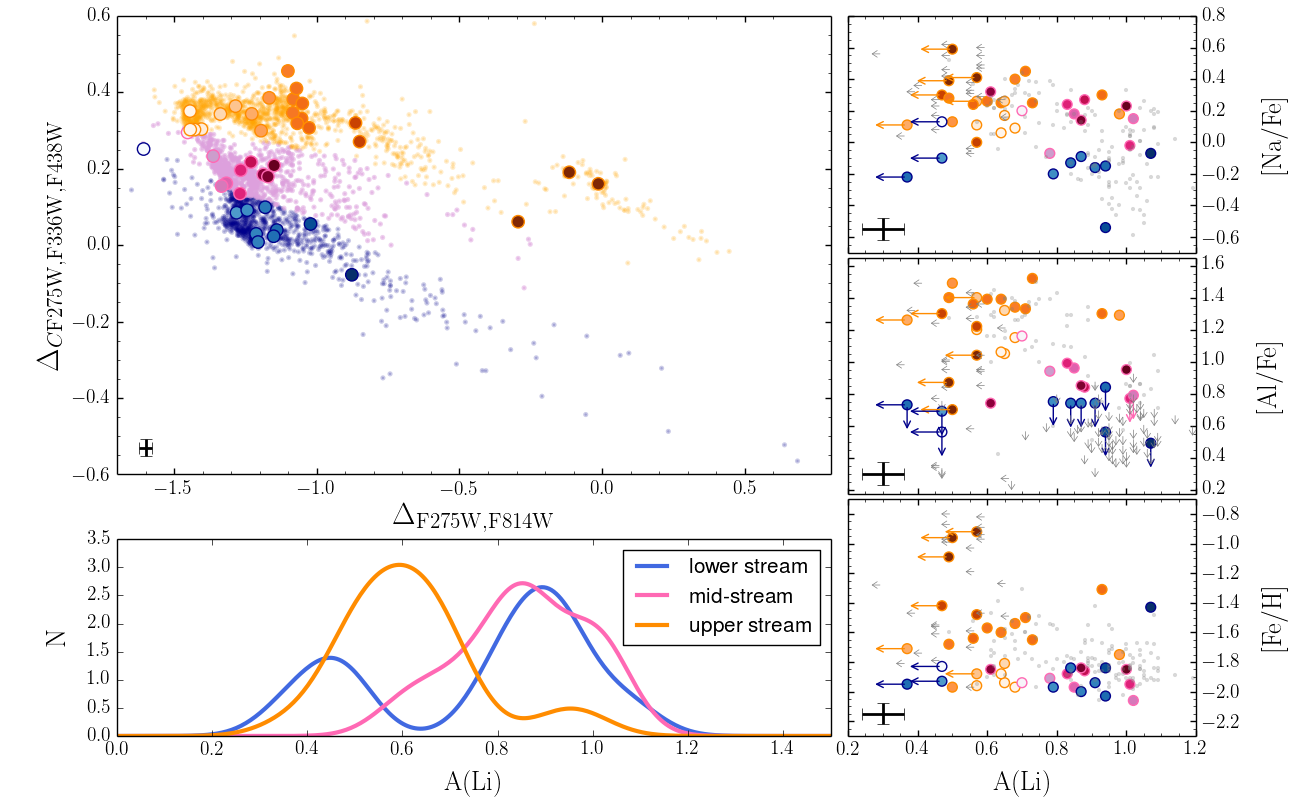}
  \caption{In the upper-left panel small grey dots represent the ChM of
      $\omega$~Centauri. Stars represented with large coloured symbols
      are those with available Li abundances (or upper limits) from
      Mucciarelli et al.\,(2018). 
      As in Figure~\ref{fig:OmegaNaO_ChM}, the lower, middle, and
      upper streams have been coloured in light blue, pink and orange
      colours, respectively.
      To each star with spectroscopic data (large dots) a gradient in
      colour has been assigned proportional to its \x\ value.  
      In the lower-left we represent the Gaussian kernel density
      distribution of Li abundances for the three streams.
      Right panels represent Na, Al, and Fe as a function of Li
      abundances (or upper limits) for all the available spectroscopic
      sample (small gray symbols), and the stars with ChM data (large
      dots), for which we used the same colour code used 
      on the map in the upper-left panel.      }
 \label{fig:OmeLi_Streams}
\end{figure*}
\end{centering}
%%%%%%%%%%%%%%%%%%%%%%%%%%%%%%%%%%%%%%%%%%%%%%%%%%%%%%%%%%%%%%%%%%%%%%%%%%%%%%%

\section{More on the chromosome map of $\omega$ Centauri}\label{sec:omegacen} 

The most complex ChM observed in the sample of clusters from the $HST$ UV Legacy
Survey is that for $\omega$~Centauri (Paper~IX). This complexity
corresponds to a very 
intricate interplay of stellar populations with different chemical
abundances in both light (C, N, O, Na) and heavy elements (including Fe
and $s$-process elements), which is observed also on the main sequence
(Bellini et al.\,2017), with up to 16 distinct stellar populations
being identified (Paper~IX).
In previous sections, for comparison purposes, we treated this GC
together with the other simpler ones.
However, given the complexity of $\omega$ Centauri, we devote this
entire section to a more careful exploration of the ChM of this cluster.

From the ChM represented in Figure~\ref{fig:mapA}, we have considered
the blue-RGB 1G 
(green) and 2G (magenta) components while lumping together all the
red-RGB stars as a single population.
However, this selection hides a higher level of complexity of stellar
populations in this remarkable cluster. Its ChM displays quite
elongated red streams asking for some more effort to be done to
characterise the chemical properties of stars in this diagram.

In this section we focus on these streams, so evident in
the ChM of $\omega$~Centauri at different \y, so that stars have been
divided into three different streams.  
In the top-left panel of Figure~\ref{fig:OmegaNaO_ChM} we represent
once again the ChM of $\omega$~Centauri, with the selection of main populations
considered here: 
\begin{itemize}
\item{a lower stream, with lower \y, painted in light-blue;}
\item{a mid-stream, with intermediate \y, shown in pink;}
\item{an upper stream with the highest \y\ values on the map,
    painted in orange.}
\end{itemize}
These so-selected populations of stars are on top of the {\it normal}
blue- and red-RGB sequences observed in all the GC maps.
In the following we discuss the chemical abundances pattern on
the ChM of $\omega$~Centauri focusing on its distinctive streams. We
start with the Na-O anticorrelation.

\subsection{Na-O anticorrelation}\label{sub:NaOStreams}

The position of the selected stream populations on the O-Na plane is
shown in the left-lower and central-lower panels of
Figure~\ref{fig:OmegaNaO_ChM}, where we plot both the sample from
Marino et al.\,(2011b) and from Johnson \& Pilachowski\,(2010), with
large and small circles, respectively. 

In Section~\ref{sec:typeII}, we have discussed that in $\omega$~Centauri
the {\it normal} sequence stars composed of blue-RGB 1G and 2G
populations define an O-Na anticorrelation, just as in Type~I GCs.
Most of the stars are enhanced in O, as typical
of primordial composition of halo field stars. Stars with similar
high O span a relatively large range in Na. 
We note here that the three streams on the red side of the ChM clearly occupy
different locations on the O-Na anticorrelation (see
Figure~\ref{fig:OmegaNaO_ChM}).  
The upper stream (orange) stars mostly occupy the section of higher Na and
lower O quadrant of the O-Na plane with [O/Fe]$<$0.0~dex. 
Mid-stream stars mostly distribute on a intermediate location in the O-Na
plane, at intermediate values of Na, without reaching too-low O values,
while lower stream stars have on average lower Na than mid-stream
ones, and higher O. 

Stars with most extreme positions on the red side of the ChM
 have the highest abundances of Na, but are not the most
depleted in O, with a few stars with
intermediate Na ([Na/Fe]$\sim$0.2~dex) and high O ([O/Fe]$>$0.5~dex). 
These stars are those defining a Na-O correlation (Marino et
al.\,2011b; D'Antona et al.\,2011).
To guide the eye, shaded contour areas have been plotted on the O-Na
anticorrelation planes (right panels of Figure~\ref{fig:OmegaNaO_ChM})
corresponding to the colours of the different streams defined on the ChM. 

\subsection{Iron enrichment}

To investigate the role of Fe in shaping the ChM of $\omega$~Centauri,
in each panel of Figure~\ref{fig:OmegaFe_ChM_0} we show stars in
different bins of [Fe/H], both from the Marino et al.\,(2011b)
sample (upper panels), and the Johnson \& Pilachowski one (lower
panels). As a general rule, stellar populations with increasing Fe
populate redder and redder regions of the map.
The [Fe/H]-\x\ correlation is well seen in the left panel of
Figure~\ref{fig:OmegaFe_ChM}, and agrees with the observations of the
other (simpler) Type~II GCs where the red RGB stars on the ChM have
higher Fe, though on lower levels.

The Fe enrichment seems generally de-coupled from the light-elements
processing, corroborating previous studies on less-complex Type~II GCs
like M\,22 (Marino et al.\,2009, 2011a), and previous analysis of
$\omega$~Centauri (Marino et al.\,2011b, 2012). 
By inspecting the streams, we observe in the right panel of
Figure~\ref{fig:OmegaFe_ChM} that all the three streams have a wide
distribution in [Fe/H]. Interestingly, the lower and mid-streams are
peaked at similar Fe abundances of [Fe/H]$\sim -$1.7~dex, with the
mid-stream displaying a minor overdensity of stars at higher Fe
($-$1.2$\lesssim$[Fe/H]$\lesssim -$1.1~dex).
The upper stream is peaked at higher Fe ([Fe/H]$\sim -$1.4~dex), and
shows hints of multiple peaks, reaching the highest metallicity
values observed in $\omega$~Centauri.
This result may suggest that, while the Na enrichment (and
O-depletion) occurs during the whole Fe enrichment phase experienced
by $\omega$~Centauri, the strongest metallicity enrichments took place
when the enrichment in high-temperature proton-capture products was maximum.

\subsection{Lithium in the streams}\label{sub:LiStreams}

Figure~\ref{fig:OmeLi_Streams}
shows again the ChM of $\omega$~Centauri, with the three streams
represented in different colours, and the stars where Li abundances,
or upper limits, are available from Mucciarelli et al.\,(2018). 
The Gaussian kernel density distribution of Li abundances for the
three streams immediately suggests that, while the lower and
mid- streams have distributions peaked around similar
relatively-high Li\footnote{Note however that Li abundances in the
  lower and mid- streams do not cover the entire range in \x: Li is
  available for \x$\lesssim -$1.2, and for \x$\lesssim -$0.9, in the
  lower and mid-stream, respectively. We cannot exclude lower Li
  abundances for stars with higher \x\ and Fe. }, 
the upper stream has lower abundances (note
that many of the Li abundances derived for the upper stream are
upper limits).

  Three stars in the lower stream have low Li
  abundances, A(Li)$\lesssim$0.5~dex, more consistent with values of
  the upper stream. On the other hand, the light element abundances of
  these stars are similar to those inferred for the remaining
  lower-stream stars, both in Na and Al (right panels in
  Figure~\ref{fig:OmeLi_Streams}). We note that, while for the other
  GCs analysed in this work the contamination from AGB stars in the
  ChMs is negligible, as they can be easily removed by inspecting {\it
  classical} CMDs, in the case of $\omega$~Centauri, the large spread
in Fe makes it difficult to clean the ChM from AGBs, and we expect
contamination from AGBs belonging to the metal-richer populations.
From the lower panel of Figure~\ref{fig:OmeLi_Streams} however all the
three stars have Fe consistent with the metal-poor population of
$\omega$~Centauri.  

A careful inspection of the upper stream reveals that the stars
with lower \x, hence metal-poorer (these stars are the 2G stars
discussed in previous sections, with the highest \y), all have Li
measurements, with A(Li)$\sim$0.7~dex. Upper limits suggesting
A(Li)$\lesssim$0.6~dex occur for larger \x\ values. The three
metal-richest stars, with the highest \x, all have
A(Li)$\lesssim$0.6~dex. 
In general, the upper stream hosts stars with higher Na and Al. Two
stars have been found to have Li similar to the lower and mid-
streams, but they follow the general upper-stream abundances in Na and Al.

\subsection{On the helium enrichment}

Our chemical analysis of the ChM of $\omega$~Centauri
suggests that the upper stream hosts the most-extreme stars in terms
of chemical properties. They are the most-processed in terms of light
elements, with the highest Na and Al abundances, and the lowest Li and
O. Although the Fe enrichment occurs in all the streams, the upper
stream Fe distribution is peaked at higher Fe, and includes the
Fe-richest stars of $\omega$~Centauri.

Among upper stream stars, those that in previous sections have been
classified as 2G (blue-RGB), have likely born from material with a {\it less degree} of
$p$-capture processing, if compared to the other upper stream stars
belonging to the red-RGB. This is suggested by their less extreme abundances of
Na, O, Al, and Li (see Sections~\ref{sub:NaOStreams}-\ref{sub:LiStreams}). 
Specifically, these stars are those coloured in magenta (e.g.\, in
Figures~\ref{fig:mapA} and \ref{fig:LiOme}), with the highest \y\ values. 

The upper-stream red-RGB stars have more extreme enrichments in Na and
Al, and depletions in Li and O. As stars with more extreme abundances
in these elements have the highest level of He enrichment (e.g.\,
Paper~III), the upper-stream red-RGB stars are likely the
most-enhanced in helium. 

Helium abundance variations in the $\omega$~Centauri sub-populations
have originally been inferred from the analysis of main sequence (MS) stars,
which suggested a 
higher He for the blue MS (e.g. Bedin et al.\,2004; Norris\,2004; Piotto et
al.\,2005; King et al.\,2012).
More recently, we have presented the He internal variations
between 1G and 2G stars appearing in the ChM of $\omega$~Centauri, and
found that the maximum He variation among blue-RGB stars (with similar
[Fe/H]$\sim -$1.7) is $\delta(Y)=$0.09 (Paper~XVI). This He difference
is between the 1G (stars coloured in green in Figures~\ref{fig:mapA}
and \ref{fig:LiOme}) and 2G (blue-RGB) upper stream stars.  
This He variation is intended for the metal-poor population of
$\omega$~Centauri corresponding to its blue-RGB.

Very interestingly, Milone et al.\,(2017b), by using optical-NIR CMDs,
have identified two main stellar Populations I and II along the entire
MS of $\omega$~Centauri, from the turn-off towards the hydrogen-burning limit. 
The two MSs are consistent with stellar populations with different
metallicity, helium and light-element abundance. Specifically, MS-I
corresponds to a metal-poor stellar population ([Fe/H]$\sim -$1.7)
with $Y\sim$0.25 and [O/Fe]$\sim$0.30; while the MS-II hosts
helium-rich stars with metallicity ranging from
[Fe/H]$\sim -$1.7 to $-$1.4~dex, and lower [O/Fe]. 
Noticeably, to match the MS-II, the helium content required for the
metal-poor isochrone, at [Fe/H]$\sim -$1.7, is $Y
\sim$0.37, while a higher He, $Y\sim$0.40, is needed for the isochrone
at [Fe/H]$\sim -$1.4. 

This result is in line with our chemical analysis of the
$\omega$~Centauri ChM, which suggests that the most He-enhanced stars
are those in the red-RGB upper stream, that have the most extreme
abundances in light elements. Stars in the blue-RGB, although
highly-enriched in He, do not reach the extreme enhancements
characterising the red-RGB upper stream stars.

\section{Discussion and conclusions}\label{sec:conclusions}

%%%%%%%%%%%%%%%%%%%%%%%%%%%%%%%%%%%%%%%%%%%%%%%%%%%%%%%%%%%%%%%%%%%%%%%%%%%%%%%
\begin{centering}
\begin{figure*}
\includegraphics[height=8.1cm]{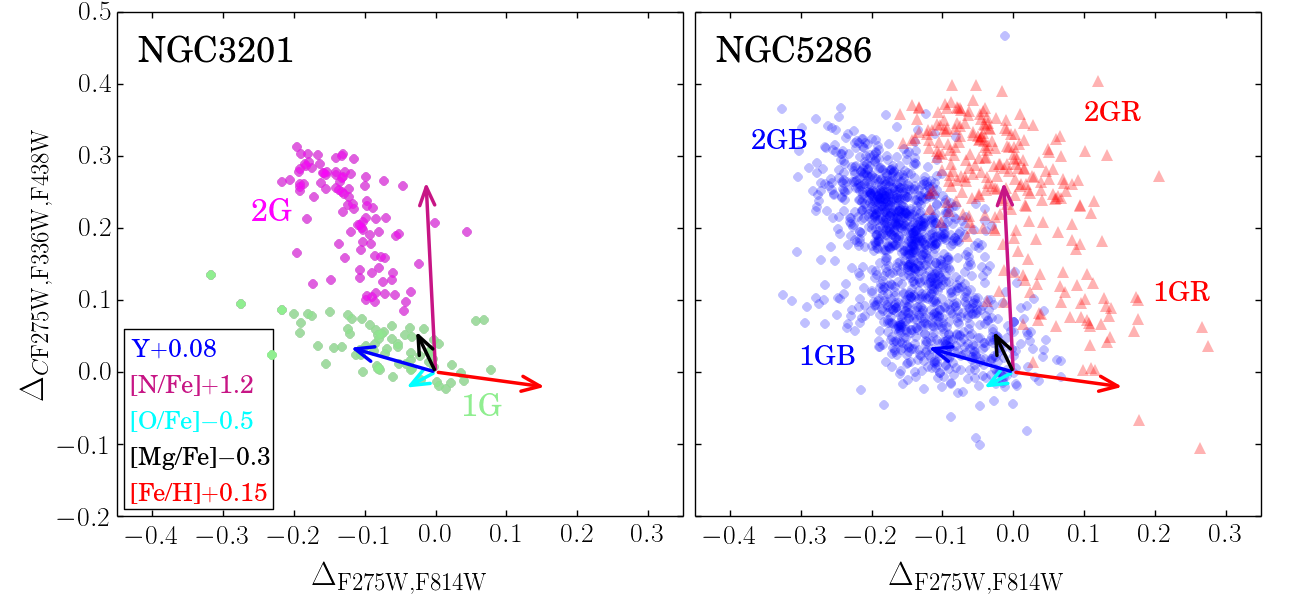}
\caption{ChMs of the Type~I GC NGC\,3201 (left) and the Type~II GC
  NGC\,5286 (right). The arrows indicate the effect of changing He, C, N, Mg,
  and O, one a at a time by the quantities quoted in the inset, on \x\ and \y.} 
\label{fig:frecce}
\end{figure*}
\end{centering}
%%%%%%%%%%%%%%%%%%%%%%%%%%%%%%%%%%%%%%%%%%%%%%%%%%%%%%%%%%%%%%%%%%%%%%%%%%%%%%%

Previous work, based on synthetic spectra, suggested that the chemical
abundances observed among different stellar populations in Galactic
GCs are strictly correlated to the distribution of stars 
on the ChMs (e.g.\,Paper XVI, see their Figure~6). This correlation has
been corroborated by direct spectroscopic measurements of stars along
the ChMs of a few GCs (e.g.\,Papers~II, III, IX; Marino et
al.\,2017). Expanding from these early investigations, 
in this paper we combined the elemental abundances of Li, N, O, Na,
Mg, Al, Si, K, Ca, Fe, and Ba and the ChMs of 29 GCs to characterize
the chemical composition of the distinct stellar populations as
identified on the ChMs, thus helping to {\it read} ChMs themselves.   

Thus, we calculated the average abundance of 1G and 2G stars for each
analysed element, providing the first determination --in a large
sample of GCs-- of the mean 
abundances of distinct stellar populations identified photometrically. 
We find strong correlations between nitrogen, oxygen and sodium
abundances and the position of stars on  the ChM. Specifically, 2G
stars identified on the ChM, have higher N and Na and lower O than
1G stars.  
This was quite expected: the \y\ axis is highly
sensitive to variations in light elements, in particular to N through
absorption by NH molecules.   
Unfortunately, nitrogen abundances are not available for a significant number
of stars and clusters, but given the known correlation between this element
and Na and its anti-correlation with O, these latter  elements can be used as a
proxy of the N variations causing the separation of stars along the
map, as proven with synthetic model atmospheres in Paper~XVI. 
We also noticed that the lithium abundance of one extreme star in
NGC\,2808 (population-E) is lower, but not zero, than in the other
stars, suggesting that, assumed that this highly-He enriched/highly-O
depleted star formed from {\it pure}
ejecta, in that material lithium was not completely destroyed.

It has emerged from previous papers of this series that two aspects of
the multiple population phenomenon are particularly intriguing: 
it appears that even 1G stars are not chemically homogeneous, as
indicated by their large spread in \x, and the split of the 1G and 2G
sequences in Type~II GCs. 

As an illustrative example, we reproduce in Figure~\ref{fig:frecce}
the ChMs of the Type~I GC NGC\,3201, with its large spread in \x\ among
1G stars, and the Type~II GC 
NGC\,5286, with its split 1G and 2G sequences. 
Overplotted on the ChMs are the vectors representing the
expected changes due to the variations in He (in mass fraction Y), N,
O, Mg, and Fe (C effects are almost negligible) as labelled 
in the left panel (see also Paper~XVI). 
The case of NGC\,3201 in Figure~\ref{fig:frecce} well illustrates how
variations in Fe and/or He (and possibly the C$+$N$+$O) 
can be responsible for variations in \x.
On the other hand, combined variations in helium and nitrogen  are
needed to populate the 2G sequence with increasing both \x\ and \y.  
A mere increase in helium seems to qualitatively reproduce
the distribution of 1G stars, although it remains mysterious how to
produce an increase in helium without a concomitant increase in
nitrogen (Paper~XVI). 
Indeed, we find that the observed colour spread in 1G stars is not
correlated to any of the analysed light elements, thus confirming that
stellar populations with different chemical abundances in light
elements occupy locations with different \y along the 2G sequence
rather than along the 1G sequence. 
We  found only a significant correlation between \x\ and the iron
abundance and this was in the case NGC\,3201 and NGC\,6254, the two
clusters with the widest 
1G range in \x, but still too few stars with data are available even
in these two clusters. 
Formally, the observed dispersion in iron of $\sim 0.1$~dex (see
Figure\ref{fig:Fe1G}) could account for the 1G spread in NGC\,3201,
with iron increasing  
with increasing \x. We conclude that a dedicated spectroscopic survey
of 1G stars in many Type I GCs is required to decide whether helium or
iron (or else?) are responsible for the 1G spread. This is a crucial
issue to solve, in the perspective of understanding how GCs formed,
along with their multiple stellar generations. 
If it will turn out that iron is responsible for the 1G spread, then
this would imply an ongoing star formation of 1G stars while at least
a few supernovae had started to pollute the interstellar medium, with
2G stars starting to form only after the 1G population was
complete. Otherwise also 2G sequence would exhibit a broadening
similar to that of the 1G sequence. 

In Paper~IX it was shown that in about 17 per cent of the analysed
clusters (then called Type~II GCs) the 1G and 2G sequences are split
as illustrated for NGC\,5286 in the right panel of
Figure~\ref{fig:frecce}. In these clusters also the RGB splits into a
blue and a red branch in e.g., the $(U-I)$ colour and this split is
used to separate the two populations. It is important to emphasise
that (in most clusters) there is a clear dichotomy between the blue
and the red populations, as evident from Figure~\ref{fig:frecce}, see
also Figures~\ref{fig:1851} and \ref{fig:GB}. 
All Type~II GCs that have been analysed spectroscopically exhibit internal
heavy-elements variations pointing to a connection between their
unusual ChM patterns and metallicity (iron) variations. 
We find that the blue-RGB stars  of Type~II GCs behave in a very
similar manner as stars 
Type~I GCs. Stars on the red-RGB have, in most cases, higher Fe, and/or
higher $s$-elements abundances. The shift on the red is likely due to
a temperature effect due to an overall higher metallicity.

\begin{centering}
\begin{figure}
\   \includegraphics[width=8.2cm]{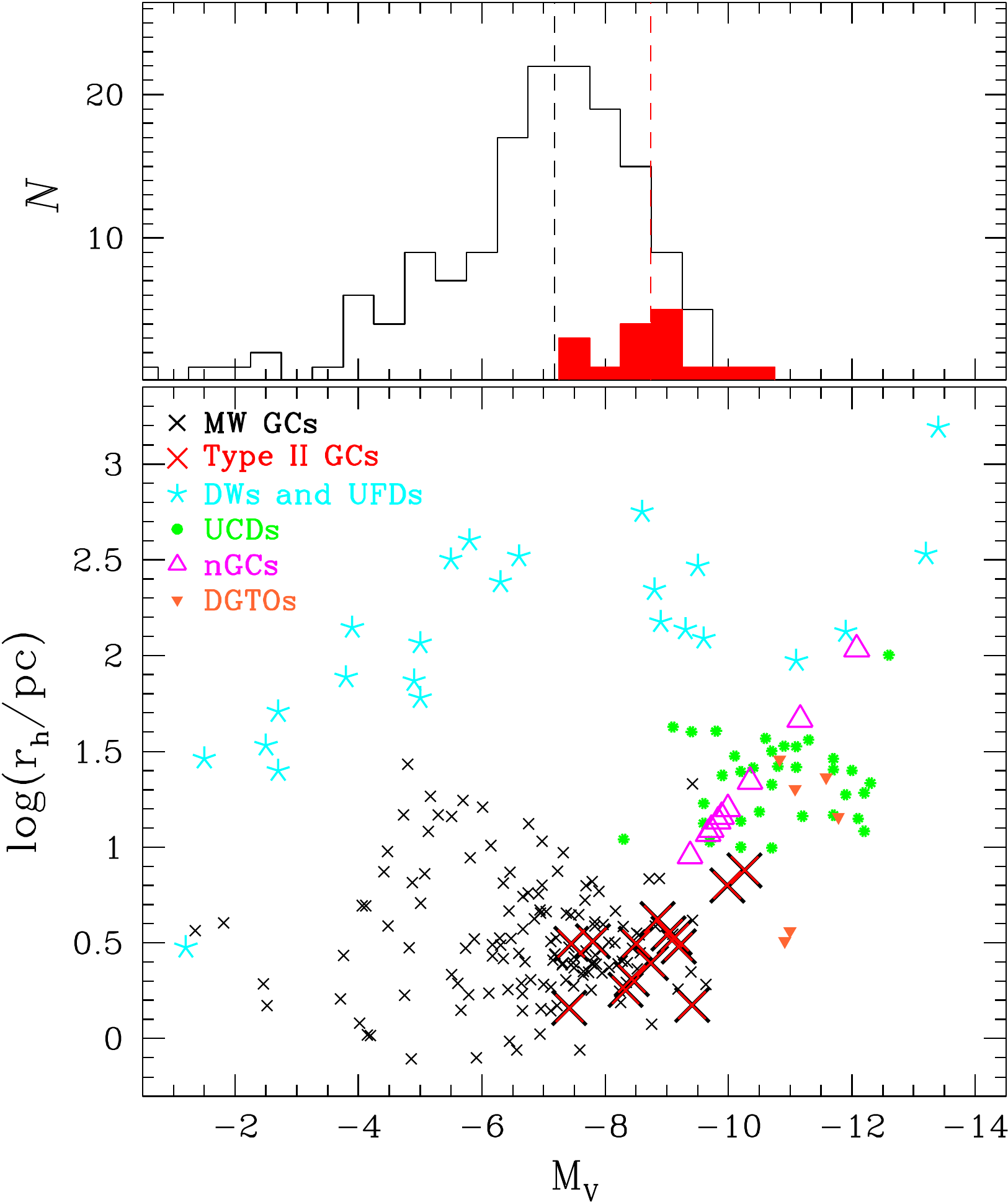}
  \caption{Absolute magnitude ($\mathrm {M_{V}}$) as a function of the half-light radius
    ($\mathrm {r_{h}}$). Different colours and symbols represent
    different class of objects: Milky Way GCs (black crosses;
    Harris\,2010), Milky Way satellites including dwarf (DW; Irwin \& 
    Hatzidimitriou\,1995; Mateo\,1998), ultrafaint dwarf galaxies
    (UFDs) and all the objects discovered by the SDSS (Willman et
    al.\,2005,\,2006; Belokurov et al.\,2006, 2007; Zucker et
    al.\,2006; Jerjen\,2010), ultracompact dwarf galaxies (UCDs;
    Brodie et al.\,2011), nucleated GCs (nGCs; Georgiev et al.\,2009),
    dwarf-globular transition objects (DGTOs; Ha{\c s}egan et
    al.\,2005). Type~II GCs, here considered as all the objects with
    internal metallicity and/or heavy-elements variations, have been
    marked with red crosses. The upper panel is the histogram
    distribution of the $\mathrm {M_{V}}$ of GCs, with Type~II GCs
    shown in red.}
 \label{fig:all_mass}
\end{figure}
\end{centering}

We have derived the average abundances of 1G and 2G stars for both
blue-RGB and red-RGB populations 
finding no evidence for a correlation between the relative 
abundances of 2G and 1G stars and the absolute magnitude (a proxy of
mass), nor with the metallicity of the host GCs\footnote{A strong
  correlation is instad found between tha GC mass and the maxium
  helium variation in the cluster (Paper~XVI).} .  
This result is consistent with previous findings of no correlation (or very 
mild correlation)  between the relative average helium abundance of 2G
and 1G stars and the cluster mass and metallicity (Lagioia et
al.\,2018; Paper~XVI). 
On the contrary, we find a correlation between the average
difference in the iron abundance of red-RGB and blue-RGB stars 
and the absolute luminosity of the host GC, which may
suggest that the internal metallicity difference between the blue and
red populations Type~II GCs increases with the cluster mass. 
Figure~\ref{fig:all_mass} shows the position of various stellar
systems in the half-light radius $\mathrm {(log r_{h}/pc)}$ versus
absolute magnitude ($\mathrm {M_{V}}$) plane. We include the position of
Milky Way GCs, classical dwarfs (DWs), Ultra Faint dwarfs (UFD),
ultracompact dwarfs (UCD), dwarf-globular transition objects (DGTO) and 
nucleated GCs (nGC). The position of Type~II GCs shows that they are
in general among the most massive GCs. 

The crucial question about Type II GCs is to understand what sequence
of events led to formation of the blue and red populations, each with
its own first and second 
generations. We see two possible options. A first and second
generation of {\it blue} stars formed as in any other Type~I
cluster. The material out of which the blue 1G and 2G stars formed was
almost completely converted into stars or any residual was
expelled. Then the cluster  re-accreted pristine gas that was enriched
in iron  
by supernovae from the blue population (of either Type~Ia or core
collapse) and then formed a new 1G and its 2G companion population. 
Alternatively, one may think to formation within a dwarf galaxy, with
the blue and red population forming at different times, while the
dwarf itself was self-enriching in iron. 
Or even the blue and red populations formed in different places, and
then merged together, as indeed speculated for NGC\,1851 (Bekki \& Yong\,2012).
Clearly the red population did not form from the gas that formed the
blue 2G stars, otherwise there would be no 1G stars (i.e. nitrogen
poor) in the red population. 

As discussed in Paper~IX, the fraction of red-RGB stars in Type~II GCs
varies from cluster to cluster, being 
NGC\,6656 one of the clusters with higher fraction with $\sim$40 per
cent of red-RGBs. 
Given the present mass of this cluster ($\sim 4\times 10^5\,
M_\odot$), the metallicity of the blue population ([Fe/H]$=-1.82$) and
the $\sim 0.15$~dex difference in iron abundance between its red and
blue populations, we estimate that  
the whole red population now contains just $\sim 1.2\, M_\odot$ of
iron more than if it had the same iron abundance of the blue
population.  
According to current understanding, a stellar generation makes $\sim
0.5\, M_\odot$ of iron from core collapse supernovae every 1,000
$M_\odot$ of gas turned into stars (e.g., Renzini \& Andreon\,2014 and
references therein). Hence, the blue population alone should have
produced over $\sim 120\, M_\odot$ of iron and therefore just $\sim 1$
per cent of it would have been sufficient to enrich the red population
to the observed level. 
Similarly, the cluster NGC\,5286 has a mass of $\sim 4.6\times 10^5\,
M_\odot$, $\sim$17 per cent  of  which in the red-RGB population, a
metallicity of the blue population [Fe/H]$=-1.77$ and again a $\sim
0.15$~dex difference in iron abundance between its red and blue
populations. Therefore, the same calculation leads to  
just $\sim 0.6\, M_\odot$ of excess iron in the red population, while
the blue population would have produced $\sim 200 \, M_\odot$ of iron,
with only $\sim 0.3$ per cent of it being sufficient to enrich the red
population to the observed level. From this point of view, there may
not be such a big difference between Type~I clusters that have not
retained much supernova ejecta at all and those Type\,II clusters
that have retained just a very small fraction of it. For example, the
same exercise for $\omega$~Centauri indicates that only $\sim 2$ per
cent of the iron from the supernovae from its metal poor population
would have been sufficient to enrich its metal rich population to the
observed level (Renzini et al.\,2015; Renzini\,2013). Thus, in this
respect there appears to be no much difference between Type~I
clusters, that have lost 100 per cent the iron produced by their
stars, and Type~II clusters that have lost just $\sim 99$ per cent. 

Finally, we have explored in details the ChM of $\omega$~Centauri, whose wide
ranges in chemical abundances of CNONa and Fe make it an ideal laboratory
to test the role of these elements in shaping the maps of GCs.
We found that \x\ is very well correlated to [Fe/H], and that the location
of the stars on the three main streams, that we have defined in the
ChM, is related to a different position on the Na-O abundances
plane. 
The upper stream is the most distinctive either in the Na-O plane, but
also in terms of Fe, Li, and likely He content.
There is no doubt that in this most massive GC of the MW galaxy
metal enrichment and star formation from  
proton-capture processed material have concomitantly taken place,
though still in a sequence of events that we have not yet been able to
decipher from  either its ChM or spectroscopic analysis. But we keep trying.

In conclusion, the \x\ spread among 1G stars ($W_{\rm 1G}$) can
be ascribed either to a helium or to an iron spread, with a spread
$\Delta Y$ giving a similar $W_{\rm 1G}$ that would be produced by a
similar spread in [Fe/H], for example, an enhancement $\Delta Y=0.1$
would produce the same $W_{\rm 1G}$ than a variation
$\Delta$[Fe/H]=$-$0.1. Here the minus sign indicates that the bluest
1G stars would be either helium rich of iron poor. Though giving similar
results, a spread in helium of $\sim 0.1$ is much more demanding in
terms of nucleosynthesis than a spread of $\sim 0.1$~dex in iron, to
the point that no viable scenario could be envisaged that would
produce such an helium spread without affecting the nitrogen abundance
(cf. Paper~XVI). For this reason we have been cautious in claiming the
1G spread being due to helium, after we first hinted at it in
Paper~III. We now have for two clusters, namely NGC\,3201 and
NGC\,6254, 
some weak evidence supporting the notion of the 1G spread being due to
iron, rather than to helium (Figure~\ref{fig:Fe1G}). However,  more systematic
and accurate measurements of the iron abundance in several clusters
with wide 1G sequences are necessary before definitely ascribe to iron
the observed spread. Iron dishomogeneities at the level of $\sim 0.1$
dex in the molecular cloud generating the 1G stars appears to be a
plausible alternative to helium and could be generated by incomplete
mixing of the supernova ejecta in the parent dwarf galaxy. 
We also notice that the correlation of the 1G spread $W_{\rm 1G}$ with
cluster mass (implied by Figure~19 in Paper~XVI) is also reminiscent of
the Type II syndrome (multiple iron abundances) being confined to the
most massive GCs, as shown in Figure~\ref{fig:all_mass}, as if
retaining supernova ejecta were more likely in the progenitors of the
most massive clusters.

The chromosome maps will be available via the
{\sc http://progetti.dfa.unipd.it/GALFOR/} web page and at the CDS
({\sc cdsarc.u-strasbg.fr}). We will continually update the files and
figures of this paper in the {\sc http://progetti.dfa.unipd.it/GALFOR/}
webpage as new spectroscopic observations come in.

%%%%%%%%%%%%%%%%%%%%%%%%%%%%%%%%%%%%% FIG 2 %%%%%%%%%%%%%%%%%%%%%%%%%%%%%%%%%%%
%%%%%%%%%%%%%%%%%%%%%%%%%%%%%%%%%%%%%%%%%%%%%%%%%%%%%%%%%%%%%%%%%%%%%%%%%%%%%%%%

\section*{acknowledgments}
\small
We warmly thank the anonymous referee for useful suggestions that
significantly impreved our work.
Support for Hubble Space Telescope proposal GO-13297 was provided by
NASA through grants from STScI, which is operated by AURA, Inc., under
NASA contract NAS 5-26555. 
This work has received funding from the European Research Council
(ERC) under the European Union's Horizon 2020 research innovation
programme (Grant Agreement ERC-StG 2016, No 716082 'GALFOR', PI:
Milone), and the European Union's Horizon 2020 research and innovation
programme under the Marie Sk{\l}odowska-Curie (Grant Agreement No 797100,
beneficiary: Marino). APM and MT acknowledge support from MIUR through
the FARE project R164RM93XW ``SEMPLICE''.
\bibliographystyle{aa}

\bibliographystyle{aa}

\clearpage

\clearpage

%%%%%%%%%%%%%%%%%%%%%%%%%%%%%%%%%%%%%%%%%%%%%%%%%%%%%%%%%%%%%%%%%%%%%%%%%%%%%%%
 \begin{table*}
 \caption{Description of the spectroscopic dataset used in this
   paper. For each cluster and element we provide the number of stars
   for which both spectroscopy and $HST$ photometry is available
   and the reference to the spectroscopic paper.} 
\begin{tabular}{ccccccccccccc}%[ht!]                                        
 \hline\hline
 CLUSTER & Reference                 &A(Li)&[N/Fe]&[O/Fe]&[Na/Fe]&[Mg/Fe]&[Al/Fe]&[Si/Fe]&[K/Fe]&[Ca/Fe]&[Fe/H]&[Ba/Fe]\\
 \hline
 %                                         Li   N   O  Na  Mg  Al  Si  K  Ca  Fe  Ba
 NGC\,104  & Carretta et al.\,2009       &---&---&  7& 10&---&---&---&---&---& 10&---\\
           & Carretta et al.\,2013       &---&  7&---&---& 10&  7& 10&---&---&---&---\\
           & Mucciarelli et al.\,2017    &---&---&---&---&---&---&---& 10&---&---&---\\
 NGC\,288  & Carretta et al.\,2009       &---&---&  9& 19&---&---&---&---&---& 19&---\\
 NGC\,362  & Carretta et al.\,2013       &---&---&  8&  9&  9&---&  9&---&  9&  9&  8\\
           & D'Orazi et al.\,2015        &  5&---&---&---&---&  6&---&---&---&---&---\\
 NGC\,1851 & Carretta et al.\,2010b      &---&---& 11& 12& 12&---& 12&---& 12& 12& 11\\
\hline
 \end{tabular}\\
 \label{tab:data}
 \end{table*}
%%%%%%%%%%%%%%%%%%%%%%%%%%%%%%%%%%%%%%%%%%%%%%%%%%%%%%%%%%%%%%%%%%%%%%%%%%%    

\onecolumn
 \begin{longtable}{ccccccccccccc}%[ht!]                                        
 \caption{ For each element and GC, we provide the average abundance for 1G stars, 2G stars, blue-RGB, and red-RGB stars. The corresponding r.m.s. and the number of used stars are also listed.}\label{tab:abb}\\ \hline\hline
          &       & 1G   &     &         & 2G   &    &         &Blue-RGB&    &         & Red-RGB& \\
\hline
\endfirsthead
\caption{continued.}\\\hline\hline
          &       & 1G   &     &         & 2G   &    &         &Blue-RGB&    &         & Red-RGB& \\
\hline                                                                   
\endhead
\endfoot
 A(Li)         &         &      &       &          &      &      &           &      &      &          &      &     \\
\hline          
NGC\,0362         &  0.91   &  --- &  1  &    1.04 & 0.06 &  3 &    1.00 & 0.08 &  4 &    0.80 & 0.00 &  1 \\ 
NGC\,2808         &  1.18   & 0.03 &  2  &    1.07 & 0.09 &  6 &    1.10 & 0.09 &  8 &     --- &  --- &  0 \\ 
NGC\,5904         &  0.99   & 0.07 &  4  &    0.82 &  --- &  1 &    0.96 & 0.10 &  5 &     --- &  --- &  0 \\ 
NGC\,6121         &  1.29   & 0.07 &  4  &    1.31 & 0.07 &  4 &    1.30 & 0.06 &  8 &     --- &  --- &  0 \\ 
\hline          
 [N/Fe]         &         &      &       &          &      &      &           &      &      &          &      &     \\
\hline          
NGC\,0104         &  0.94   & 0.05 &  2  &    1.12 & 0.03 &  5 &    1.07 & 0.10 &  7 &     --- &  --- &  0 \\ 
NGC\,6121         &  0.93   & 0.15 &  3  &    1.25 & 0.18 &  5 &    1.13 & 0.23 &  8 &     --- &  --- &  0 \\
\hline
 \end{longtable}
%%%%%%%%%%%%%%%%%%%%%%%%%%%%%%%%%%%%%%%%%%%%%%%%%%%%%%%%%%%%%%%%%%%%%%%%%%%    

\twocolumn

\onecolumn
\onecolumn
\scriptsize
\begin{longtable}{ccrrcrrcrrcrr}%[ht!]                                        
\caption{For each cluster we provide the Spearman's rank correlation coefficient between the abundance of the various elements and $\Delta_{\rm F275W, F814W}$, ($r1$), and between the abundance and $\Delta_{C \rm F275W, F336W, F438W}$, ($r2$). We have considered four groups including all stars, blue-RGB stars, 1G stars and 2G stars. We provide the number of used stars $N$ for each group.}\label{tab:cor}\\ \hline\hline
  ID    &      & All             &       &      &   blue-RGB  &       &      &     1G   &         &      &        2G   &         \\                
        & $N$  &        $r1$     & $r2$  & $N$  &        $r1$ & $r2$  & $N$  &     $r1$ & $r2$    & $N$  &        $r1$ & $r2$    \\ 
\hline
\endfirsthead
\caption{continued.}\\\hline\hline
  ID    &      & All             &       &      &   blue-RGB  &       &      &     1G   &         &      &        2G   &         \\                
        & $N$  &        $r1$     & $r2$  & $N$  &        $r1$ & $r2$  & $N$  &     $r1$ & $r2$    & $N$  &        $r1$ & $r2$    \\ 
\hline                                                                   
\endhead
\endfoot
A(Li)           &    &                  &                  &    &                  &                  &    &                  &                  &    &                  &                     \\  
\hline
NGC\,0362       &  5 & $-$0.50$\pm$0.40 & $-$0.40$\pm$0.50 &  4 &        ---       &          ---     &  1 &          ---     &          ---     &  3 &          ---     &        ---          \\
NGC\,2808       &  8 &    0.79$\pm$0.17 & $-$0.79$\pm$0.14 &  8 &    0.79$\pm$0.17 & $-$0.79$\pm$0.14 &  2 &          ---     &          ---     &  6 &    0.77$\pm$0.17 & $-$0.77$\pm$0.23    \\
NGC\,5139       & 41 & $-$0.15$\pm$0.15 & $-$0.34$\pm$0.16 & 13 &    0.54$\pm$0.21 & $-$0.54$\pm$0.30 & 3 &      ---      &      ---      & 10 &    0.66$\pm$0.15 & $-$0.84$\pm$0.12 \\
NGC\,5904       &  5 &    0.90$\pm$0.10 & $-$0.90$\pm$0.20 &  5 &    0.90$\pm$0.10 & $-$0.90$\pm$0.20 &  4 &          ---     &          ---     &  1 &          ---     &        ---          \\
NGC\,6121       &  8 &    0.29$\pm$0.38 &    0.33$\pm$0.31 &  8 &    0.29$\pm$0.38 &    0.33$\pm$0.29 &  4 &          ---     &          ---     &  4 &          ---     &        ---          \\   
\hline
[N/Fe]          &    &                  &                  &    &                  &                  &    &                  &                  &    &                  &                     \\  
\hline
NGC\,0104       &  7 & $-$0.36$\pm$0.39 &    0.93$\pm$0.11 &  7 & $-$0.36$\pm$0.39 &    0.93$\pm$0.11 &  2 &          ---     &          ---     &  5 &    0.00$\pm$0.60 &    0.90$\pm$0.20    \\
\hline
 \end{longtable}
%%%%%%%%%%%%%%%%%%%%%%%%%%%%%%%%%%%%%%%%%%%%%%%%%%%%%%%%%%%%%%%%%%%%%%%%%%%    

\twocolumn
  
\end{document}